\documentclass[preprint,prd,aps,showpacs,showkeys,nofootinbib]{revtex4}
\usepackage{cancel}
\usepackage{ulem}
\usepackage{color}
\usepackage{CJK}
\usepackage{graphicx}
\usepackage{dcolumn}
\usepackage{bm}
\usepackage[colorlinks=true,linkcolor=blue,citecolor=blue, urlcolor=blue]{hyperref}

\begin{document}


\title{On the possibility of mixed axion/neutralino dark matter in specific SUSY DFSZ axion models}

\author{Zhong-Jun Yang$^{1}$\footnote{zj\_yang@cqu.edu.cn}, Tai-Fu Feng$^{1,2}$\footnote{fengtf@hbu.edu.cn}, Xing-Gang Wu$^{1}$\footnote{wuxg@cqu.edu.cn}}

\affiliation{$^1$ Department of Physics, Chongqing Key Laboratory for Strongly Coupled Physics, Chongqing University, Chongqing 401331, China\\
$^2$ Department of Physics, Key Laboratory of High-precision Computation
and Application of Quantum Field Theory of Hebei Province, Hebei University, Baoding 071002, China}
\date{\today}

\begin{abstract}
We introduce four supersymmetric (SUSY) axion models in which the strong CP problem and the $\mu$ problem are solved with the help of the Peccei-Quinn mechanism and the Kim-Nilles mechanism, respectively. The axion physics enriches the SUSY model by introducing axion as a dark matter candidate and, therefore, the lightest supersymmetric particle (LSP) could just be a part of the total dark matter. For this reason, axion relieves the tensions between SUSY models and numerous experimental measurements, such as the dark matter direct detection experiments and the precise measurements of anomalous magnetic moment of the muon $a_\mu$. In the present paper, we investigate the constraints imposed by the latest $a_\mu$ measurements and LUX-ZEPLIN (LZ) experiment on the relic density of the Higgsino-like LSP. Additionally, we consider the constraints arising from the cosmology of saxions and axinos, and their impacts on the parameter space of our models are carefully examined. For the axion constituting the remaining portion of dark matter, we find that the conventional misalignment mechanism can successfully account for the correct dark matter relic density observed by the Planck satellite.
\end{abstract}


\maketitle
\newpage
\tableofcontents
\newpage

\section{Introduction}
The discovery of the $125~\rm{GeV}$ Higgs boson at the Large Hadron Collider (LHC) in 2012 \cite{125GeV1,125GeV2} is an important moment in the history of human exploration, and the Standard Model (SM) of particle physics becomes the most successful theory to date. However, this elegant theory is still plagued by several problems, and for this reason many theoretical physicists are calling for the new physics (NP) beyond the SM. The Minimal Supersymmetric Standard Model (MSSM), which is a famous SUSY extension of the SM, could solve the so-called hierarchy problem that besets the SM, and simultaneously unifies the gauge coupling constants. Unfortunately, even in this model, the strong CP problem in the QCD sector has not been well understood yet. We were puzzled by this problem for a long time until Peccei and Quinn put forward their genius idea which is called the Peccei-Quinn (PQ) mechanism~\cite{PQM}. Reviews on this mechanism can be found in Refs.~\cite{review1, review2, review3, review4, review5, review6, review7, review8, AC, landscape}. Under the global $\rm{U(1)_{PQ}}$ symmetry, a pseudo-Nambu-Goldstone boson axion denoted by $A$ in our work appears~\cite{PQWW1,PQWW2}, and this particle provides a large number of valuable research topics in phenomenology and cosmology. Particularly, this particle can serve as the dark matter we have been looking for for many years. Since this mechanism was proposed, various models based on the SM and involving PQ mechanism, such as PQWW~\cite{PQWW1,PQWW2}, DFSZ~\cite{DFSZ1,DFSZ2}, and KSVZ~\cite{KSVZ1,KSVZ2}, have sprung up. Many of them, however, were quickly ruled out by experiments, while the invisible axion models are still alive. In a SUSY model, such an invisible axion could also be introduced by establishing DFSZ or KSVZ type interactions. The DFSZ type with the Kim-Nilles mechanism \cite{KNM}, for example, is particularly attractive, since it's also a solution to the $\mu$ problem. In MSSM, the Higgsino mass parameter $\mu$ is introduced through the superpotential term $\mu H_u H_d$, where $H_u$ and $H_d$ are the two Higgs doublet superfields. Theoretically, the natural value of this SUSY-conserving $\mu$ parameter would seem to be of the order of the Planck scale. For consistency with phenomenology, however, $\mu$ is required to be of the order of the weak scale. This is known as the SUSY $\mu$ problem. To solve this problem, the Kim-Nilles mechanism first forbids the $\mu$ term via PQ symmetry, and then regenerates it once the PQ symmetry is spontaneously broken.

Axion also brings lots of new considerations and restrictions to the model. For instance, the domain wall (DW) number of the model and the quality of axion are needed to be taken into account. The DW problem~\cite{landscape} could be understood by analyzing the periodic potential of axion. As the temperature of Universe falls to a value $T \sim\Lambda_{\rm QCD}$, the axion potential is lifted by the non-perturbative QCD effects and the axion mass switches on. Since the axion $A$, being an angular variable, can take values in the interval $[0, 2\pi v_{A})$ and the period of axion potential is $2\pi v_{A}/N_{\rm{DW}}$, there would be $N_{\rm{DW}}$ degenerate vacua. This leads to different vacua in different patches of the Universe, and then DWs will form at the boundaries between regions of different vacua. These DWs give nothing but disappointing cosmological events such as another inflation. In the pre-inflationary scenario~\cite{landscape}, however, the model is free from topological defects, so there would be no DW problem. Otherwise, the DW number $N_{\rm{DW}}$ needs to be one to avoid this cosmological problem. The quality problem \cite{landscape,QP1,QP2,QP3,QP4,QP5,QP6} originates from the fact that the PQ symmetry, which is a global symmetry, is not fundamental in a quantum field theory. As we know, global symmetry is not respected by quantum gravitational effects~\cite{QGE1,QGE2}. Therefore, this PQ symmetry has to be preserved to a great degree of accuracy as long as we consider it to be the key to solving the strong CP problem.

In the framework of SM, there is also no candidate for dark matter. Up to date, the dark matter relic density is about $\Omega_{\rm{DM}} h^2\approx0.12$ which is observed by the Planck satellite~\cite{Planck}. For the sake of explaining this observation, the dark matter candidate must appear in the NP models. In SUSY models, for example, the LSP, which is a weakly interacting massive particle (WIMP), can be the dark matter candidate. This WIMP (typically the lightest neutralino) could give a very appropriate annihilation cross section for cold dark matter freeze-out production mechanism, and this is known as the ``WIMP miracle''. However, the dark matter direct detection experiments have ruled out a large amount of parameter space that was once considered promising. Besides, the latest measurements of the anomalous magnetic moment of the muon $a_\mu$ given by the Brookhaven National Laboratory (BNL) E821 measurement \cite{Muong-2:2006rrc} and the Fermilab National Accelerator Laboratory Muon Experiment (FNAL) \cite{g-2 SM EXP} show that the discrepancy between the measured and SM predicted $a_\mu$ is $4.2\sigma$, which is also a problem that needs to be solved in the NP models. It is worth noting that the result of Budapest-Marseille-Wuppertal (BMW) lattice collaboration~\cite{Kotov:2023wug,Toth:2022lsa,BMW:2022rdk,Budapest-Marseille-Wuppertal:2017sdk,Budapest-Marseille-Wuppertal:2018ivi}, which includes both QED leading contribution and QCD leading-order hadronic vacuum polarization (LO-HVP) contribution, indicates that the discrepancy between the standard model's prediction for the muon $g-2$ and its experimental value can be markedly alleviated and, therefore, the corrections from NP models are not necessary. Given that this result still awaits independent confirmation by other research collaborations, we do not intend to give a further discussion in this work.

Taking into consideration all the aforementioned points, we introduce four SUSY axion models, in which all these problems can be overcome or alleviated. The remainder of this paper is organized as follows: In Sec.~\ref{Sec II}, we provide an introduction to our models, along with a discussion on the quality problem. In Sec.~\ref{Sec III}, we commence with an analysis of the constraints arising from the $a_\mu$ measurements, dark matter direct detection experiments and LHC on the abundance of the LSP, which, in this paper, is only a small part of the total dark matter. Subsequently, we explore the constraints imposed by the cosmology of the saxions and axinos on the parameter space of the models, and discuss the potential cosmological moduli problem~\cite{Coughlan:1983ci,Banks:1993en,deCarlos:1993wie,Endo:2006zj}. Lastly, we take the constraints from the effective coupling constant $g_{A\gamma\gamma}$ on the axion decay constant $f_A$ into account. Sec.~\ref{Sec IV} is reserved as a summary of our findings. In the Appendix, we give a brief review of the conventional misalignment mechanism (CMM)~\cite{CMM1,CMM2,CMM3} and discuss the feasibility of the kinetic misalignment mechanism (KMM)~\cite{KMM1,KMM2} in our models.
\section{Axion models}\label{Sec II}
\subsection{Superpotential}
In our SUSY model, the $\mu$ term is absent, since the Kim-Nilles mechanism have been
used to solve the $\mu$ problem. The superpotential of the model is given by
\begin{eqnarray}
W_{\rm{P Q_{1}}} & = & \frac{\lambda_{\mu}}{M_{P}} X^{2} H_{u} H_{d}+y_{u} H_{u} q \bar{u}-y_{d} H_{d} q \bar{d}-y_{e} H_{d} \ell \bar{e}+y_{\bar{\nu}} H_{u} \ell \bar{\nu}+M_{\bar{\nu}} \bar{\nu} \bar{\nu},
\end{eqnarray}
where $M_P=2.4\times10^{18}\rm{GeV}$ is the reduced Plank scale. $X$ is a gauge-singlet chiral superfield, and $H_{u}$ as well as $ H_{d} $ are the MSSM Higgs doublet chiral superfields. The $q$ $\overline u$ $\overline d$ $\ell$ $ \overline e$ are the same as the MSSM quark and lepton chiral superfields. The last two terms are the seesaw terms of gauge-singlet neutrino superfield $ \bar{\nu} $. In order to stabilize the potential at large field values, we need to add the $ \lambda $ terms involving only the gauge-singlet chiral superfields into the superpotential. The cubic $ \lambda $ term $ \lambda X^{3} $ would explicitly violate the PQ symmetry, so that we need at least two gauge-singlet chiral superfields to stabilize the potential. The cubic $ \lambda $ terms such as $ \lambda X^{2} Y $, where $ Y $ is also a gauge-singlet chiral superfield like $ X $, will lead to a very small dimensional parameter $ a_{\lambda} $ (corresponding to the soft breaking term of the $ \lambda $ term). Typically, such a dimensional parameter should be of the order of $ m_{\text {soft }} $. In this sense, a quadratic $ \lambda $ term is more appropriate, since in this case there is no such concern. In literatures, e.g., Refs.~\cite{Bae:2019dgg,Babu:2002ic,Baer:2018avn}, the $ \lambda $ term can take the form $ \frac{\lambda}{M_{P}} X^{3} Y, \frac{\lambda}{M_{P}} X^{2} Y^{2} $ or $ \frac{\lambda}{M_{P}} X Y^{3} $, and a lot of researches have been done based on them. In this work, we consider that the $ \lambda $ terms involve three gauge-singlet chiral superfields $ X, Y $ and $ Z $, as it is also a possible scenario even though not the most economical~\footnote{In the work~\cite{S P M} carried out by Bhattiprolu and Martin, which mainly focuses on solving SUSY $\mu$ problem and the axion quality problem using two fields ($X$ and $Y$), the authors state that ``For simplicity and economy, we restrict our attention to only two such fields. Although it is possible to have more than two, this would seem to make it harder to find solutions to the axion quality problem to be discussed shortly.". This motivates our present work, and we have tried the case of adding only one additional filed $Z$, e.g. one of the purposes of our work is to supplement their research to some extent. Additionally, we are interested in whether the model with $Z$ field is workable or not, and many other aspects related to the axion have also been addressed.}. The $ \lambda $ terms we assumed have the following form
\begin{eqnarray}
W_{\rm{P Q_{2}}} & = & \frac{\lambda_{1}}{M_{P}} X^{\alpha_{\mathrm{w}}} Y^{4-{\alpha_{\mathrm{w}}}}+\frac{\lambda_{2}}{M_{P}} X^{\beta_{\mathrm{w}}} Z^{4-{\beta_{\mathrm{w}}}} \text {, }
\end{eqnarray}
where the indexes $ \alpha_{\mathrm{w}} $ and $ \beta_{\mathrm{w}} $ could be $1,2$ or $3 $, the subscript $``\rm{w}"$ denotes the superpotential $W$. If $ \alpha_{\mathrm{w}} $ or $ \beta_{\mathrm{w}} $ is $4 $, the PQ symmetry would be explicitly violated like the $ \lambda X^{3} $ case. To avoid repetitive discussions, we study three cases that are different from each other: $ (3,2),(3,1) $ and $ (2,1) $ for $ (\alpha_{\mathrm{w}}, \beta_{\mathrm{w}}) $. Here we have assumed that the PQ charges of $ X, Y $ and $ Z $ are different from each other, since there would be more terms appearing in $ W_{\rm{P Q_{2}}} $ if that is not the case. There does exist one exception that $ \alpha_{\mathrm{w}}=\beta_{\mathrm{w}}=3 $, however, such a model predicts a massless axino which is not what we expected. In the following, we will refer to these models with $ (3,2),(3,1) $ and $ (2,1) $ as base models $\rm{B_I}, \rm{B_{II}}$ and $\rm{B_{III}}$, respectively.

Up until now, we can write down the soft supersymmetry-breaking terms
\begin{eqnarray}
\mathcal{L}_{\mathrm{soft}}  = &&\left(\frac{a_{\mu}}{M_{P}} X^{2} H_{u} H_{d}+\frac{a_{1}}{M_{P}} X^{\alpha_{\mathrm{w}}} Y^{4-\alpha_{\mathrm{w}}}+\frac{a_{2}}{M_{P}} X^{\beta_{\mathrm{w}}} Z^{4-\beta_{\mathrm{w}}}+\text { h.c. }\right)\nonumber \\
&&-m_{X}^{2}|X|^{2}-m_{Y}^{2}|Y|^{2}-m_{Z}^{2}|Z|^{2}+\cdots.
\end{eqnarray}
For the VEV of $X$ is of the order of $10^{9-12}~\rm{GeV}$, we can chose the reasonable values of $\lambda_\mu$ and $a_\mu$ to solve the $\mu$ problem:
\begin{eqnarray}
\mu = \frac{\lambda_\mu}{M_P} \langle X^2 \rangle,
b = \frac{a_\mu}{M_P} \langle X^2 \rangle.
\end{eqnarray}

We can also add pairs of vectorlike quarks and leptons superfields, which are denoted by capital letters (such as $Q+\overline Q$) to the model, which could be used to solve the DW problem. In this study, we discuss models without as well as with vectorlike superfields. In the latter case, we just add superpotential $W_{\rm{PQ_3}}$ to model $\rm{B_I}$ and denote this extension as model $\rm{E_I}$ (we do not add any vectorlike superfields to the model $\rm{B_{II}}$ or $\rm{B_{III}}$ in this work):
\begin{eqnarray}
W_{\rm{P Q_{3}}} & = & \lambda_{Q} X Q \bar{Q}+\lambda_{U} X U \bar{U}+\lambda_{E} X E \bar{E}+\frac{\lambda_{D}}{ M_{P}} X^{2} D \bar{D}+\frac{\lambda_{L}}{ M_{P}} X^{2} L \bar{L} .
\end{eqnarray}
The information about color and the other quantum number of them is shown in Table \ref{tab:additionalfields}. Besides, these additional chiral superfields in ${\bf 5}+{\bf \overline 5} = D + \overline D + L + \overline L$ and ${\bf 10} + {\bf \overline{10}} = Q + \overline Q + U + \overline U + E + \overline E$ representations of the $\rm{SU(5)}$ grand unified theory~\cite{Georgi:1974sy} can preserve the unification of gauge couplings as in the MSSM~\cite{S P M}. Nevertheless, we do not assume that $ \rm{SU(5)}$ is the unbroken gauge group in the ultraviolet.
\begin{table}[htb]
\begin{center}
\begin{minipage}[]{0.95\linewidth}
\caption{Vectorlike pairs of chiral superfields $\Phi + \overline \Phi$ added to the model and their Standard Model gauge transformation properties.
These pairs will carry non-zero net PQ charges.\label{tab:additionalfields}}
\end{minipage}

\vspace{0.2cm}

\begin{tabular}{|c c|}
\hline
~Superfields~  &  ~$\rm{SU(3)_C \bigotimes SU(2)_L \bigotimes U(1)_Y}$~\\
\hline
\hline
~$Q + \overline{Q}$~  &  ~$({\bf 3}, {\bf 2}, 1/6)$ + $({\bf \overline{3}}, {\bf 2}, -1/6)$~\\
~$U + \overline{U}$~  &  ~$({\bf 3}, {\bf 1}, 2/3)$ + $({\bf \overline{3}}, {\bf 1}, -2/3)$~\\
~$E + \overline{E}$~  &  ~$({\bf 1}, {\bf 1}, -1)$ + $({\bf 1}, {\bf 1}, 1)$~\\
~$D + \overline{D}$~  &  ~$({\bf 3}, {\bf 1}, -1/3)$ + $({\bf \overline{3}}, {\bf 1}, 1/3)$~\\
~$L + \overline{L}$~  &  ~$({\bf 1}, {\bf 2}, -1/2)$ + $({\bf 1}, {\bf 2}, 1/2)$~\\
\hline
\end{tabular}
\end{center}
\end{table}
\subsection{Axion properties in our models}
Following Refs.~\cite{landscape,S P M}, in general QCD axion models, the PQ symmetry can be spontaneously broken by the non-zero vacuum expectation values (VEVs) of some complex scalars $\phi_s$ with PQ charge $Q_s$. Before the breaking of PQ symmetry, the Lagrangian is invariant under the PQ transformation $\phi_s \rightarrow e^{i Q_s \alpha} \phi_s$. These complex scalars can be parameterized as $\phi_s = \frac{1}{\sqrt{2}} (v_s + \rho_s) e^{i a_s/v_s}$, where $\rho_s$ and $a_s$ are radial and angular fields respectively. The scalar part of PQ current can then be written as
\begin{eqnarray}
j^{\rm \mu}_{\text{PQ}} = - i {\sum_s} Q_s
\bigl ( \phi_s^\dag \partial^\mu \phi_s - \phi_s \partial^\mu \phi_s^\dag \bigr )={\sum_{s}} Q_s v_s \partial^\mu a_s=v_A \partial^\mu A,
\end{eqnarray}
where $v_A^2 = {\sum_{s}} Q_s^2 v_s^2$ and the axion field $A = \frac{1}{v_A} \sum_{s} Q_s v_s a_s$. Note that the ``physical" PQ charges of scalars are ``unique'' and need to be deduced by imposing an orthogonality condition
\begin{eqnarray}
{\sum_{s}} Y_s Q_s v_s^2 = 0,
\label{eq:Z_ortho}
\end{eqnarray}
where $Y_s$ is the $\rm{U(1)_Y}$ charge of the scalar $\phi_s$. This condition ensures that there is no kinetic mixing between the physical axion and the $Z$ boson. The anomalous divergence of the PQ current can be written as
\begin{eqnarray}
\partial_{\mu}j^{\mu}_{\rm{PQ}} &=&
\frac{g_3^2}{16 \pi^2} N G^a_{\mu \nu} \widetilde{G}^{a \mu \nu}+ \frac{e^2}{16 \pi^2} E F_{\mu \nu} \widetilde{F}^{\mu \nu},
\label{eq:PQdivergence}
\end{eqnarray}
where $N$ is the $\rm{U(1)_{\text{PQ}}}$-$\rm{SU(3)_C}$-$\rm{SU(3)_C}$ anomaly coefficient and $E$ is $\rm{U(1)_{\text{PQ}}}$-$\rm{U(1)_{ EM}}$-$\rm{U(1)_{EM}}$ anomaly coefficient. Then, the effective Lagrangian of axion is
\begin{eqnarray}
\mathcal{L}_{A} &\supset&
\frac{1}{2} \partial_\mu A \partial^\mu A
+ \frac{g_3^2}{16 \pi^2} \frac{A}{v_A} N G^a_{\mu \nu} \tilde{G}^{a \mu \nu}
+  \frac{e^2}{16 \pi^2} \frac{A}{v_A} E F_{\mu \nu} \tilde{F}^{\mu \nu}+\cdots \nonumber\\
&&=\frac{1}{2} \partial_\mu A \partial^\mu A
+ \frac{g_3^2}{32 \pi^2} \frac{A}{f_A} G^a_{\mu \nu} \tilde{G}^{a \mu \nu}
+ \frac{E}{N} \frac{e^2}{32 \pi^2} \frac{A}{f_A} F_{\mu \nu} \tilde{F}^{\mu \nu}+\cdots,
\phantom{xxxx}
\end{eqnarray}
where $f_A \equiv \frac{v_A}{2N}$ is the axion decay constant. Under the $\rm{U(1)_{\text{PQ}}}$ transformation:
\begin{eqnarray}
a_s \rightarrow a_s + \alpha Q_s v_s , A  \rightarrow A + \alpha v_A ,
\label{eq:axionshift}
\end{eqnarray}
where $\alpha$ is an infinitesimal parameter, the strong CP problem could be solved. The low-energy axion interaction Lagrangian under the $\Lambda_{\rm QCD}$ scale at which the axion mass switches on can be read as
\begin{eqnarray}
\mathcal{L}_{A} &=& \frac{1}{2} \partial^\mu A \partial_\mu A - \frac{1}{2} m_A^2 A^2 +
 \frac{1}{4} g_{A \gamma \gamma} A F_{\mu \nu} \tilde{F}^{\mu \nu}
+\cdots, \phantom{xxx}
\label{eq:LA_int}
\end{eqnarray}
where $g_{A \gamma \gamma} = \frac{\alpha_{em}}{2 \pi {f_A}} (\frac{E}{N}-1.92)$ and $\alpha_{em}$ is the electromagnetic coupling constant. The axion mass is given in terms of the axion decay constant $f_A$ by
\begin{eqnarray}
m_A \simeq 5.7 \frac{10^{12} \, {\text{GeV}}}{f_A} ~\mu{\rm eV}.
\label{eq:mAintermsoffA}
\end{eqnarray}

In our models, the scalars that acquire VEVs are parameterized as:
\begin{eqnarray}
&&X = \frac{1}{\sqrt{2}}(v_X+\rho_X) e^{i a_X/v_X}
,
Y = \frac{1}{\sqrt{2}}(v_Y+\rho_Y) e^{i a_Y/v_Y}
,Z = \frac{1}{\sqrt{2}}(v_Z+\rho_Z) e^{i a_Z/v_Z}
,\nonumber\\
&&H^0_u = \frac{1}{\sqrt{2}}(v_u+\rho_u) e^{i a_u/v_u}
,
H^0_d = \frac{1}{\sqrt{2}}(v_d+\rho_d) e^{i a_d/v_d}
,
\end{eqnarray}
here $H^0_u$ and $H^0_d$ are neutral MSSM Higgs scalars, and $a_X, a_Y, a_Z, a_u$ and $a_d$ are pseudo-scalar bosons that contribute to the axion. Choosing the PQ charges of $X$ to be $Q_{X}=1$, then we get $Q_Y =\frac{\alpha_{\mathrm{w}}}{\alpha_{\mathrm{w}}-4}$ and $Q_Z =\frac{\beta_{\mathrm{w}}}{\beta_{\mathrm{w}}-4}$. Considering the first term of $ W_{P Q_1}$, the PQ charges of the MSSM Higgses should satisfy
\begin{eqnarray}
Q_{H_u} + Q_{H_d} &=& -2.
\label{eq:Higgs PQ charge condition}
\end{eqnarray}
Imposing the orthogonality condition of Eq.~(\ref{eq:Z_ortho}), these models yield $\frac{Q_{H_d}}{Q_{H_u}} = \frac{v_u^2}{v_d^2} \equiv \tan^2 \beta$, where $s_\beta \equiv \sin \beta = v_u/v$, $c_\beta \equiv \cos \beta = v_d/v$, and $v = \sqrt{v_u^2 + v_d^2}$. Combining Eq.~(\ref{eq:Higgs PQ charge condition}), we get
\begin{eqnarray}
Q_{H_u} = - 2 c_\beta^2,
\qquad
Q_{H_d} = - 2 s_\beta^2.
\end{eqnarray}

In SUSY models, each left-handed Weyl fermion $\Psi_f$ possesses a PQ charge $Q_{f}$ determined by the interaction terms with the PQ charged scalars, and the PQ transformations of fermions are $\Psi_f \rightarrow e^{i Q_f \alpha} \Psi_f$. For the neutrino superfield $ \bar{\nu} $, the PQ charge is zero due to the last term of $ W_{P Q_{1}} $, and so that $ Q_{\ell}=-Q_{H_{u}}=2 c_{\beta}^{2} $. In Table \ref{tab:PQ charges}, all the PQ charges of MSSM superfields are shown. The PQ charges $ Q_{\Phi \bar{\Phi}} $ in model $\rm{E_I}$ can also be determined without difficulty, and they are listed in Table \ref{tab:vectorlike PQ charges}.
\begin{table}[htb]
\begin{center}
\begin{minipage}[]{0.95\linewidth}\caption{The charges of the Peccei-Quinn symmetry of the superfields. \label{tab:PQ charges}}
\end{minipage}

\vspace{0.3cm}

\begin{tabular}{|c|c|c|c|c|c|c|c|c|c|c|c|}
\hline &  ~$X$~  &  ~$Y$~  &  ~$Z$~  &  ~$H_{u}$~  &  ~$H_{d}$~  &  ~~$q$~~  &  ~~$\ell$~~  &  ~$\bar{u}$~  &  ~$\bar{d}$~  &  ~$\bar{e}$~  &  ~~$\bar{\nu}$~~  \\
\hline \hline $\rm{PQ}$ & $1$ &  $\frac{\alpha_{\mathrm{w}}}{\alpha_{\mathrm{w}}-4}$  &  $\frac{\beta_{\mathrm{w}}}{\beta_{\mathrm{w}}-4}$  &  $-2 c_{\beta}^{2}$  &  $-2 s_{\beta}^{2}$  &  $Q_{q}$  &  $2 c_{\beta}^{2}$  &  $2 c_{\beta}^{2}-Q_{q}$  &  $2 s_{\beta}^{2}-Q_{q}$  &  $2 s_{\beta}^{2}-2 c_{\beta}^{2}$  & $0$ \\
\hline
\end{tabular}
\end{center}
\end{table}
\begin{table}[htb]
\begin{center}
\begin{minipage}[]{0.95\linewidth}\caption{The charges of the Peccei-Quinn symmetry of the vectorlike superfields. \label{tab:vectorlike PQ charges}}
\end{minipage}

\vspace{0.3cm}

\begin{tabular}{|c|c|c|c|c|c|}
\hline
& $Q_{Q \bar{Q}}$ & $Q_{U \bar{U}}$ & $Q_{E \bar{E}}$ & $Q_{L \bar{L}}$ & $Q_{D \bar{D}}$ \\
\hline \hline
$\rm{PQ}$ & $-1$ & $-1$ & $-1$ & $-2$ & $-2$ \\
\hline
\end{tabular}
\end{center}
\end{table}

In addition to the information about PQ charges of superfields, the anomaly coefficients $N$ and $E$ of our models can be deduced here~\cite{S P M}
\begin{eqnarray}
N &=&n_g(\frac{1}{2} 2Q_{q}+\frac{1}{2} Q_{\overline u}+\frac{1}{2} Q_{\overline d}) + \frac{1}{2} (2{\textstyle\sum} Q_{Q\overline Q}) + \frac{1}{2} {\textstyle\sum} Q_{U \overline U} +
\frac{1}{2} {\textstyle\sum} Q_{D \overline D},
\label{eq:Nforextendedmodel}
\\
E &=&n_g[3(\frac{2}{3})^2 (Q_{q}+ Q_{\overline u})+3(-\frac{1}{3})^2 (Q_{q}+ Q_{\overline d})+(-1)^2 (Q_{l}+ Q_{\overline e})]
+(\pm1)^2(Q_{\widetilde{H_u}}+Q_{\widetilde{H_d}})
\nonumber\\
 &&+ 3[(\pm\frac{2}{3})^2+(\pm\frac{1}{3})^2] {\textstyle\sum} Q_{Q\overline Q} + 3(\pm\frac{2}{3})^2 {\textstyle\sum} Q_{U \overline U}
+ 3(\pm\frac{1}{3})^2 {\textstyle\sum} Q_{D \overline D}\nonumber\\
&&+(\pm1)^2 {\textstyle\sum} Q_{L \overline L}
+ (\pm1)^2{\textstyle\sum} Q_{E \overline E},
\phantom{xxxx}
\label{eq:Eforextendedmodel}
\end{eqnarray}
where $n_g=3$ is the number of chiral quark and lepton generations. Now we have
\begin{eqnarray}
f_A \>=\> \frac{v_A}{2N}, v_A =\left [Q_X^2 v_X^2 + Q_Y^2 v_Y^2+ Q_Z^2 v_Z^2 + 4 s_\beta^2 c_\beta^2 v^2\right ]^{1/2},
\label{eq:fA}
\end{eqnarray}
with the contribution of $ 4 s_{\beta}^{2} c_{\beta}^{2} v^{2} $ being numerically negligible due to the condition $v_X, v_Y, v_Z \gg v_u, v_d$ which is required in an invisible axion model.

After the PQ symmetry breaking, a discrete subgroup $Z_{N_{\rm DW}} (= e^{2 k \pi i/N_{\rm DW}}$, $k = 0, 1, \ldots, N_{\rm DW} - 1)$ is left unbroken. $N_{\rm DW}$ is the DW number that corresponds to the number of inequivalent degenerate minima of the axion potential. With the above definition of anomaly coefficient $N$, the DW number can be computed \cite{DW} as
\begin{eqnarray}
N_{\rm{DW}} \equiv \textrm{minimum integer} \left (2 N \sum_{s} \frac{n_s Q_s v_s^2}{v_A^2}\right ),
\label{eq:DWnumber}
\end{eqnarray}
where $n_s \in {Z}$. For the models $\rm{B_{I}},\rm{B_{II}}$ and $\rm{B_{III}}$, we have $ N=3, E=6 $, and the domain wall number $ N_{\rm{D W}}$ are $2N, 6N$, and $6N$ respectively. We also obtain that $ N=\frac{1}{2}$ and $ E=-\frac{2}{3} $ for model $\rm{E_{I}}$, in this case the DW number $ N_{\rm{D W}}=1 $, which means there is no DW problem.
\subsection{Discrete $R$ symmetries $Z^R_n$ to protect $\rm{U(1)_{P Q}}$}
Since the PQ symmetry is an ungauged symmetry, it is not respected by quantum gravitational effects. As a result, there are some allowed higher-dimensional terms which may spoil the PQ symmetry in superpotential
\begin{eqnarray}
W &=& \frac{\kappa}{M_P^{p-3}} X^{i} Y^{j} Z^{p-i-j},
\label{eq:XaYbsuperpotential}
\end{eqnarray}
with dimensionless parameter $\kappa$. Terms like this may result in the failure of the solution to the strong CP problem, because they can displace the QCD parameter $\theta$ away from $0$. Numerous studies, e.g., Refs.~\cite{Babu:2002ic,Babu:2002tx,Choi:2022fha,Baer:2018avn}, have been performed and many solutions have been put forth to address this quality problem. In these solutions, the scheme resorting to a discrete $R$ symmetry $Z^R_n$, which is widely used to protect the global $\rm{U(1)_{P Q}}$ symmetry, is very attractive and, therefore, is also employed in this work.

If the superpotential of a model respects an Abelian discrete $R$ symmetry $Z_{n}^R$, which can be a subgroup of an anomaly-free continuous $\rm{U(1)}$ symmetry that is spontaneously broken by a scalar field VEV of charge $n$~\cite{charge n}, and such a discrete symmetry can forbid PQ-violating terms up to some mass dimension, the PQ symmetry can be seen as an accidental consequence of the discrete symmetry and the model perhaps give a high quality axion. Under such an assumption, the $Z_{n}$ $\times$ $\rm{U(1)_{Y}}$ $\times$ $\rm{U(1)_{Y}}$, $Z_{n}$ $\times$ $\rm{SU(2)_{L}}$ $\times$ $\rm{SU(2)_{L}}$ and $Z_{n}$ $\times$ $\rm{SU(3)_{C}}$ $\times$ $\rm{SU(3)_{C}}$ anomalies, which denoted by $A_1$ $A_2$ and $A_3$ respectively, should satisfy an anomaly-free condition involving the Green-Schwarz (GS) mechanism~\cite{GSM}:
\begin{eqnarray}
\frac{A_1 + m_1 n}{5 k_1}
\,=\,
\frac{A_2 + m_2 n}{k_2}
\,=\,
\frac{A_3 + m_3 n}{k_3}
\,=\,
\rho_{\rm GS}.
\label{GS}
\end{eqnarray}
In Eq.~(\ref{GS}), $\rho_{\rm GS}$ is a constant, and $m_1$, $m_2$, $m_3$ are integers. $k_1$ can be arbitrary, and $k_2$, $k_3$ should be positive integer Kac-Moody levels. In this paper, we take an assumption that $k_1$ $=$ $k_2$  $=$ $k_3$ $=$ $1$ under which gauge coupling unification can be achieved \cite{S P M}. Then, anomaly-free condition Eq.~(\ref{GS}) can be simplified as
\begin{eqnarray}
\frac{1}{5}A_1 = A_2 = A_3 \quad\mbox{(mod $n$)}.
\end{eqnarray}
The expressions of $A_1$, $A_2$ and $A_3$ read as
\begin{eqnarray}
A_{1}  = && n_{g}\left(z_{q}+3 z_{\ell}+8 z_{\bar{u}}+2 z_{\bar{d}}+6 z_{\bar{e}}-20 R\right)+3\left(z_{H_{u}}+z_{H_{d}}-2 R\right)\nonumber \\
&&+\Delta_{Q \bar{Q}}+3 \Delta_{L \bar{L}}+8 \Delta_{U \bar{U}}+2 \Delta_{D \bar{D}}+6 \Delta_{E \bar{E}}, \nonumber\\
A_{2}  = && 4 R+n_{g}\left(3 z_{q}+z_{\ell}-4 R\right)+z_{H_{u}}+z_{H_{d}}-2 R+3 \Delta_{Q \bar{Q}}+\Delta_{L \bar{L}}, \nonumber\\
A_{3}  = && 6 R+n_{g}\left(2 z_{q}+z_{\bar{u}}+z_{\bar{d}}-4 R\right)+2 \Delta_{Q \bar{Q}}+\Delta_{U \bar{U}}+\Delta_{D \bar{D}},\label{A1A2A3}
\end{eqnarray}
and we have taken the same convention as Ref.~\cite{S P M}. The gauginos and the anticommuting coordinates $\theta_{\alpha}$ have $ Z_{n}^{R} $ charge $ R $, and the superpotential has a total charge $ 2 R \ (\bmod \ n) $. Unlike the continuous $ R $ symmetries, for discrete $ R $ symmetries, $ R $ is not always 1 because we have taken all $ Z_{n}^{R} $ charges for all superfields to be integers $ (\bmod \ n) $. In Eq.~(\ref{A1A2A3}), we have also considered the contributions from the vectorlike pairs of chiral superfields $ \Phi+\bar{\Phi} $ if they are added to the model, and the contribution $ \Delta_{\Phi \bar{\Phi}} $ is equal to $ z_{\Phi}+z_{\bar{\Phi}}-2 R $. For $Z_n^R$ charges of MSSM chiral superfields, we can reset $z_q$ (the $Z_n^R$ charge of the quark doublet chiral superfield $q$) to be $z_q^{'}$  = 0, since we can redefine all of the $Z_n^R$ charges by adding a multiple of 6Y, where Y is the weak hypercharge. After this treatment, the $Z_n^R$ charges of other MSSM chiral superfields are changed synchronously. We can do such a shift because the $ \rm{U(1)_{6 Y} }$ is an anomaly-free symmetry, and the $\rm{ U(1)_{6 Y} }$ charges of MSSM superfields are shown in Table \ref{tab:U16Y}.
\begin{table}
\begin{center}
\begin{minipage}[]{0.95\linewidth}\caption{The charges of the anomaly-free $\rm{U(1)_{6 Y} }$ symmetry of the MSSM superfields. \label{tab:U16Y}}
\end{minipage}

\vspace{0.3cm}

\begin{tabular}{|c|c|c|c|c|c|c|c|c|}
\hline & ~$H_{u}$~ & ~$H_{d}$~ & ~~$q$~~ & ~~$\ell$~~ & ~~$\bar{u}$~~ & ~~$\bar{d}$~~ & ~~$\bar{e}$~~ & ~~$\bar{\nu}$~~ \\
\hline \hline $6 Y$ & $3$ & $-3$ & $1$ & $-3$ & $-4$ & $2$ & $6$ & $0$ \\
\hline
\end{tabular}
\end{center}
\end{table}
\begin{table}
\begin{center}
\begin{minipage}[]{0.95\linewidth}\caption{The redefined $ Z_{n}^{R} $ charges, in terms of two integers $ h $ and $ x $, of the MSSM superfields in our models. \label{tab:znR charges}}
\end{minipage}

\vspace{0.3cm}

\begin{tabular}{|c|c|c|c|c|c|c|c|c|c|}
\hline &  ~$X$~  &  ~$H_{u}$~  &  $H_{d}$  &  ~$q$~  &  $\ell$  &  $\bar{u}$  &  $\bar{d}$  &  $\bar{e}$  &  ~$\bar{\nu}$~  \\
\hline \hline $z^{\prime}$  &  $x$  &  $h$  &  $2 R-2 x-h$  & $0$ &  $R-h$  &  $2 R-h$  &  $h+2 x$  &  $2 h+2 x-R$  & $ R $ \\
\hline
\end{tabular}
\end{center}
\end{table}

From the superpotential $ W_{\rm{P Q_{1}}} $, we can obtain the redefined $ Z_{n}^{R} $ charges of the MSSM superfields, and they are shown in Table \ref{tab:znR charges}. Using the charges of Table \ref{tab:znR charges}, for the models $\rm{B_{I}},\rm{B_{II}},\rm{B_{III}}$ as well as $\rm{E_{I}}$, one can check that the anomaly-free conditions Eq.~(\ref{GS}) all can be simplified as
\begin{eqnarray}
&&8 x+3 h+5 R   =  0(\bmod\ \mathrm{n}),\nonumber \\
&&12 x+9 h-21 R   = 0(\bmod\ \mathrm{n}) .
\label{condition gs}
\end{eqnarray}
In the cases that $W_{\rm{P Q_{3}}} $ is absent, we consider two ultraviolet completions of the theory: $\rm{ S U(5) }$ GUT and Pati-Salam $\rm{ S U(4)_{C} \otimes S U(2)_{L} \otimes U(1)_{R}} $ theory~\cite{Pati:1974yy}.

In the first case, the discrete gauge charges of the multiplets are required to be consistent with $\rm{SU(5)}$, therefore they need satisfy the following conditions:
\begin{eqnarray}
&&z_{q}   =  z_{\bar{u}} = z_{\bar{e}}(\bmod\ \mathrm{n}),\nonumber \\
&&z_{\ell}   =  z_{\bar{d}}(\bmod\ \mathrm{n}).
\end{eqnarray}
Equivalently, this implies that there exists an integer $ m $ satisfying the following conditions:
\begin{eqnarray}
&&z_{q}^{\prime}+m  =  z_{\bar{u}}^{\prime}+(-4) m  =  z_{\bar{e}}^{\prime}+6 m(\bmod\ \mathrm{n}),\nonumber \\
&&z_{\ell}^{\prime}+(-3) m  =  z_{\bar{d}}^{\prime}+2 m(\bmod\ \mathrm{n}).
\end{eqnarray}
Simplify these equations by using the charges of Table \ref{tab:znR charges}, then we can obtain:
\begin{eqnarray}
&&2 R-h-5 m   =  0(\bmod\ \mathrm{n}),\nonumber \\
&&R-2 h-2 x-5 m   =  0(\bmod\ \mathrm{n}).
\label{condition 1}
\end{eqnarray}
The $ Z_{n}^{R} $ symmetries that satisfy these conditions (Eqs.~(\ref{condition gs}) and (\ref{condition 1})) have been listed in Table \ref{tab:SU5}. From this table, we can see that the largest $p$ is 7, and the allowed PQ-violating terms in scalar potential may include the $\frac{\varphi^9}{M_P^5}$ term, where $\varphi$ can be $X, Y$ or $Z$. So that the axion quality problem may not be solved in some sense.
\begin{table}
\begin{center}
\begin{minipage}[]{0.95\linewidth}\caption{Some $ Z_{n}^{R} $ symmetries can be made consistent with $ \rm{S U(5)} $ embedding. Since the $\rm{U(1)_{6Y}}$ charge of $X$ is 0, we have $z_X=z_X^{\prime}=x$, and so do the superfields $Y$ and $Z$.\label{tab:SU5}}
\end{minipage}

\vspace{0.3cm}

\begin{tabular}{|c|c|c|c|c|c|c|c|c|}
\hline ~$\alpha_{\mathrm{w}}$~ & ~$\beta_{\mathrm{w}}$~ & ~$R$~ & ~$z_{X}$~ & ~$z_{Y}$~ & ~$z_{Z}$~ & ~~$h$~~ & ~~$n$~~ & ~~$p$~~ \\
\hline 3 & 2 & 1 & 11 & 17 & 14 & 1 & 24 & 7 \\
3 & 2 & 1 & 23 & 5 & 14 & 1 & 24 & 7 \\
3 & 1 & 1 & 11 & 17 & 21 & 1 & 24 & 6 \\
3 & 1 & 1 & 23 & 5 & 9 & 1 & 24 & 6 \\
2 & 1 & 1 & 11 & 14 & 5 & 1 & 24 & 7 \\
2 & 1 & 1 & 23 & 14 & 17 & 1 & 24 & 7 \\
\hline
\end{tabular}
\end{center}
\end{table}
\begin{table}
\begin{center}
\begin{minipage}[]{0.95\linewidth}\caption{Some $ Z_{n}^{R} $ symmetries can be made consistent with a Pati-Salam $\rm{ S U(4)_{C} \otimes S U(2)_{L} \otimes U(1)_{R}} $ embedding. Under these symmetries, the allowed higher dimensional terms all can satisfy $ p \geq 8 $. \label{SU4C}}
\end{minipage}

\vspace{0.3cm}

\begin{tabular}{|c|c|c|c|c|c|c|c|c|c|c|c|c|c|c|}
\hline
\multicolumn{5}{|c|}{ $\alpha_{\mathrm{w}}=3, \beta_{\mathrm{w}}=2, R=1, x=-3$ } & \multicolumn{5}{|c|}{ $\alpha_{\mathrm{w}}=3, \beta_{\mathrm{w}}=1, R=1, x=-3$ } & \multicolumn{5}{|c|}{ $\alpha_{\mathrm{w}}=2, \beta_{\mathrm{w}}=1, R=1, x=-3 $} \\
\hline
 ~$z_{Y}$~  &  ~$z_{Z}$~  &  ~~$h$~~  &  ~~$n$~~  &  ~$p$~  &  ~$z_{Y}$~  &  ~~$z_{Z}$~~  &  ~~$h$~~  &  ~~$n$~~  &  ~$p$~  &  ~$z_{Y}$~  &  ~$z_{Z}$~  &  ~~$h$~~  &  ~~$n$~~  &  ~$p$~  \\
\hline 11 & 4 & 31 & 37 & 8 & 11 & 31 & 21 & 44 & 8 & 4 & 31 & 21 & 44 & 9 \\
11 & 4 & 19 & 38 & 9 & 11 & 21 & 45 & 58 & 8 & 4 & 18 & 39 & 49 & 11 \\
11 & 24 & 33 & 40 & 9 & 11 & 41 & 26 & 59 & 8 & 4 & 37 & 24 & 53 & 9 \\
11 & 4 & 20 & 41 & 9 & 11 & 43 & 27 & 62 & 8 & 4 or 32 & 39 & 25 & 56 & 11 \\
11 & 4 or 26 & 21 & 44 & 8 & 11 & 24 & 51 & 67 & 9 & 4 & 21 & 45 & 58 & 10 \\
11 & 27 & 37 & 46 & 8 & 11 & 51 & 31 & 74 & 10 & 4 & 22 & 47 & 61 & 10 \\
11 & 4 & 39 & 49 & 11 & 11 & 53 & 32 & 77 & 11 & 4 & 43 & 27 & 62 & 8 \\
11 & 4 & 41 & 52 & 10 & 11 & 28 & 59 & 79 & 13 & 36 & 23 & 49 & 64 & 8 \\
11 & 4 & 43 & 55 & 9 & 11 & 59 & 35 & 86 & 8 & 4 & 45 & 28 & 65 & 10 \\
11 & 4 or 32 & 25 & 56 & 11 & 11 & 31 & 65 & 88 & 12 & 4 & 24 & 51 & 67 & 9 \\
\hline
\end{tabular}
\end{center}
\end{table}

On the other hand, considering the $ \rm{S U(4)_{C} \otimes S U(2)_{L} \otimes U(1)_{R}} $ embedding, we obtain the required conditions that are:
\begin{eqnarray}
z_{q} & = & z_{\ell}(\bmod\ n),\nonumber \\
z_{\bar{u}} & = & z_{\bar{\nu}}(\bmod\ n),\nonumber \\
z_{\bar{d}} & = & z_{\bar{e}}(\bmod\ n).
\end{eqnarray}
Similarly, this implies that there exists an integer $ m $ satisfying the following conditions:
\begin{eqnarray}
&&z_{q}^{\prime}+m  =  z_{\ell}^{\prime}+(-3) m(\bmod\ \mathrm{n}), \nonumber\\
&&z_{\bar{u}}^{\prime}+(-4) m  =  z_{\bar{\nu}}^{\prime}(\bmod\ \mathrm{n}), \nonumber\\
&&z_{\bar{d}}^{\prime}+2 m  =  z_{\bar{e}}^{\prime}+6 m(\bmod\ \mathrm{n}).
\end{eqnarray}
Using the charges of Table \ref{tab:znR charges}, we can obtain:
\begin{eqnarray}
R-h-4 m & = & 0(\bmod\ \mathrm{n}).
\label{condition 2}
\end{eqnarray}
The $ Z_{n}^{R} $ symmetries that satisfy these conditions (Eqs.~(\ref{condition gs}) and (\ref{condition 2})) have been listed in Table \ref{SU4C}, and the largest $p$ are 11 or 13 for these models. As declared in Ref.~\cite{S P M}, if PQ-violating superpotential terms with $p=(8,9,10,11$ or $12)$ are present, one typically should have $f_A \lesssim 4 \times 10^9, 3 \times 10^{10}, 10^{11}, 4 \times 10^{11}$ or $10^{12} ~\rm{GeV}$ respectively. However, there are so many possibilities that $f_A$ can get rid of these constraints. The corresponding coupling(s) $\kappa$, for example, may happen to have a small magnitude so that $f_A$ could be larger than $4 \times 10^{11} ~\rm{GeV}$, even if $p=11$ is present. Following Ref.~\cite{S P M}, we will therefore not commit to a specific requirement for $f_A$, with the understanding that smaller $f_A$ is safer in some sense.

In the extended model $\rm{E_I}$, the $ Z_{n}^{R} $ symmetries for the base model $\rm{B_I}$ also work and give the same $ p $. However, we have no insight into the ultraviolet complexity other than the solution to the domain wall problem. By the way, some superpotential terms that violate baryon or lepton number may predict very rapid proton decay, such as $ H_{u} \ell, q \ell \bar{d}, \ell \ell \bar{e}, \bar{u} \bar{d} \bar{d}, \frac{1}{M_{P}} q q q \ell $ and $ \frac{1}{M_{P}} \bar{u} \bar{u} \bar{d} \bar{e} $. We note that all of the $ Z_{n}^{R} $ symmetries we find can safely forbid these superpotential terms.

\subsection{Low-energy effective theory}
In our models, there are some new particles other than the particles within MSSM, including two pseudoscalar particles $ A_{i}^{\prime} $, three scalar particles $ S_{i} $ and three majorana spinors $ \tilde{a}_{i} $ with nonzero masses.

The pseudoscalars $A^\prime_i$ as well as the axion $A$ are made up of $a_X$, $a_Y$ and $a_Z$. The mass squared mixing matrix is diagonalized by the matrix $Z^{A}$:
\begin{eqnarray}
Z^{A}\left(\begin{array}{lll}
M_{a_{X} a_{X}}^{2} & M_{a_{X} a_{Y}}^{2} & M_{a_{X} a_{Z}}^{2} \\
M_{a_{Y} a_{X}}^{2} & M_{a_{Y} a_{Y}}^{2} & M_{a_{Y} a_{Z}}^{2} \\
M_{a_{Z} a_{X}}^{2} & M_{a_{Z} a_{Y}}^{2} & M_{a_{Z} a_{Z}}^{2}
\end{array}\right) Z^{A^{T}}=diag\left(0, m_{A_{1}^{\prime}}^{2}, m_{A_{2}^{\prime}}^{2}\right).\label{}
\end{eqnarray}
The axion mass eigenstate $ A \approx \frac{Q_{X} v_{X}}{v_{A}} a_{X}+\frac{Q_{Y} v_{Y}}{v_{A}} a_{Y}+\frac{Q_{Z} v_{Z}}{v_{A}} a_{Z} $, so that we arrive at $ Z_{11}^{A}=\frac{Q_{X} v_{X}}{v_{A}} $, $ Z_{12}^{A}=\frac{Q_{Y } v_{Y}}{v_{A}} $ and $ Z_{13}^{A}=\frac{Q_{Z} v_{Z}}{v_{A}} $. Similarly, the scalar particles $ S_{1}, S_{2} $ and $ S_{3} $, which we will call saxions in the following, are made up of $ \rho_{X}, \rho_{Y} $ and $ \rho_{Z} $. The mass squared mixing matrix is diagonalized by the matrix $ Z^{S} $:
\begin{eqnarray}
Z^{S}\left(\begin{array}{ccc}
M_{\rho_{X} \rho_{X}}^{2} & M_{\rho_{X} \rho_{Y}}^{2} & M_{\rho_{X} \rho_{Z}}^{2} \\
M_{\rho_{Y} \rho_{X}}^{2} & M_{\rho_{Y} \rho_{Y}}^{2} & M_{\rho_{Y} \rho_{Z}}^{2} \\
M_{\rho_{Z} \rho_{X}}^{2} & M_{\rho_{Z} \rho_{Y}}^{2} & M_{\rho_{Z} \rho_{Z}}^{2}
\end{array}\right) Z^{S^{T}}=diag\left(m_{S_{i}}^{2}\right), i=1-3 .\label{}
\end{eqnarray}
Finally, majorana spinors $\tilde{a}_{1}$, $\tilde{a}_{2}$ and $\tilde{a}_{3}$ are the so-called axinos. In the basis $(\tilde{a}_{X}$, $\tilde{a}_{Y}$  and $\tilde{a}_{Z})$, the axino mass mixing matrix is diagonalized by $Z^{\tilde{a}}$:
\begin{eqnarray}
Z^{\tilde{a}}\left(\begin{array}{lll}
M_{\tilde{a}_{X} \tilde{a}_{X}} & M_{\tilde{a}_{X} \tilde{a}_{Y}} & M_{\tilde{a}_{X} \tilde{a}_{Z}} \\
M_{\tilde{a}_{Y} \tilde{a}_{X}} & M_{\tilde{a}_{Y} \tilde{a}_{Y}} & M_{\tilde{a}_{Y} \tilde{a}_{Z}} \\
M_{\tilde{a}_{Z} \tilde{a}_{X}} & M_{\tilde{a}_{Z} \tilde{a}_{Y}} & M_{\tilde{a}_{Z} \tilde{a}_{Z}}
\end{array}\right) Z^{\tilde{a}^{T}}=diag\left(m_{\tilde{a}_{i}}\right), i=1-3 .
\end{eqnarray}
All the matrix elements above as well as the tadpole equations involving $ X, Y $ and $ Z $ are given in the Appendix.~\ref{APPENDIX A}.

In order to study the phenomenology, we can write the low-energy superpotential of the model as
\begin{eqnarray}
W &=& \mu(1+ \frac{\zeta_{\hat{\mathcal{S}}_i}}{v_A/\sqrt{2}} \hat{\mathcal{S}}_i )\hat{H}_u \hat{H}_d.
\end{eqnarray}
The supermultiplets $\hat{\mathcal{S}}_i$, usually the one we're interested in, have a relationship with the superfields $\hat{X}\sim~(\frac{v_X+\rho_X+i a_X}{\sqrt{2}},\tilde{a}_X)$, $\hat{Y}\sim~(\frac{v_Y+\rho_Y+i a_Y}{\sqrt{2}},\tilde{a}_Y)$ and $\hat{Z}\sim~(\frac{v_Z+\rho_Z+i a_Z}{\sqrt{2}},\tilde{a}_Z)$:
\begin{eqnarray}
\left(\begin{array}{c}
  \hat{X} \\
   \hat{Y}\\\hat{Z}
    \end{array}\right)&=&\left(\begin{array}{c}
   \frac{v_A}{\sqrt{2}}\frac{Z^{A}_{11}}{Q_X} \\
   \frac{v_A }{\sqrt{2}}\frac{Z^{A}_{12}}{Q_Y}\\ \frac{v_A}{\sqrt{2}}\frac{Z^{A}_{13}}{Q_Z}
    \end{array}\right)
 + \left(\begin{array}{ccc}
Z^{\hat{\mathcal{S}}}_{11}&Z^{\hat{\mathcal{S}}}_{12}&Z^{\hat{\mathcal{S}}}_{13}\\ Z^{\hat{\mathcal{S}}}_{12}&Z^{\hat{\mathcal{S}}}_{22}&Z^{\hat{\mathcal{S}}}_{23}\\Z^{\hat{\mathcal{S}}}_{13}&Z^{\hat{\mathcal{S}}}_{23}&Z^{\hat{\mathcal{S}}}_{33}
    \end{array}\right)^{T} \left(\begin{array}{c}
  \hat{\mathcal{S}}_1 \\
  \hat{\mathcal{S}}_2\\
  \hat{\mathcal{S}}_3
    \end{array}\right)\nonumber\\
    &=&\left(\begin{array}{c}
 \frac{v_A}{\sqrt{2}}\frac{Z^{A}_{11}}{Q_X} \exp\{\sum_{i=1}^{3}\frac{Z^{\hat{\mathcal{S}}}_{i1}}{Z^{A}_{11}}\frac{Q_X \hat{\mathcal{S}}_i}{v_A/\sqrt{2}} \}\\
 \frac{v_A}{\sqrt{2}}\frac{Z^{A}_{12}}{Q_Y} \exp\{\sum_{i=1}^{3}\frac{Z^{\hat{\mathcal{S}}}_{i2}}{Z^{A}_{12}}\frac{Q_Y \hat{\mathcal{S}}_i}{v_A/\sqrt{2}} \}\\
 \frac{v_A}{\sqrt{2}}\frac{Z^{A}_{13}}{Q_Z} \exp\{\sum_{i=1}^{3}\frac{Z^{\hat{\mathcal{S}}}_{i3}}{Z^{A}_{13}}\frac{Q_Z \hat{\mathcal{S}}_i}{v_A/\sqrt{2}} \}
    \end{array}\right)
    .\label{}
      \end{eqnarray}
In this way, the effective superpotential coefficients can be expressed as $ \zeta_{\hat{\mathcal{S}}_{i}}=2 \frac{Z_{i 1}^{\hat{\mathcal{S}}}}{Z_{11}^{A}} Q_{X} $. When $ Z^{\hat{\mathcal{S}}}=Z^{A} $, the $ \hat{\mathcal{S}}_{1} $ is just the axion supermultiplets $ \hat{A} \sim\left(\frac{S+i A}{\sqrt{2}}, \tilde{a}\right) $, and the corresponding $ \zeta_{\hat{A}}=2 Q_{X} $. Here $ S $ and $ \tilde{a} $ are typically not the mass eigenstates. To be concrete, they are
\begin{eqnarray}
S&=&\sum_{i=1}^{3} Z_{1 i}^{A} Z_{1 i}^{S} S_{1}+\sum_{i=1}^{3} Z_{1 i}^{A} Z_{2 i}^{S} S_{2}+\sum_{i=1}^{3} Z_{1 i}^{A} Z_{3 i}^{S} S_{3} \equiv C_{S_{1}} S_{1}+C_{S_{2}} S_{2}+C_{S_{3}} S_{3} ,\\
\tilde{a}&=&\sum_{i=1}^{3} Z_{1 i}^{A} Z_{1 i}^{\tilde{a}} \tilde{a}_{1}+\sum_{i=1}^{3} Z_{1 i}^{A} Z_{2 i}^{\tilde{a}} \tilde{a}_{2}+\sum_{i=1}^{3} Z_{1 i}^{A} Z_{3 i}^{\tilde{a}} \tilde{a}_{3} \equiv C_{\tilde{a}_{1}} \tilde{a}_{1}+C_{\tilde{a}_{2}} \tilde{a}_{2}+C_{\tilde{a}_{3}} \tilde{a}_{3}.
\end{eqnarray}
For $ Z^{\hat{\mathcal{S}}}=Z^{S} $, the $ \hat{\mathcal{S}}_{i} $ is the saxion supermultiplets $ \hat{S}_{i} \sim\left(\frac{S_{i}}{\sqrt{2}}\right) $, and the $ \hat{\mathcal{S}}_{i} $ is the axino supermultiplets $ \hat{\tilde{a}}_{i} \sim\left(\tilde{a}_{i}\right) $ for $ Z^{\hat{\mathcal{S}}}=Z^{\tilde{a}} $. Based on these symbolic conventions, the superpotential could be rewritten as
\begin{eqnarray}
W & = & \mu\left\{1+\frac{1}{v_{A} / \sqrt{2}}\left(\zeta_{\hat{A}} \hat{A}+\zeta_{\hat{S}_{i}} \hat{S}_{i}+\zeta_{\hat{\tilde{a}}_{i}} \hat{\tilde{a}}_{i}\right)+\cdots\right\} \hat{H}_{u} \hat{H}_{d},\label{effW}
\end{eqnarray}
noticing that we just put them together for simplicity, and usually only one or two terms exist at the same time depending on the particles we studied.

Finally, the K$\ddot{\rm{a}}$hler potential gives interactions between the axion and the saxions:
\begin{eqnarray}
\mathcal{L}&=& (1+\frac{\sqrt{2}\xi}{v_A/\sqrt{2}}S)[\frac{1}{2}\partial^\mu A \partial_\mu A+\frac{1}{2}\partial^\mu S \partial_\mu S],\label{eq:LSAA}
\end{eqnarray}
where $\xi=\Sigma_i Q_i^3 v_i^2/v_A^2$ is expected to be $\mathcal{O}(1)$ but could also be as small as zero \cite{SAA}. Using Eq.~(\ref{eq:LSAA}), we find the partial width for $S_i\rightarrow AA$ to be
\begin{eqnarray}
\Gamma(S_i\rightarrow AA)\approx\frac{( C_{S_i} \xi)^2 m_{S_i}^3}{64\pi(v_A/\sqrt{2})^2}.\label{eq:GMSiAA}
\end{eqnarray}
\section{Dark matter candidates in our models}\label{Sec III}
In this section, we examine the Higgsino-like neutralino and axion as potential candidates for dark matter. Firstly, we assume that the Higgsino-like neutralino is the LSP and discuss the constraints from the anomalous magnetic moment of the muon $a_\mu$ as well as the LZ-2022 direct detection experiment \cite{LZ:2022lsv} on the relic density of the LSP. Secondly, we investigate the scenario where the axion constitutes the remaining portion of dark matter and satisfies the observed relic density via the conventional misalignment mechanism (CMM). We take into account the constraints originating from the cosmology of saxions/axinos and study their impacts on the parameter space of our models. In addition, we also give a discussion on the potential cosmological moduli problem. Lastly, we give the experimental limitations arising from $g_{A\gamma\gamma}$ on axion decay constant $f_A$ for different models.
\subsection{Constraints from $a_\mu$, LZ-2022 and LHC}
In our models, the mixings between fields that belong to and do not belong to the MSSM are very tiny, and the interactions between particles belonging to and not belonging to the MSSM are also $\frac{1}{f_A}$ suppressed. These newly introduced particles affect low-energy phenomenology, but the impacts are extremely small and can be safely ignored. Therefore, our models are almost identical to the MSSM at low energy if we do not consider the axion portion.

In the framework of MSSM, the LSP, which we taken as the lightest neutralino $\tilde{\chi}^0_1$ (also denoted by $\tilde{\chi}$ in the following), can be the dark matter candidate. The masses and mixing of neutralinos are determined by the parameters $M_1, M_2, \mu$ and $\tan\beta$, and that of charginos is determined by the parameters $M_2, \mu$ and $\tan\beta$. As a result, the LSP can be classified as bino-like, wino-like, Higgsino-like or other well-tempered types (e.g., wino-Higgsino LSP has been studied in Refs.~\cite{Iwamoto:2021aaf}), determined by the parameters mentioned above. The relic density can be obtained via computing the cross section of the annihilation processes or the coannihilation processes with the next-to-lightest supersymmetric particle (NLSP). In the case of bino-like LSP, it turns out that there would be too large relic density to explain the Planck observation. As the LSP is wino-like, considering the effect of ``Sommerfeld enhancement" \cite{SOMMERFELD}, wino-like LSP providing the full amount of dark matter needs $m_{\tilde{\chi}}$ to be as large as roughly $2.9 ~\rm{TeV}$. In the limit of pure Higgsino, the parameter $\mu$ should also be as large as about $1.1~\rm{TeV}$.

In addition, when we consider the latest experimental data of anomalous magnetic moment of the muon $a_\mu:=(g-2)_\mu/2$, more constraints should be imposed on the parameter space. The SM prediction of muon anomaly is $a^{\rm{sm}}_\mu = 116591810 (43) \times 10^{-11} (0.37 \rm{ppm})$~\cite{g-2 SM 1,g-2 SM 2,g-2 SM 3}, and the latest averaged experiment value of muon anomaly is $a^{\rm{exp}}_\mu = 116592061 (41) \times 10^{-11} (0.35 \rm{ppm})$~\cite{g-2 SM EXP}. Therefore, the deviation between experiment and SM prediction is $\Delta a_\mu = a^{\rm{exp}}_\mu - a^{\rm{sm}}_\mu= 251 (59) \times 10^{-11}$, which is $4.2\sigma$. This anomaly is an important hint of the existence of the NP models. Within MSSM, both wino-like and Higgsino-like LSP cannot fulfill the $(g - 2)_\mu$ constraints because their masses are too heavy unless they are only responsible for a small part of the total relic density of dark matter. This is consistent with our idea that the dark matter is a mixture of LSP and axion. In this study, we consider the Higgsino-like LSP as the candidate of dark matter. In this case, the parameter $\mu$ is much smaller than $M_1$ and $M_2$. Such a low $\mu$ is also favored by naturalness arguments in some literatures \cite{Bae:2019dgg,Baer:2018avn,Baer,Chan,BaerTata}. $\mu\sim 100 - 300 ~\rm{GeV}$ (the lower the more natural), for example, is claimed in Ref.~\cite{Bae:2019dgg}, and $\mu\sim 100 - 350 ~\rm{GeV}$ (where $\mu \gtrsim 100 ~\rm{GeV}$ is required to accommodate LEP2 limits from chargino pair production searches) is claimed in Ref.~\cite{Baer:2018avn}. Notably, in these literatures, the topic of ``naturalness'' mainly focuses on the fine-tuning of parameters in the potential minimization condition
\begin{eqnarray}
\frac{M_Z^2}{2}=\frac{m_{H_d}^2-m_{H_u}^2\tan^2\beta}{\tan^2\beta-1}-\mu^2.\label{eq:mu2}
\end{eqnarray}
Naively, if $\mu\gg M_Z$, the first term on the right side of Eq.~(\ref{eq:mu2}) must also be large, so that these two terms may largely cancel to give $M_Z$. In this sense, one can have a suspicion that the models with $\mu\gg M_Z$ would be fine tuned. It is also worthy to mention that a larger $\mu$ could also be found in massive literatures. In particular, the parameter $\mu$ as large as $10^5,10^6,10^7~\rm{GeV}$ have been used in the numerical analysis for a high-scale SUSY in Ref.~\cite{Choi}. With our assumption that the LSP is Higgsino-like, we prefer a $\mu$ with low value.

In our numerical analysis, the SUSY particle spectrum is generated by using the $\verb"SPheno-4.0.5"$~\cite{Porod:2011nf,Porod:2003um}, and the parameters are given at the low SUSY scale with the renormalization scale $M=1~\rm{TeV}$. Additionally, we use the $\verb"GM2Calc-2.1.0"$~\cite{GM2} to calculate the contribution to $a_\mu$ up to two-loop order, and the relic density of LSP is evaluated with $\verb"MicrOMEGAs-5.2.13"$~\cite{MOMG1,MOMG2,MOMG3,MOMG4}. These codes are numerically reliable. To find the parameter space not excluded by experiments, we scan the parameter space. The following are some of the assumptions we made:

1. The two-loop level SM-like Higgs boson mass $m_h$ is in the region $[123,127]~\rm{GeV}$. This is reasonable because the uncertainty of this theoretical prediction can be as large as 2$-$3 $\rm{GeV}$~\cite{Athron:2016fuq,Allanach:2018fif,Bahl:2019hmm}.

2. For scalar leptons, we take the trilinear coupling $A_l$ to be zero for all three generations, so the mixing is significant only for the scalar taus.

3. The mass of the MSSM pseudoscalar particle $A^0$ is above $1.5~\rm{TeV}$, and the masses of scalar quarks are larger than $3~\rm{TeV}$.

\begin{figure}[!htb]
\begin{center}
\begin{minipage}[c]{0.48\textwidth}
\includegraphics[width=3.1in]{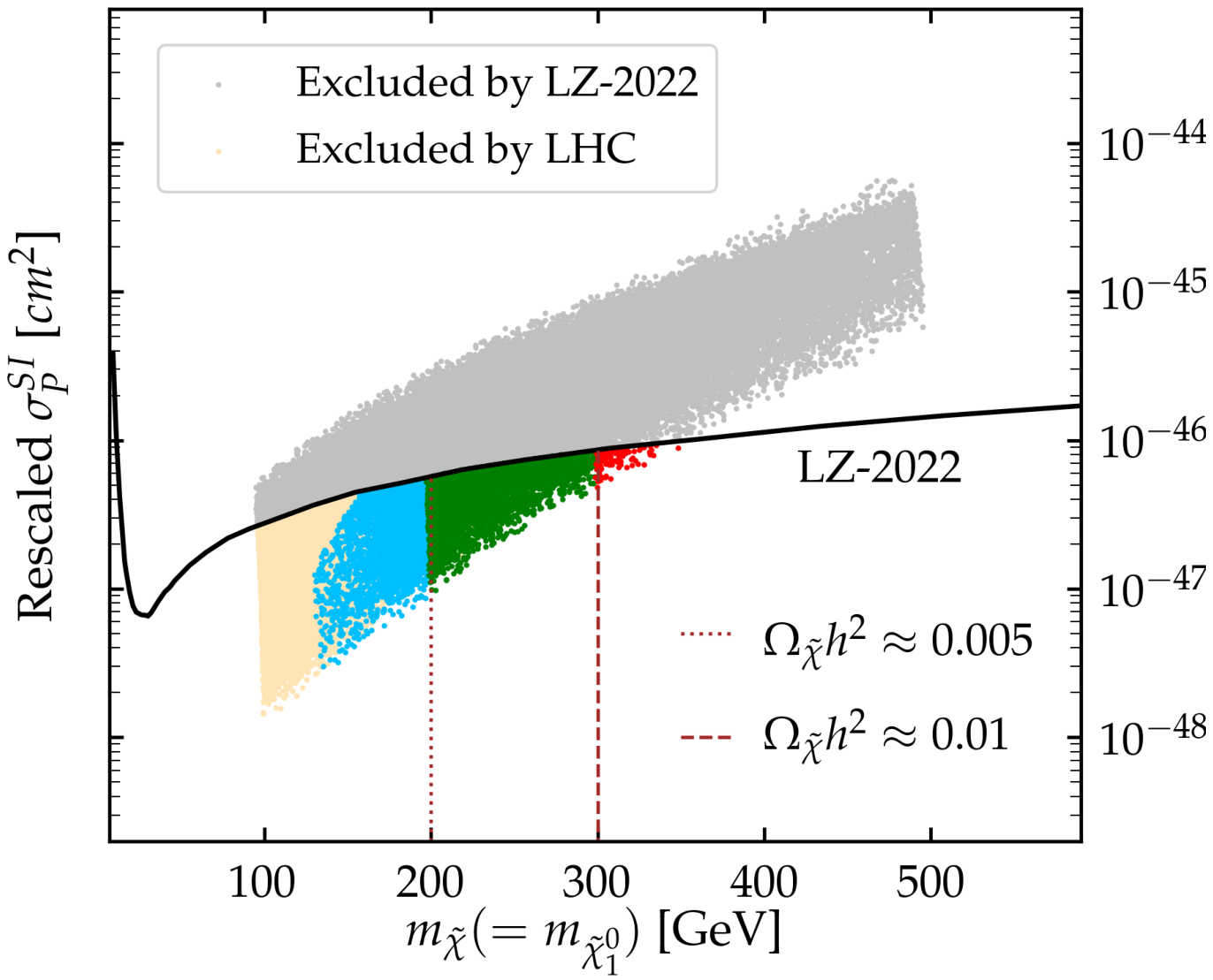}
\end{minipage}%
\begin{minipage}[c]{0.48\textwidth}
\includegraphics[width=3.1in]{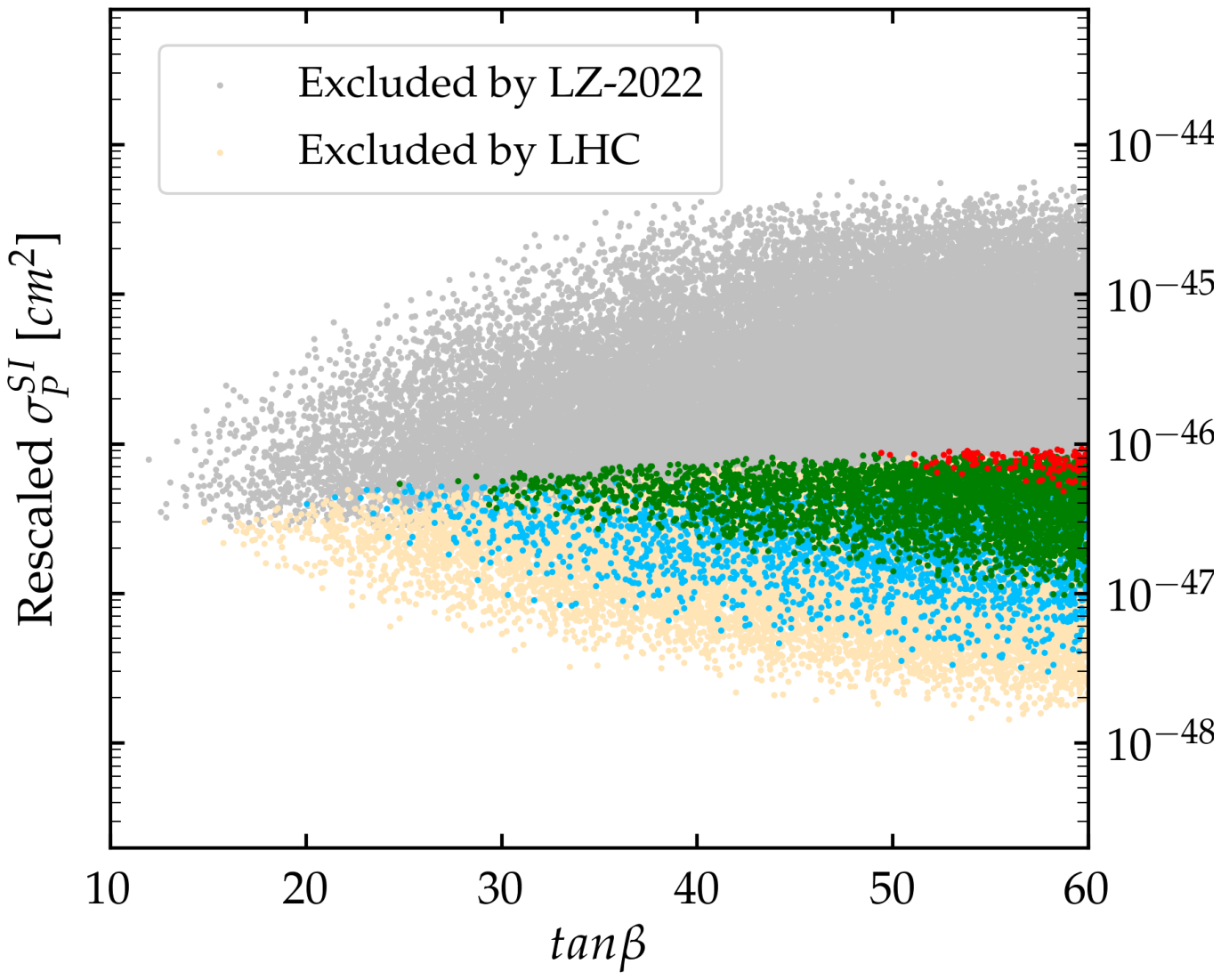}
\end{minipage}
\caption[]{\label{gm2fig1} The constraints on the mass of the Higgsino-like LSP $m_{\tilde{\chi}}$ (left panel) and parameter $\tan\beta$ (right panel) from the latest $a_\mu$ measurements, LZ-2022 direct detection experiment \cite{LZ:2022lsv} and LHC. The light gray points are excluded by the bound of the LZ-2022 direct detection experiment which is plotted by black solid line in left panel, and light orange points are excluded by the LHC. The blue, green and red points are surviving and denote the parameter $\mu\in$ $[100,200]$, $[200,300]$ and $[300,400]$ $\rm{GeV}$ respectively. In addition, the relic density of $\tilde{\chi}$ with $\mu= 200$ and $300 ~\rm{GeV}$ are also shown with brown dotted and dashed lines respectively. All of them are small comparing with the total dark matter abundance which is $0.12$, and this is consistent with our purpose that LSP is a small part of the total dark matter.}
\end{center}
\end{figure}

\begin{figure}[!htb]
\begin{center}
\begin{minipage}[c]{0.48\textwidth}
\includegraphics[width=3in]{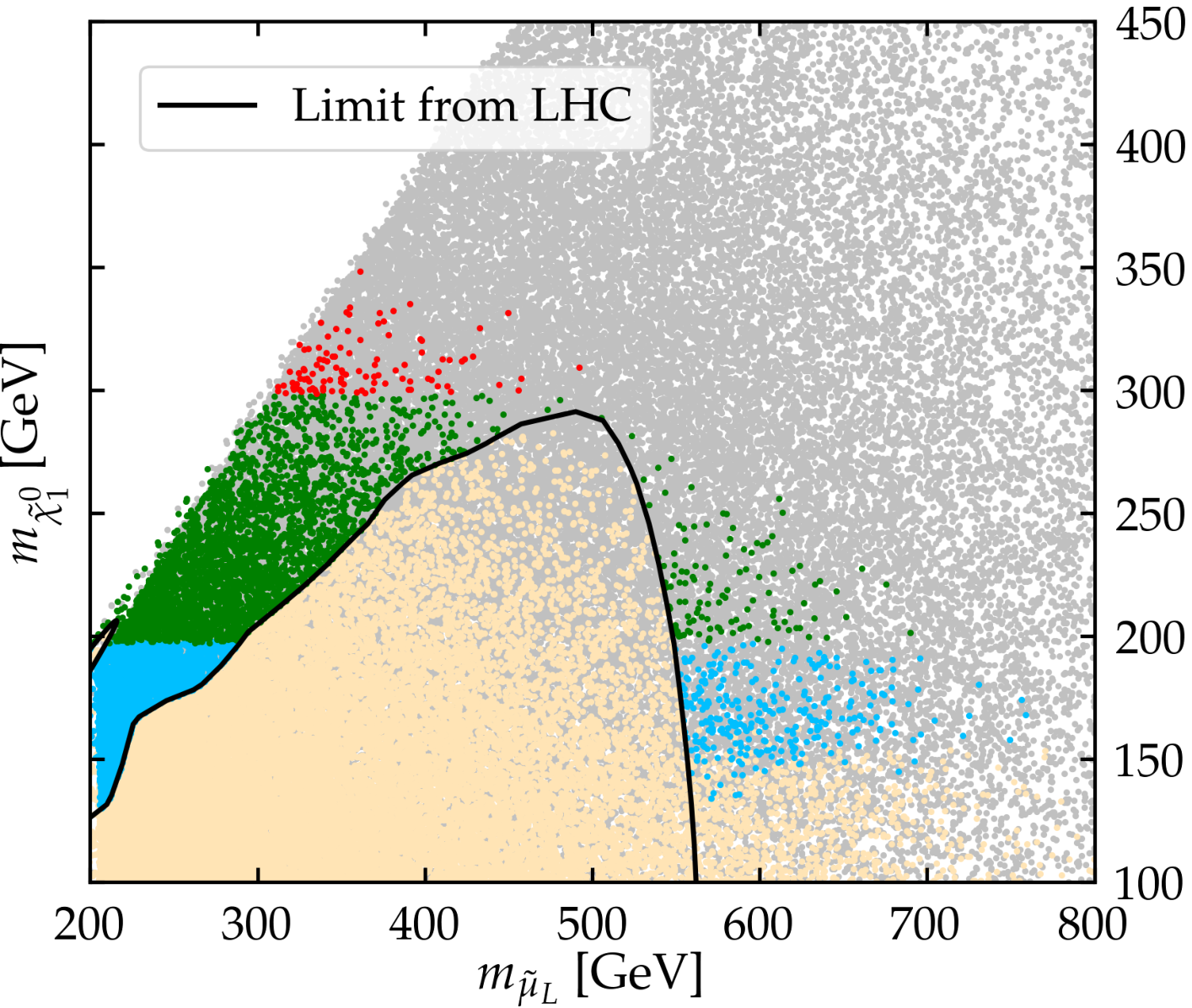}
\end{minipage}%
\begin{minipage}[c]{0.48\textwidth}
\includegraphics[width=3in]{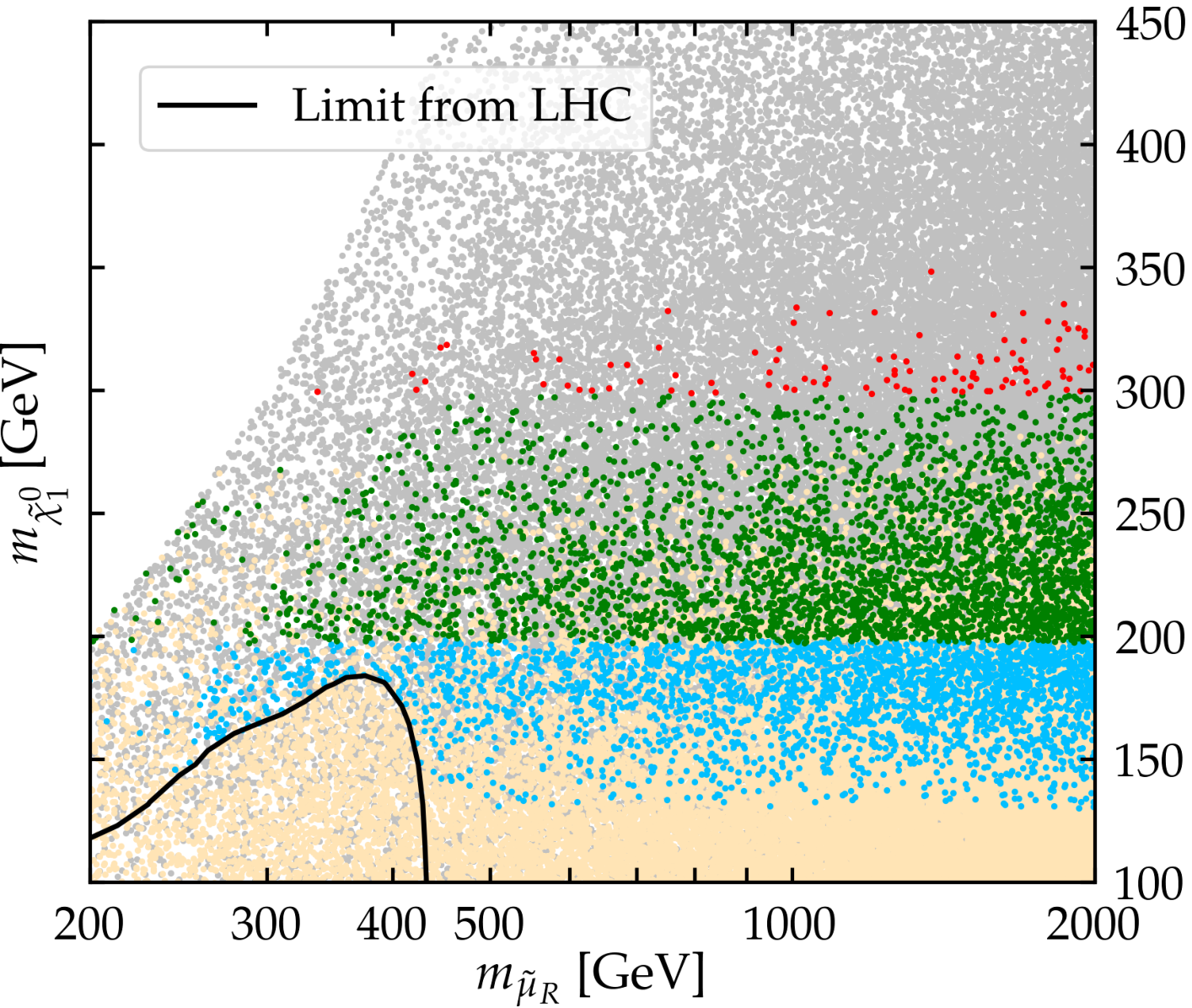}
\end{minipage}
\caption[]{\label{gm2fig2} The samples displayed in Fig.~\ref{gm2fig1} on the planes of $m_{\tilde{\mu}_L}$ versus $m_{\tilde{\chi}^0_1}$ (left panel) and $m_{\tilde{\mu}_R}$ versus $m_{\tilde{\chi}^0_1}$ (right panel). The black solid line in left (right) panel is the observed exclusion limits for a direct smuon $\tilde{\mu}_L$ ($\tilde{\mu}_R$) production recorded by the ATLAS detector at the LHC at $\sqrt{s} = 13 ~\rm{TeV}$ with $139 ~\rm{fb}^{-1}$ \cite{ATLAS:2019lff}. The colors of the points have the same meaning as them in Fig.~\ref{gm2fig1}.}
\end{center}
\end{figure}

\begin{figure}[!htb]
\begin{center}
\begin{minipage}[c]{0.48\textwidth}
\includegraphics[width=3in]{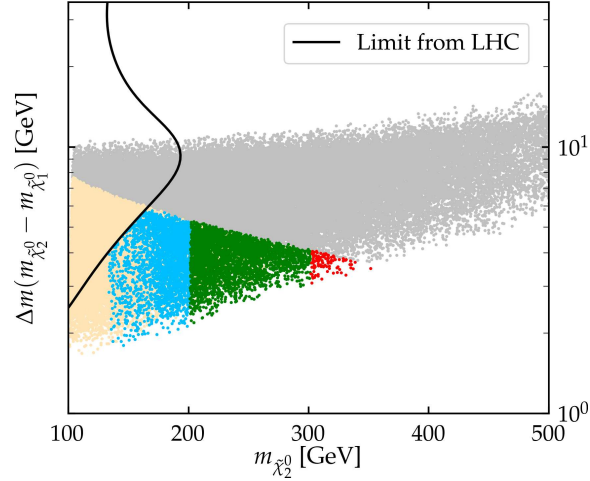}
\end{minipage}%
\caption[]{\label{gm2fig3} Same as the samples of Fig.~\ref{gm2fig1}, but showing $m_{\tilde{\chi}^0_2}$ versus $\Delta m(=m_{\tilde{\chi}^0_2}-m_{\tilde{\chi}^0_1})$. The black solid line is the observed exclusion limits for a direct Higgsino production recorded by the ATLAS detector at the LHC at $\sqrt{s} = 13 ~\rm{TeV}$ with $139 ~\rm{fb}^{-1}$ \cite{ATLAS:2019lng}. The color coding is as in Fig.~\ref{gm2fig1}.}
\end{center}
\end{figure}

In the MSSM, the one-loop correction to $a_\mu$ mainly comes from $\tilde{\chi}^\pm-\tilde{\nu}_{\mu}$ and $\tilde{\chi}^0-\tilde{\mu}$, hence the mass of scalar muon is a sensitive parameter. Considering the MSSM mass relation $M^2_{L_{ii}}=m^2_{\tilde{\nu}_i}-M_W^2\cos(2\beta)$~\cite{g-2 one,g-2 two} by which the sneutrino and slepton masses are related, in this work, we scan the following parameter space:
\begin{eqnarray}
&&\tan\beta \in[2,60],~\mu\in[100,500]~\rm{GeV},\nonumber\\
&&M_1\in[0.8,6]~\rm{TeV},~M_2\in[0.8,6]~\rm{TeV},\nonumber\\
&&M_{L_{22}}\in[\rm{Max}[\sqrt{\mu^2+M_W^2},0.2~\rm{TeV}],2~\rm{TeV}],\nonumber\\
&&M_{R_{22}}\in[\rm{Max}[\sqrt{\mu^2+M_W^2},0.2~\rm{TeV}],2~\rm{TeV}],\nonumber
\end{eqnarray}
where $M_{L}^2$ and $M_{R}^2$ are the mass squared parameters of soft terms in MSSM. The resulting parameter space that can satisfy the current experimental limitations are shown in Fig.~\ref{gm2fig1}-\ref{gm2fig3}. In these figures, all points can guarantee that the lightest neutralino $\tilde{\chi}^0_1$ is the LSP, and can satisfy the constraint from $a_\mu$ within $2\sigma$ deviation.

In Fig.~\ref{gm2fig1} left panel, the constraints on the mass of the Higgsino-like LSP $m_{\tilde{\chi}}$ are shown. Despite the fact that the LZ-2022 bound is very strict, there are still parameter spaces that are not excluded. The LSP-proton spin-independent scattering cross section $\sigma_{P}^{\rm{SI}}$ has been rescaled as $\frac{\Omega_{\tilde{\chi}} h^2}{\Omega_{\rm{DM}} h^2}\sigma_{P}^{\rm{SI}}$ since the LSP is just a small part of the total dark matter. The mass of the LSP, which is approximately equal to the parameter $\mu$, is restricted in the range $[100,400]~\rm{GeV}$. Besides, we also show the relic density of LSP in this panel for the parameter $\mu= 200$ and $300 ~\rm{GeV}$ respectively. As a crude estimate, the relic density of LSP is limited to no more than 0.013.

The restrictions imposed on $\tan\beta$ when the rescaled $\sigma_{P}^{\rm{SI}}$ is under the LZ-2022 bound are given in the right panel in Fig.~\ref{gm2fig1}. As the parameter $\mu$ increases, corresponding $\sigma_{P}^{\rm{SI}}$ rises concurrently. For smaller $\mu$, the restriction placed on $\tan\beta$ is lax. When the parameter $\mu$ is large, however, e.g., $\mu>300~\rm{GeV}$, $\tan\beta$ needs to remain in a small interval of large value to prevent the parameter space from being excluded. This behavior is easy to understand, since the squarks in s-channel are heavy and, therefore, the most important contribution to $\sigma_{P}^{\rm{SI}}$ comes from the exchange of the SM-like Higgs in t-channel, whose coupling could be expressed as \cite{Chxx}
\begin{eqnarray}
&& C_{h\tilde{\chi}\tilde{\chi}} \propto (1+\sin2\beta)(\tan^2\theta_W\frac{M_W}{M_1-\mu}+\frac{M_W}{M_2-\mu}).
\end{eqnarray}
The first term in the second bracket of the above formula can be omitted numerically, so that small $\mu$ and large $\tan\beta$ can result in a small $\sigma_{P}^{\rm{SI}}$ that would not be excluded by the direct detection experiment.

For the limits from the LHC, we consider the electroweak production of supersymmetric particles in $\sqrt{s}=13~ \rm{TeV}$ $pp$ collisions with the ATLAS detector. The most stringent limitations on the parameter space we assumed come from the smuon production process $pp \rightarrow \tilde{\mu}\tilde{\mu}$ \cite{ATLAS:2019lff} and the Higgsino production process $pp \rightarrow \tilde{\chi}^0_2 \tilde{\chi}^\pm_1, \tilde{\chi}^0_2 \tilde{\chi}^0_1, \tilde{\chi}^+_1 \tilde{\chi}^-_1$(Higgsino) \cite{ATLAS:2019lng}, where $\tilde{\chi}^0_2$ is next-to-lightest neutralino and $\tilde{\chi}^\pm_1$ is the lightest chargino. In Fig.~\ref{gm2fig1}, the points colored by the light orange are excluded by these LHC constraints. We replot the samples displayed in Fig.~\ref{gm2fig1} on the planes of $m_{\tilde{\mu}_L}$ versus $m_{\tilde{\chi}^0_1}$ and $m_{\tilde{\mu}_R}$ versus $m_{\tilde{\chi}^0_1}$ in the left and right panels in Fig.~\ref{gm2fig2} respectively. From this figure, we find that the parameter $M_{R_{22}}$ is a relatively insensitive parameter compared with $M_{L_{22}}$, since the dominant SUSY contribution to $a_\mu$ is from wino-Higgsino-$\tilde{\mu}_L$ loop in this Higgsino-like LSP scenario. In order to escape the constraint from the LZ-2022 direct detection experiment successfully, one would expect that the parameters $M_1$ and $M_2$ are decoupled($C_{h\tilde{\chi}\tilde{\chi}}$ originates from gaugino Yukawa couplings of the form $h^\dag \tilde{h} \tilde{b}$ and $h^\dag \tilde{h} \tilde{w}$). In this case, for the sake of explaining the latest measurements of $a_\mu$, one needs relatively light masses of loop particles that appear in the involved Feynman diagrams. For these reasons, the resulting viable parameter space in Fig.~\ref{gm2fig2} are as expected. The constraints from LHC on $M_{R_{22}}$ is also much looser than that on $M_{L_{22}}$. Overall, we can place an upper limit on $m_{\tilde{\mu}_L}$ of about $750 ~\rm{GeV}$. Besides, considering the constraint from LHC Higgsino production, we also show the samples of Fig.~\ref{gm2fig1} on the plane of $m_{\tilde{\chi}^0_2}$ versus $\Delta m(=m_{\tilde{\chi}^0_2}-m_{\tilde{\chi}^0_1})$ in Fig.~\ref{gm2fig3}. Since the maximum of the excluded $m_{\tilde{\chi}^0_2}$ is smaller than $200~\rm{GeV}$, this constraint is quite loose and we can take no account of this limitation if the LSP is heavier than $200~\rm{GeV}$.

Finally, three sample points are shown in Table \ref{TmMSSM}. Combining the constraints on these parameters that are shown in Fig.~\ref{gm2fig1}-\ref{gm2fig3}, we just take the parameter $\mu=250~\rm{GeV}$ for safety in the following discussion, and the corresponding relic density $\Omega_{\tilde{\chi}}h^2$ is approximately $0.007$.
\begin{table}
\begin{center}
\begin{minipage}[]{0.95\linewidth}\caption{Typical mass (in GeV) of some particles in the Higgsino-like LSP scenario, as well as the rescaled $\sigma_{P}^{\rm{SI}}$ and the SUSY contribution to muon anomaly $a_\mu^{\rm{SUSY}}$. For $\mu=250~\rm{GeV}$, the relic density of LSP $\Omega_{\tilde{\chi}}h^2$ is approximately $0.007$. \label{TmMSSM}}
\end{minipage}

\vspace{0.3cm}

\begin{tabular}{|c | c | c | c | c | c | c | c| c| c| }
\hline
  ~$\mu$~ & ~$\tan\beta$~ & ~$M_1$~ & ~$M_2$~ & ~$m_{\tilde{\chi}_1^0}$~ & ~$m_{\tilde{\chi}_1^\pm}$~ & ~$m_{\tilde{\chi}_2^0}$~ & ~$m_{\tilde{\mu}_L}$~ & Rescaled $\sigma_{P}^{\rm{SI}}(cm^2)$ & ~$a_\mu^{\rm{SUSY}}\times10^{10}$~
\\
\hline\hline
 ~$200$~ & ~$32.9$~ & ~$4347.9$~ & ~$1376.4$~ & ~$197.2$~ & ~$199.3$~ & ~$202.4$~ & ~$262.7$~ & ~$5.52\times10^{-47}$~ & ~$17.0$~
\\
\hline
 ~$250$~ & ~$37.1$~ & ~$2976.8$~ & ~$1641.7$~ & ~$246.9$~ & ~$248.9$~ & ~$251.6$~ & ~$308.4$~ & ~$7.07\times10^{-47}$~ & ~$13.7$~
\\
\hline
 ~$300$~ & ~$48.5$~ & ~$5037.1$~ & ~$1797.2$~ & ~$297.0$~ & ~$298.7$~ & ~$301.1$~ & ~$357.9$~ & ~$8.31\times10^{-47}$~ & ~$14.0$~
\\
\hline
\end{tabular}
\end{center}
\end{table}

\subsection{Constraints from cosmology}
Considering the current limits on $f_A$, we find the red-giant bound~\cite{gae1,gae2} on the axion-electron coupling $g_{Aee} < 1.3 \times 10^{-13}$ sets the most stringent astrophysical constraint which gives \cite{S P M}
\begin{eqnarray}
f_A>\frac{\sin^2\beta}{|N|}(3.9\times10^{9}~\rm{GeV}).
\end{eqnarray}
Therefore, the lower bound on axion decay constant $f_A$ for large $ \tan \beta$ is $f_{A}>\frac{7.8 \times 10^{9}}{2 N} ~\mathrm{GeV} $. The upper bounds on $ f_{A} $ in literature usually take a value at which the initial misalignment angle is 1. In this work, however, we restrict our attention to $f_{A}\leq10^{12}~\rm{GeV}$, since the larger the $ f_{A} $ is, the larger the $ p $ is required for a high-quality axion model.

In Appendix.~\ref{APPENDIX B}, we give a brief review on the conventional misalignment mechanism (CMM) for axion production, and also give a discussion about the possibility that the relic density of axion is produced via the kinetic misalignment mechanism (KMM). It then turns out that the KMM not works since our models can not provide a flat enough scalar potential.

Given the failure of the KMM in our models, we assume the cold axion is produced via the CMM in the following discussion. We primarily concentrate on the cosmology of saxions and axinos, and there are at least three concerns to consider: 1. Is there any effects on the relic density of LSP due to the decays of saxions and axinos? 2. Is there a cosmological era dominated by saxions or axinos? 3. Whether the relativistic axion from the decays of saxions or other channel satisfies the constraints from dark radiation or not?

In order to discuss these concerns, we firstly study the decays of saxions and axinos. Thanks to the superpotential shown in Eq.(\ref{effW}), the saxions and axinos with heavy mass can decay to the SM or MSSM particles. The partial widths of saxions decaying to higgses or gauge bosons are~\cite{Kyu Jung Bae 2 one,Kyu Jung Bae 2 two}
\begin{eqnarray}
\Gamma\left(S_{i} \rightarrow h h\right) \approx&& \Gamma\left(S_{i} \rightarrow Z Z\right) \approx \frac{1}{2} \Gamma\left(S_{i} \rightarrow W^{+} W^{-}\right)\nonumber\\
\approx&& \frac{\zeta_{\hat{S}_{i}}^{2}}{16 \pi}\left(\frac{\mu^{2}}{v_{A} / \sqrt{2}}\right)^{2}\left(1-\frac{m_{A^{0}}^{2} \cos ^{2} \beta}{\mu^{2}}\right)^{2} \frac{1}{m_{S_{i}}},\nonumber\\
\Gamma\left(S_{i} \rightarrow h H\right) \approx&& \frac{1}{2} \Gamma\left(S_{i} \rightarrow Z A^{0}\right) \approx \frac{1}{2} \Gamma\left(S_{i} \rightarrow W^{+} H^{-}\right) \approx \frac{1}{2} \Gamma\left(S_{i} \rightarrow W^{-} H^{+}\right)\nonumber\\
\approx&& \frac{\zeta_{\hat{S}_{i}}^{2}}{32 \pi}\left(\frac{m_{A^{0}}^{2} \cos \beta}{v_{A} / \sqrt{2}}\right)^{2} \frac{1}{m_{S_{i}}},\nonumber\\
\Gamma\left(S_{i} \rightarrow H H\right) \approx&& \Gamma\left(S_{i} \rightarrow A^{0} A^{0}\right) \approx \frac{1}{2} \Gamma\left(S_{i} \rightarrow H^{+} H^{-}\right)\nonumber\\
\approx&& \frac{\zeta_{\hat{S}_{i}}^{2}}{16 \pi}\left(\frac{\mu^{2}}{v_{A} / \sqrt{2}}\right)^{2}\left(1+\frac{m_{A^{0}}^{2} \cos ^{2} \beta}{\mu^{2}}\right)^{2} \frac{1}{m_{S_{i}}},\nonumber\\
\Gamma\left(S_{i} \rightarrow g g\right) \approx&& \frac{\left(\alpha_{s} C_{S_{i}}\right)^{2} m_{S_{i}}^{3}}{64 \pi^{3}\left(v_{A} / \sqrt{2}\right)^{2}}.\label{decay1}
\end{eqnarray}
Saxions and axinos can also decay to charginos or neutralinos, and corresponding decay widths are
\begin{eqnarray}
&&\Gamma(S_i\rightarrow neutralinos)\approx \Gamma(S_i\rightarrow charginos)\approx\frac{ \zeta_{\hat{S}_i}^2 }{64\pi}(\frac{\mu}{v_A/\sqrt{2}})^2m_{S_i},\nonumber\\
&&\Gamma(\tilde{a}_i\rightarrow neutralinos+Z)\approx\frac{ \zeta_{\hat{\tilde{a}}_i}^2 }{32\pi}(\frac{\mu}{v_A/\sqrt{2}})^2m_{\tilde{a}_i},\nonumber\\
&&\Gamma(\tilde{a}_i\rightarrow neutralinos+higgses)\approx2\times3\times\frac{ \zeta_{\hat{\tilde{a}}_i}^2 }{64\pi}(\frac{\mu}{v_A/\sqrt{2}})^2m_{\tilde{a}_i},\nonumber\\
&&\Gamma(\tilde{a}_i\rightarrow charginos^\pm+W^\mp)\approx\Gamma(\tilde{a}_i\rightarrow charginos^\pm+H^\mp)\approx\frac{ \zeta_{\hat{\tilde{a}}_i}^2 }{16\pi}(\frac{\mu}{v_A/\sqrt{2}})^2m_{\tilde{a}_i}.\label{decay2}
\end{eqnarray}
Other decay channels have been omitted because they are not the primary decay channels in our assumed parameter space, and they can be found in the literature~\cite{Kyu Jung Bae 2 one,Kyu Jung Bae 2 two}. Using Eq.~(\ref{eq:GMSiAA}), we can calculate the branching ratio of the decay process $S_i\rightarrow AA$, which is denoted by $BR_ {S_i} $
\begin{eqnarray}
B R_{S_{i}} & = & \Gamma\left(S_{i} \rightarrow A A\right) / \Gamma_{\text {total }}\left(S_{i}\right),
\end{eqnarray}
and it would be used to study the dark radiation constraints in the following. To calculate the decay temperature of saxions and axinos, we assume that decays occurred in the radiation dominated period, hence the decay temperature $T_D$ are given by
\begin{eqnarray}
T_D\approx (\frac{90}{\pi^2 g_*(T_D)})^{\frac{1}{4}}\sqrt{\Gamma M_P}.
\end{eqnarray}
In order to guarantee that the decays of the saxions and axinos do not affect the relic density of $\tilde{\chi}$ we given above, saxions and axinos should decay before the LSP freeze-out. When parameter $\mu=250~\rm{GeV}$, corresponding LSP freeze-out temperature $T_{fr}^{\tilde{\chi}}$ is approximated as $10~\rm{GeV}$, this implying that $T_D>10~\rm{GeV}$ is needed.

Secondly, we consider the probability that there exists a cosmological era dominated by saxions or axinos. In early Universe, the reheating temperature $T_R$, the temperature of radiation after inflaton decay, is a very crucial factor that affects the evolution of the Universe. If $T_R$ is larger than the saxions (axinos) decoupling temperature $T_{dcp}$, saxions (axinos) were in the thermal equilibrium. The decoupling temperature $T_{dcp}$ can be approximated as~\cite{Kyu Jung Bae 1,TDCP1,TDCP2,Baer:2011hx}
\begin{eqnarray}
&&T^{S}_{dcp}\approx1.4\times 10^{9} (\frac{f_A}{10^{11}~\rm{GeV}})^2~ \rm{GeV},\nonumber\\
&&T^{\tilde{a}}_{dcp}\approx 10^{11} (\frac{0.1}{\alpha_s})^3(\frac{f_A}{10^{12}~\rm{GeV}})^2 ~\rm{GeV},
\end{eqnarray}
and the yield of saxions and axinos in this case are given by~\cite{Kyu Jung Bae 1}
\begin{eqnarray}
Y_{S_i}^{eq}\approx1.2\times 10^{-3},~Y_{\tilde{a}_i}^{eq}\approx1.8\times 10^{-3}.
\end{eqnarray}
For reheating temperature being smaller than the saxions (axinos) decoupling temperature $T_{dcp}$, saxions (axinos) would never be in thermal equilibrium. In this case, the saxions or axinos could be produced via thermal production or coherent oscillation. In addition, the yield of them may be $T_R$ -dependent or -independent, depending on the model we studied. We can compare the $T_D$ with the saxion/axino-radiation equality temperature $T_e=\frac{4}{3}m (Y^{\rm{TP}}+Y^{\rm{CO}})$ to determine whether the saxions- or axinos- dominated Universe exists or not, where $Y^{\rm{TP}}$ is the yield from thermal production and $Y^{\rm{CO}}$ is the yield from coherent oscillation. If the saxion/axino decay temperature $T_D$ is large than $T_e$, such a cosmological era never occurred.

Finally, we consider constraints from dark radiation. Relativistic axion could be produced from the thermal scattering or saxions decay, and gives contribution to the effective number of neutrinos~\cite{Kyu Jung Bae 1}
\begin{eqnarray}
\Delta N_{\rm{eff}}\approx \Delta N_{\rm{eff}}^{\rm{TP}}+\sum_i \frac{18}{r}BR_{S_i} g_*(T_D^{S_i})^{-\frac{1}{3}}\frac{m_{S_i}(Y_{S_i}^{\rm{CO}}+Y_{S_i}^{\rm{TP}})}{T_D^{S_i}}.\label{eq:DNeff}
\end{eqnarray}
In Eq.~(\ref{eq:DNeff}), $\Delta N_{\rm{eff}}^{\rm{TP}}$ is the contribution from thermally-produced hot axion, and the maximum of it could be estimated as about $9.5\times10^{-3}$ \cite{Kyu Jung Bae 1} which is much smaller than the bound of $\Delta N_{\rm{eff}}^{\rm{exp}}=0.17$~\cite{BBN} and can be safely neglected. The factor $r\simeq\max[1,\sum (\frac{T_e^{\tilde{a}_i}}{T_D^{\tilde{a}_i}}+\frac{T_e^{S_i}}{T_D^{S_i}})]$ represents the entropy dilution caused by saxions and axinos decay~\cite{Kyu Jung Bae 1}. We set $r$ to be 1 in this work, since the $T_e\ll T_D$ are satisfied for all the saxions and axinos in the parameter space of our interest. Besides, a factor $r>1$ provides a smaller $\Delta N_{\rm{eff}}$, which would make the dark radiation constraint less significant.

\subsubsection{Numerical analysis for the model $\rm{E_{I}}$}
In model $\rm{E_{I}}$, we firstly make a numerical analysis based on a hypothesis that the parameters satisfy formulae $ m_{X}=m_{Y}=m_{Z}=m_{\rm{x y z}} $, $ \lambda_{1}=\lambda_{2}=\lambda $, $ a_{1}=a_{2}=a_{\lambda} $ and $ r_{p} \equiv\left(\frac{a_{\lambda}}{\lambda}\right)^{2} / m_{\rm{x y z}}^{2} $, and then give a more general analysis by scanning parameter space for all these seven parameters.

The thermal production of saxions and axinos are via the anomaly interaction~\cite{Kyu Jung Bae 2 one,Kyu Jung Bae 2 two,Baer:2011hx}:
\begin{eqnarray}
\mathcal{L}_{anomaly}=\frac{\alpha_s}{8\pi f_A}(S G^a_{\mu\nu}G^{a\mu\nu}-i\bar{\tilde{a}}\frac{[\gamma^\mu,\gamma^\nu]}{2}\gamma_5 \tilde{g}^a G^{a}_{\mu\nu}).
\end{eqnarray}
These couplings lead to thermally produced densities of saxions and axinos which are dependent on the reheating temperature $T_R$. Unlike the pure DFSZ SUSY axion model such as $\rm{B_{I}}$, $\rm{B_{II}}$ or $\rm{B_{III}}$, where the saxion and axino abundances from thermal production are independent on $T_R$ since the heaviest PQ-charged and gauge-charged matter supermultiplets are the MSSM Higgs doublets $H_u, H_d$ \cite{Kyu Jung Bae 3 one,Kyu Jung Bae 3 two}. The heaviest PQ-charged and gauge-charged matter supermultiplets, in model $\rm{E_{I}}$, are Vectorlike quarks just as in the case of the SUSY KSVZ model. Therefore, the yields of saxion and axino are given by~\cite{Kyu Jung Bae 1,YTP}
\begin{eqnarray}
&&Y_S^{\rm{TP}}\approx1.33\times10^{-5}g_3^6\log(\frac{1.01}{g_3}) (\frac{10^{12}~\rm{GeV}}{f_A})^2(\frac{T_R}{10^8 ~\rm{GeV}}),\nonumber\\
&&Y_{\tilde{a}}^{\rm{TP}}\approx2\times10^{-7}g_3^6\log(\frac{1.211}{g_3}) (\frac{10^{11}~\rm{GeV}}{f_A})^2(\frac{T_R}{10^4 ~\rm{GeV}}).
\end{eqnarray}
The abundance of saxions could also be produced via coherent oscillations. However, the yield $Y_{S_{i}}^{\rm{C O}}$ is quite small for the initial amplitude $s_0\sim f_A\leq10^{12}~ \rm{GeV}$, so we can safely neglect it \cite{Kyu Jung Bae 1}.

\begin{figure}[!htb]
\begin{center}
\begin{minipage}[c]{0.48\textwidth}
\includegraphics[width=3in]{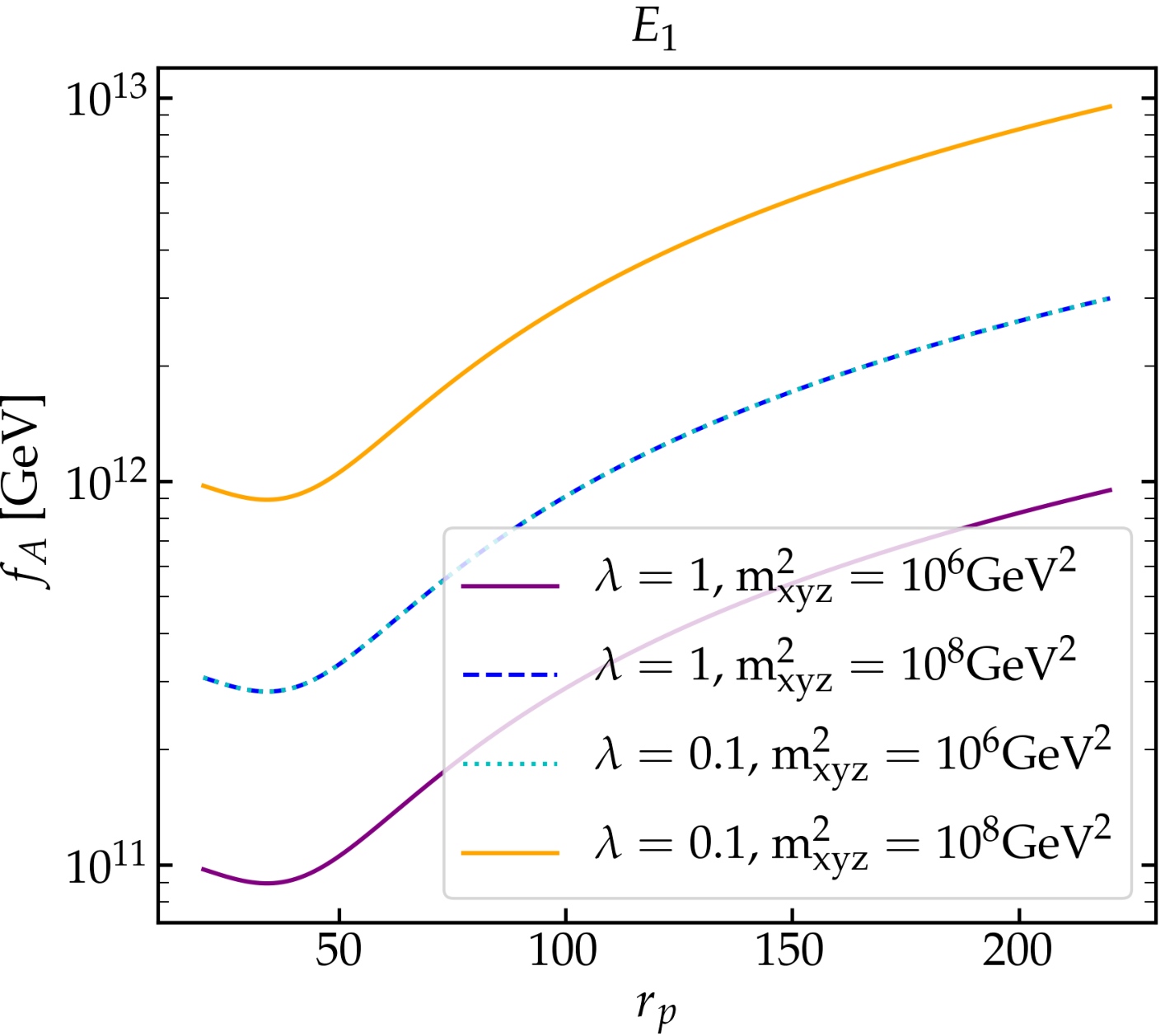}
\end{minipage}%
\begin{minipage}[c]{0.48\textwidth}
\includegraphics[width=3in]{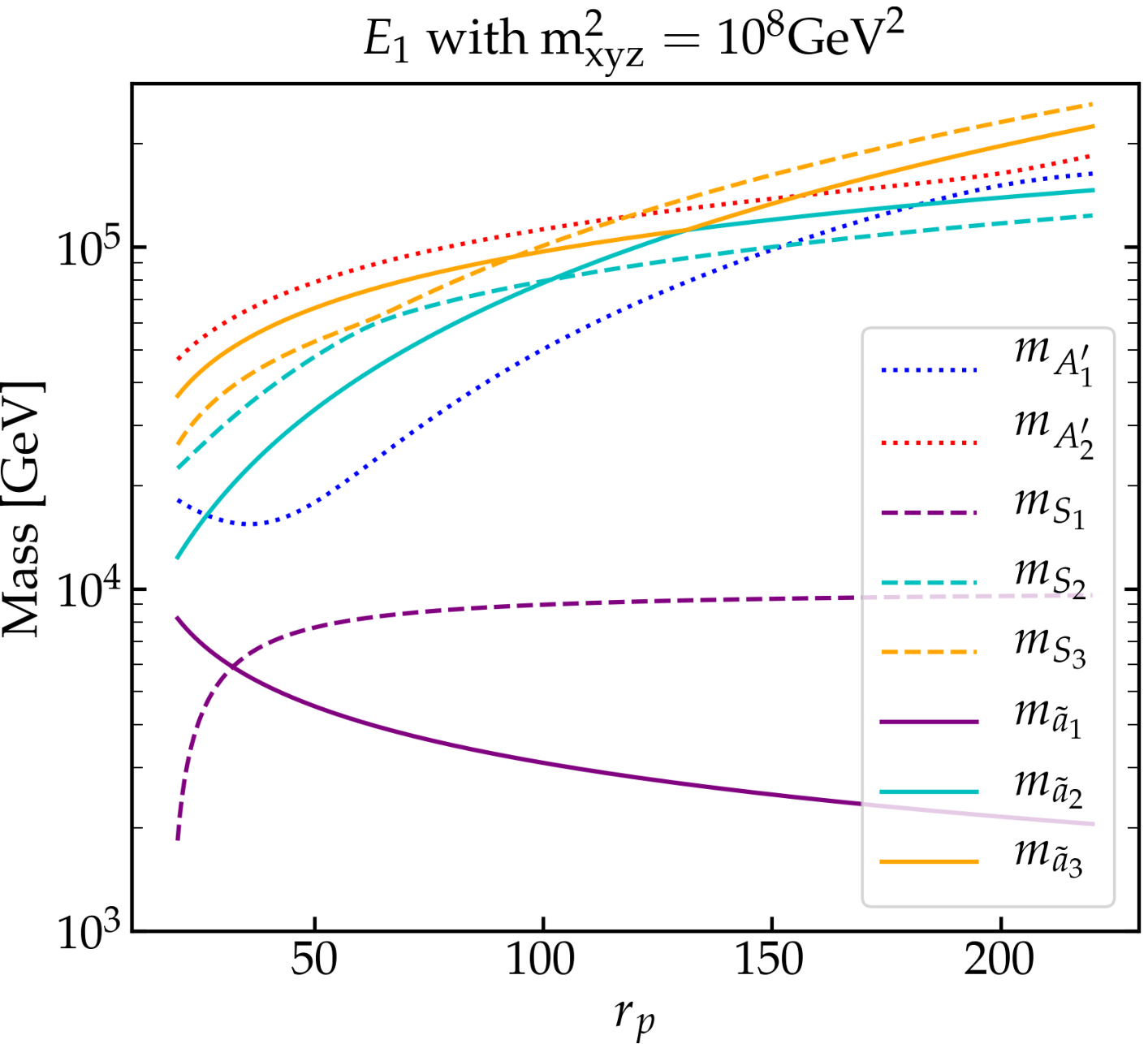}
\end{minipage}
\caption[]{\label{fig:rp} The $f_A$-$r_p$ (left panel) and $m_i$-$r_p$ (right panel) curve charts under the assumption that there are three free parameters $ \lambda $, $r_{p} $ and $ m_{\rm{xyz}}^2 $ in model $\rm{E_I}$.}
\end{center}
\end{figure}

Under the assumption that there are three free parameters $ \lambda $, $r_{p} $ and $ m_{\rm{xyz}}^2 $, we investigate the allowed parameter space for model $\rm{E_I}$. To begin, the $f_A$-$r_p$ and $m_i$-$r_p$ curve charts are shown in Fig.~\ref{fig:rp}, where $i$ represent $A^\prime_i$, $S_i$ and $\tilde{a}_i$.

From the left panel of Fig.~\ref{fig:rp}, we find that the appropriate parameters for $f_{A}\leq10^{12}~\rm{GeV}$ primarily fall within the ranges where $m^2_{\rm{xyz}}$ is small and $\lambda$ is large (e.g., $m^2_{\rm{xyz}}=10^{6}~\rm{GeV^2}$ and $\lambda=1$). Combining Eq.~(\ref{S2}), this behavior can be easily understood even though this equation is only an approximate formula. We can also find that $f_A$ may receive the smallest value if the parameter $r_p$ is approximately equal to $40$ and such a $r_p$ is very attractive due to that in this case the quality problem is more susceptible to avoidance.

In the right panel of Fig.~\ref{fig:rp}, parameter $r_p$ being close to $40$ is also very appropriate because we expect that all these particles have relatively heavy masses and, in this way, they can successfully decay to SM or MSSM particles. When $r_p$ take a larger or smaller value, on the other hand, the lightest particle that shown in Fig.~\ref{fig:rp}, which could be the axino $\tilde{a}_1$ or saxion $S_1$, is usually much lighter than that of $r_p\approx40$ case. As a result, some decay channels in Eqs.~(\ref{decay1}) and (\ref{decay2}) may close and the relic density of LSP may change since the decay temperature $T_D$ may be much lower than the LSP freeze-out temperature $T^{\tilde{\chi}}_{fr}$. Besides, we note that the masses of these particles are positively correlated with the parameter $m_{\rm{xyz}}^2$, and curves in the right panel in Fig.~\ref{fig:rp} would be translated up or down in the coordinate system by adjusting parameter $m_{\rm{xyz}}^2$ (for simplicity, we did not show this in Fig.~\ref{fig:rp}). Given this reason, $m_{\rm{xyz}}^2\geq10^8~ \rm{GeV}^2$ is needed if we expect the masses of these particles are heavier than about $5 ~\rm{TeV}$.

\begin{figure}[!htb]
\begin{center}
\begin{minipage}[c]{0.48\textwidth}
\includegraphics[width=3in]{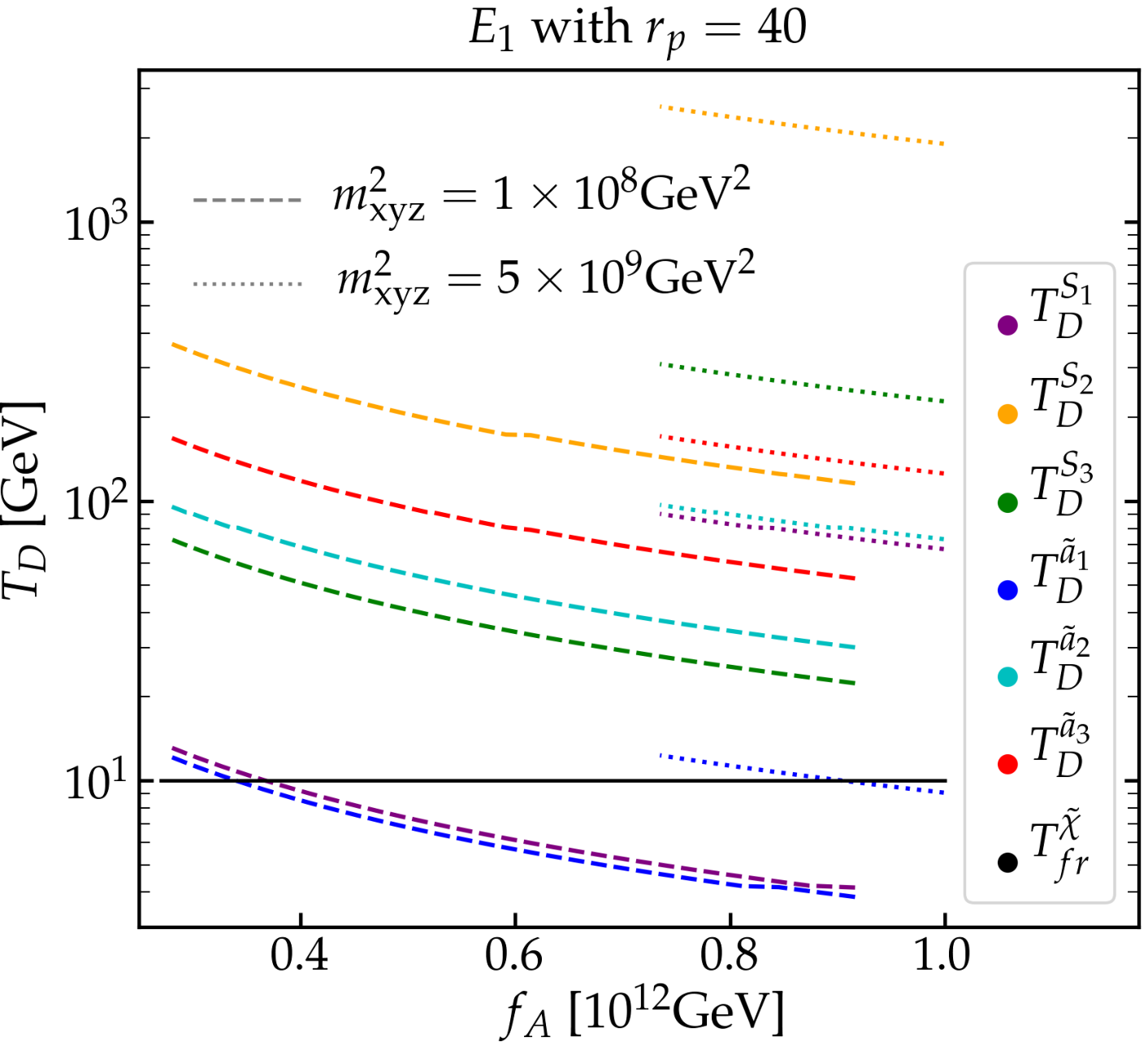}
\end{minipage}%
\begin{minipage}[c]{0.48\textwidth}
\includegraphics[width=3.05in]{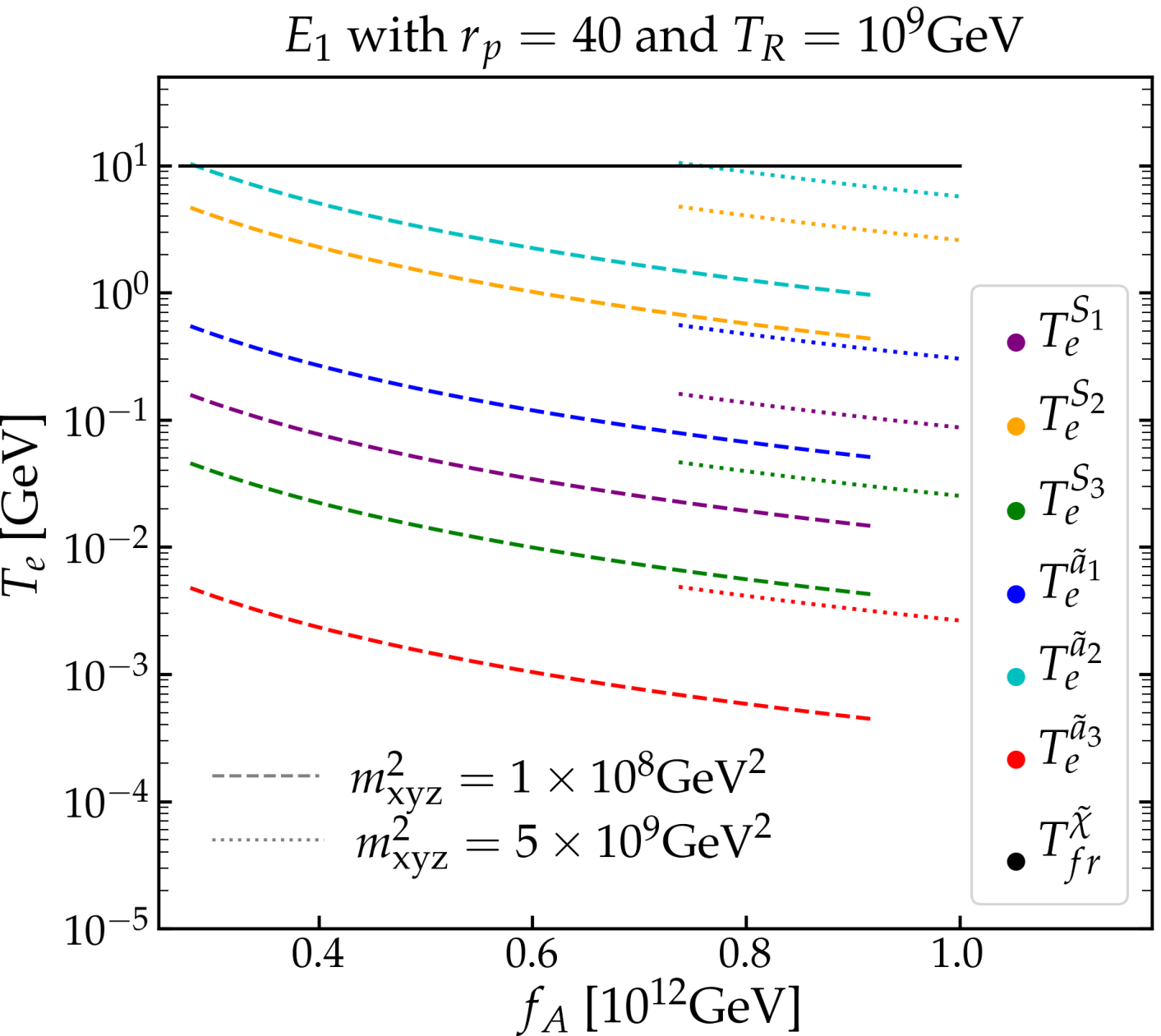}
\end{minipage}
\caption[]{\label{fig:TDTe} The dependence of the decay temperature $T_D$ (left panel) and the saxion/axino-radiation equality temperature $T_e$ (right panel) on $f_A$. Lines colored purple, orange, green, blue, cyan and red belong to $S_1, S_2, S_3, \tilde{a}_1, \tilde{a}_2$ and $\tilde{a}_3$ respectively. The dashed (dotted) line represents that the parameter $m_{\rm{xyz}}^2$ is equal to $1\times10^8~ \rm{GeV}^2$ ($5\times10^9 \rm{GeV}^2$), and the black solid line represents the LSP freeze-out temperature $T_{fr}^{\tilde{\chi}}$ which is approximately $10~\rm{GeV}$.}
\end{center}
\end{figure}

Secondly, we plot the dependence of the decay temperature $T_D$ (left panel) and the saxion/axino-radiation equality temperature $T_e$ (right panel) on $f_A$ in Fig.~\ref{fig:TDTe}, where we have used the parameters $r_p=40$, $\lambda\in[0.1,1]$ and the saxions(axinos) thermal production yield $Y_{S_i}^{\rm{TP}}=\mathrm{Min}[Y_{S_i}^{eq},|C_{S_i}|^2 Y_{S}^{\rm{TP}}]$ ($Y_{{\tilde{a}_i}}^{\rm{TP}}=\mathrm{Min}[Y_{{\tilde{a}_i}}^{eq},|C_{{\tilde{a}_i}}|^2 Y_{\tilde{a}}^{\rm{TP}}]$). In this work, we are interested in the circumstance that $T_R$ is smaller than $10^9~\rm{GeV}$ and, therefore, the parameter $T_R$ for analysis in the right panel is taken to be $10^9~\rm{GeV}$. We did not draw curves for smaller $T_R$ in Fig.~\ref{fig:TDTe}, since smaller $T_R$ would cause $T_e$ to become smaller and the restrictions on parameters would be more relaxed.

As can be seen in Fig.~\ref{fig:TDTe}, $T_e$ are always smaller than $T_D$ in both cases where $m_{\rm{xyz}}^2$ have different values. However, in the case of $m_{\rm{xyz}}^2=10^8~\rm{GeV}^2$, the decay temperatures of $S_1$ and $\tilde{a}_1$ would be smaller than $T_{fr}^{\tilde{\chi}}$ if $f_A$ is larger than about $3.4\times10^{11}~ \rm{GeV}$. In order not to affect the relic density of LSP we assumed, for a large $f_A$, a large $m_{\rm{xyz}}^2$ is also necessary. Based on these analysis, there always exist viable parameter space for $f_A\in[2.7\times10^{11},10^{12}]~\rm{GeV}$, in which the saxions and axinos can decay before LSP freeze-out and saxions- or axinos- dominated era would never occur.

\begin{figure}[!htb]
\begin{center}
\includegraphics[width=3.5in]{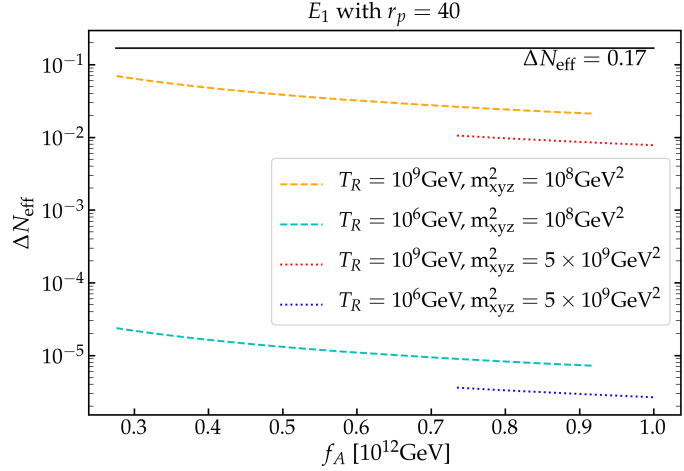}
\caption[]{\label{fig:DNeff} The constraint from dark radiation. The black solid line is the upper bound of $\Delta N_{\rm{eff}}^{\rm{exp}}$~\cite{BBN}, and the dashed (dotted) line represents that the parameter $m_{\rm{xyz}}^2$ is equal to $1\times10^8 ~\rm{GeV}^2$ ($5\times10^9 ~\rm{GeV}^2$). Lines colored red (or orange) and blue (or cyan) belong to $T_R=10^9~\rm{GeV}$ and $T_R=10^6~\rm{GeV}$ respectively.}
\end{center}
\end{figure}

Finally, we give the constraint from dark radiation in Fig.~\ref{fig:DNeff}. We use the parameters $r_p=40$, $\lambda\in[0.1,1]$ and $m_{\rm{xyz}}^2$ being equal to $1\times10^8 ~\rm{GeV}^2$ or $5\times10^9 ~\rm{GeV}^2$. For $T_R \in [10^6,10^9] ~\rm{GeV}$, it is clear that the dark radiation constraint can be easily satisfied for all $f_A\in[2.7\times10^{11},10^{12}]~\rm{GeV}$.

As a more general case, there are seven free parameters that are actually not restricted by the relational expressions we have imposed. Therefore, the parameter space is much larger than what we discussed above. Clearly, it is difficult to analyze such a complicated parameter space shortly, and scanning the appropriate parameter space with necessary restrictions would be more economical and efficient. Thus, we present the results here resorting to the scans of the following parameter space:
\begin{eqnarray}
&&\lambda_1,\lambda_2\in[0.01,10],\nonumber\\
&&a_1,a_2\in[10^3,10^5]~\rm{GeV},\nonumber\\
&&m_{X}^2,m_{Y}^2,m_{Z}^2\in[10^6,10^{10}]~\rm{GeV}^2.
\end{eqnarray}
The restrictions that should be imposed on these parameters are given as follows: 1. The VeVs of $X, Y$ and $Z$ are in $[10^9,10^{12}]~\rm{GeV}$, so that the resulted axion is an invisible axion. 2. The masses of saxions and axinos are heavier than $2~\rm{TeV}$, so that the dominant decay channels to MSSM or SM particles can safely open. 3. The decay temperatures $T_D$ of saxions (axinos) are larger than the $T_e$ of them, so that the saxions- or axinos- dominated era would never occur. 4. The decay temperatures $T_D$ of saxions (axinos) are larger than the LSP freeze-out temperature $T_{fr}^{\tilde{\chi}}$, so that the relic density of LSP would not be affected by saxions or axinos decays. 5. The resulting effective number of neutrinos is under the upper bound given by Ref.~\cite{BBN}.

\begin{figure}[!htb]
\begin{center}
\begin{minipage}[c]{0.48\textwidth}
\includegraphics[width=3in]{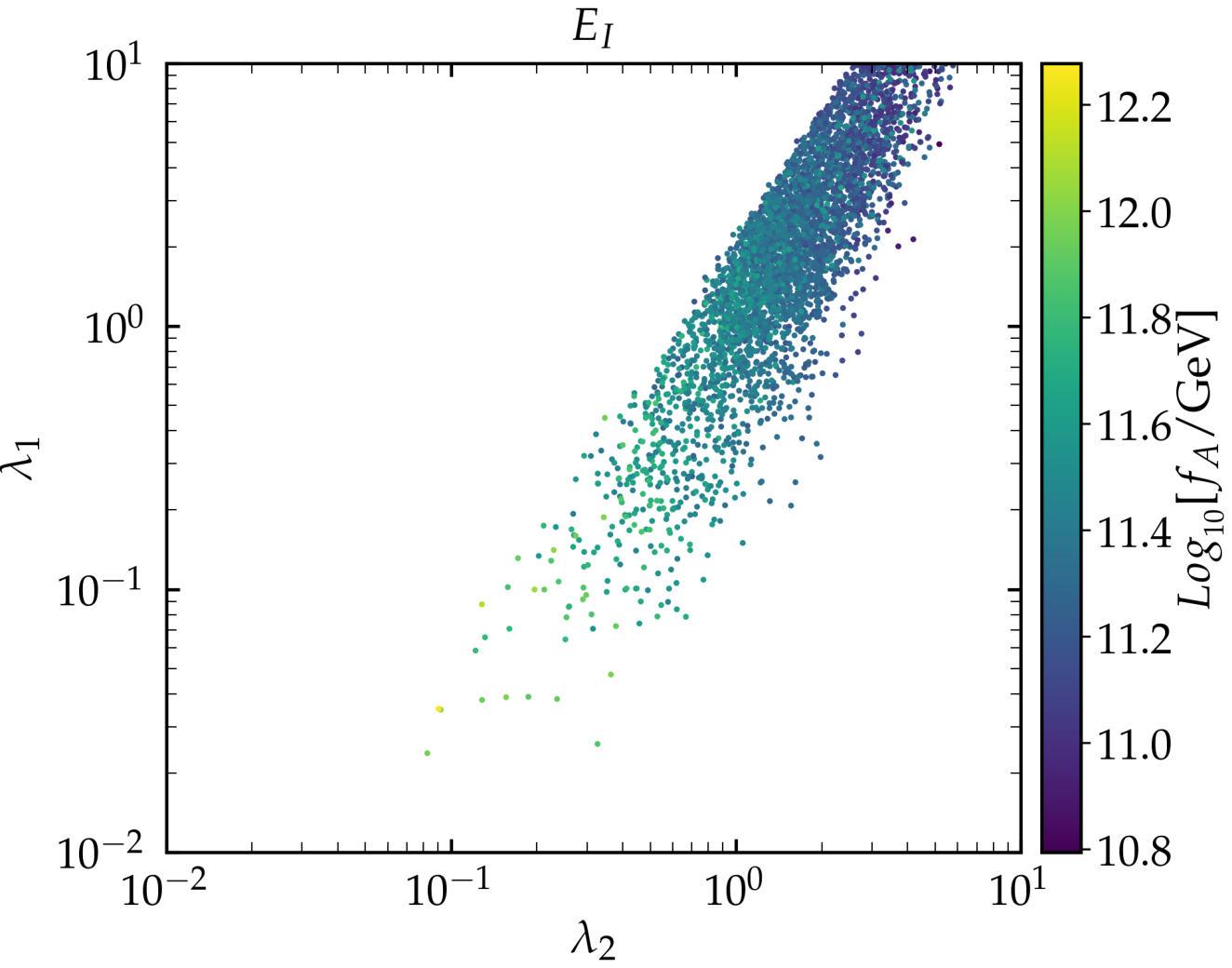}
\end{minipage}%
\begin{minipage}[c]{0.48\textwidth}
\includegraphics[width=3in]{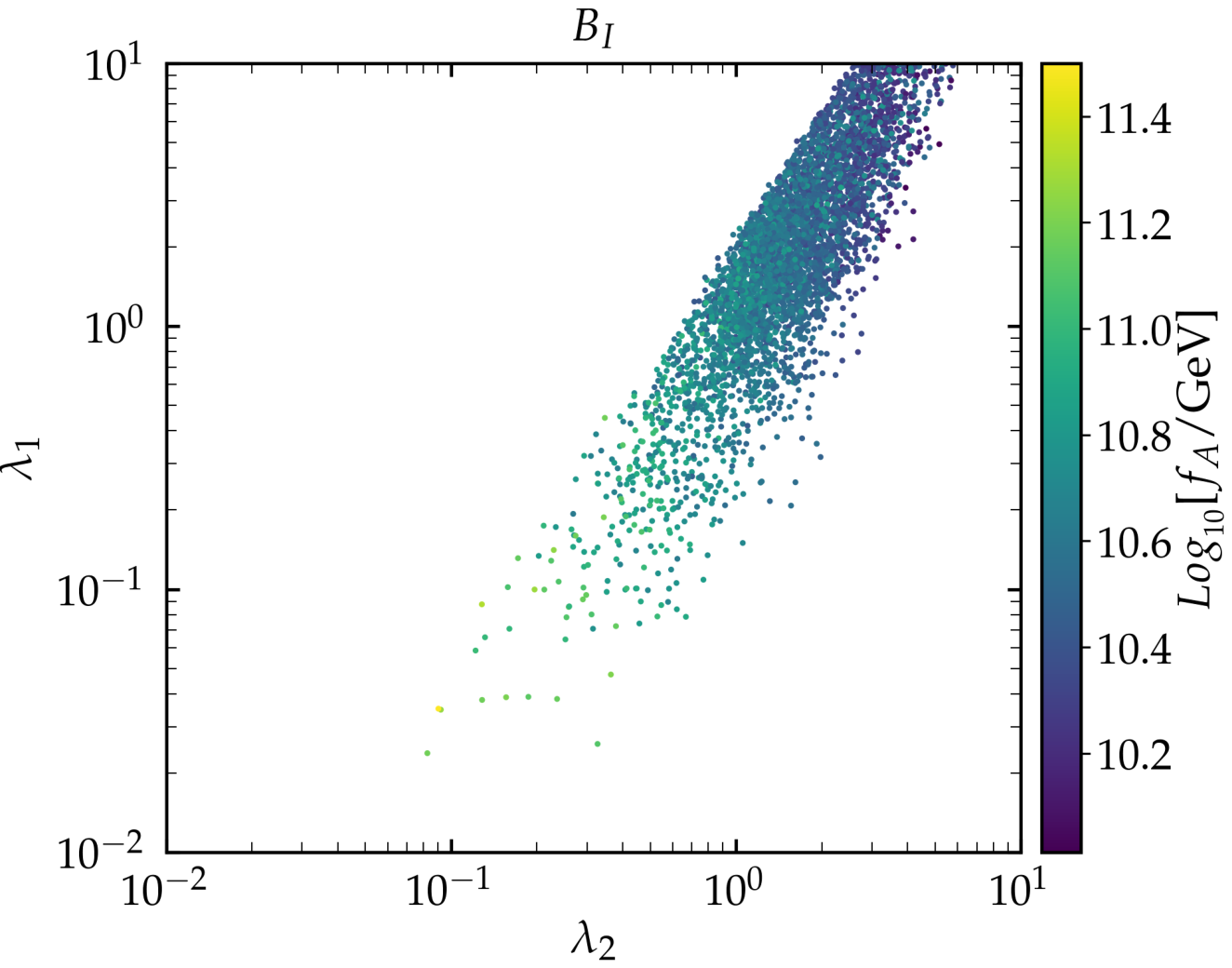}
\end{minipage}\\
\begin{minipage}[c]{0.48\textwidth}
\includegraphics[width=3in]{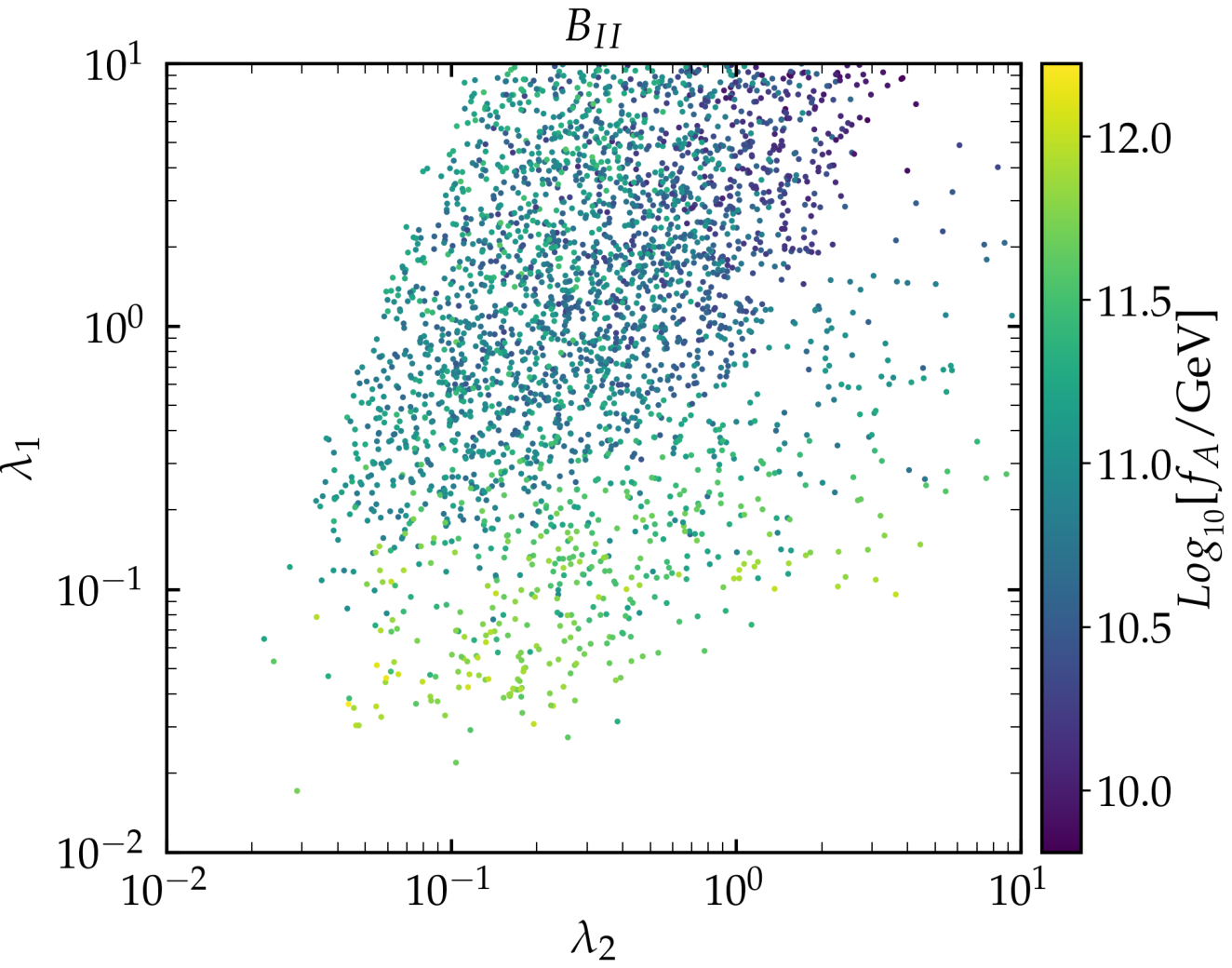}
\end{minipage}%
\begin{minipage}[c]{0.48\textwidth}
\includegraphics[width=3in]{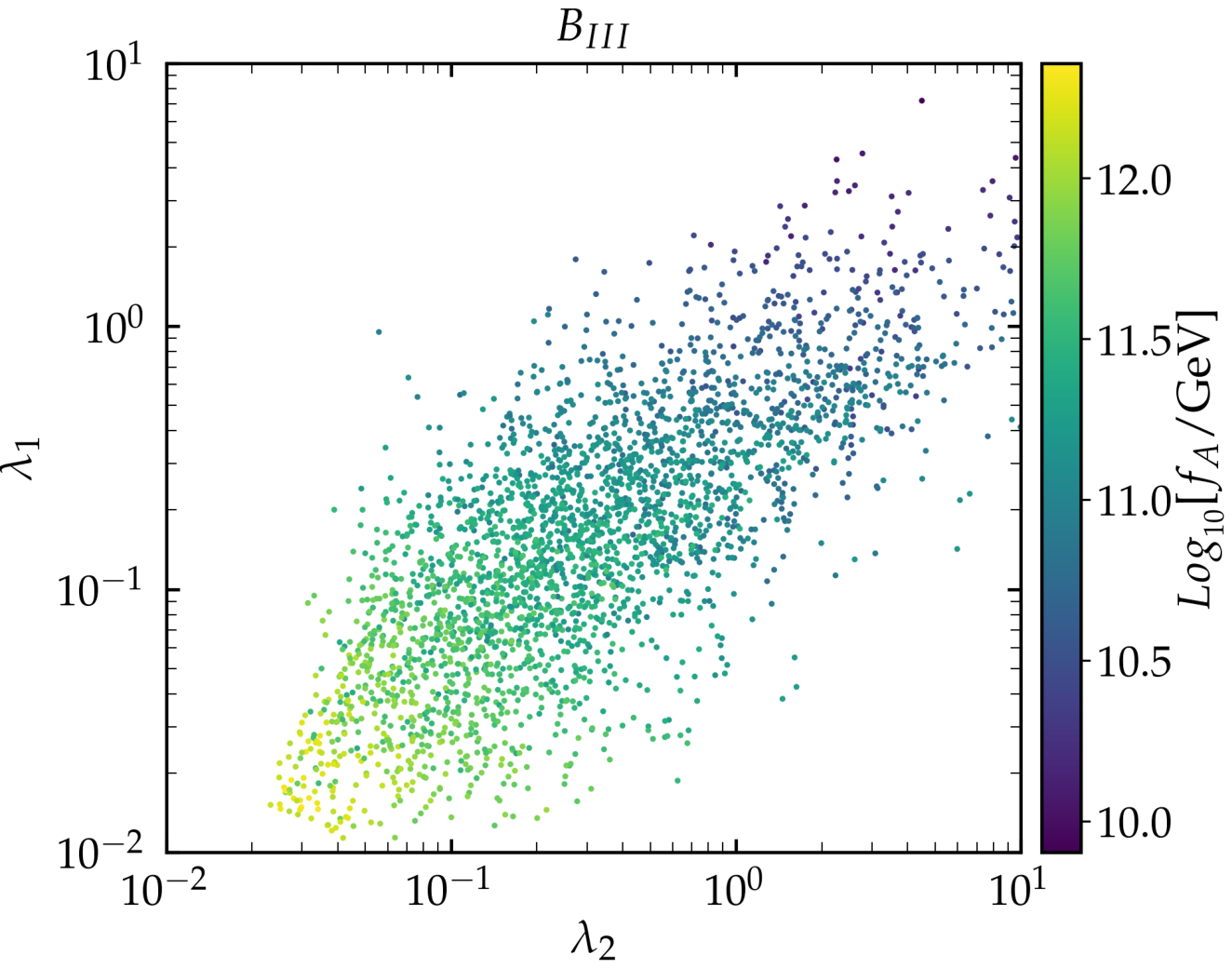}
\end{minipage}
\caption[]{\label{fig:L1L2} The parameter space of $\lambda_1$ versus $\lambda_2$ for different models. The sampling points are colored with various hues according to the value of $f_A$.}
\end{center}
\end{figure}
Under these restrictions, the viable parameter space for $\lambda_i$ has been plotted in Fig.~\ref{fig:L1L2} ($\rm{E_I}$). In this picture, the sampling points are colored with various hues according to the value of $f_A$. For $\lambda_1,\lambda_2\in[0.1,1]$, as one expects, the smaller $f_A$ gather in the corner of $\lambda_1\sim\lambda_2\sim1$ just like the case of $\lambda_1 = \lambda_2 = \lambda$. In addition, we can also find that the resulted $f_A$ can be much smaller than $2.7\times10^{11}~\rm{GeV}$, so that the smallest $f_A$ that this model can offer could be roughly obtained by such a scan on parameter space. In model $\rm{E_I}$, since the DW number $N_{\rm{DW}}=1$, the axion relic density can be given by Eq.~(\ref{OmegaPIA}) corresponding to post-inflationary scenario. With the assumption that the relic density of axion $\Omega_A h^2$ is approximately $0.113$, the axion decay constant $f_A\approx 1.11(2.59)\times10^{11}~ \rm{GeV}$ is required for $\langle \theta_i^2 \rangle\simeq 2.96^2(1.81^2)$, and this can easily be achieved in the assumed parameter space. On the other hand, the PQ symmetry may be broken before or during the inflation. In such a pre-inflationary scenario, the relic density of axion $\Omega_A h^2\approx0.113$ is more easily to be realized, and we plot the misalignment angle required in Fig.~\ref{fig:PREIA}.
\begin{table}[!htb]
\begin{center}
\begin{minipage}[]{0.95\linewidth}\caption{Sample points for different models. The cases 1 and 2 in this table correspond to the smallest $f_A$ in all samples with parameters $\lambda_{1,2}\in[0.1,1]$ and $[1,10]$ respectively. In the following, we take the values of $f_A$ in case 2 as the lower limits of the available $f_A$ for each model.\label{sample points}}
\end{minipage}

\vspace{0.3cm}

\begin{tabular}{|c | c | c | c | c | c | c | c| c| }
\hline
Models & \multicolumn{2}{|c|}{ $\rm{E_I}$ } & \multicolumn{2}{|c|}{ $\rm{B_I}$ } & \multicolumn{2}{|c|}{ $\rm{B_{II}}$ } & \multicolumn{2}{|c|}{ $\rm{B_{III}}$} \\
\hline
  Samples & ~case\ 1~ & ~case\ 2~ & ~case\ 1~ & ~case\ 2~ & ~case\ 1~ & ~case\ 2~ & ~case\ 1~ & ~case\ 2~
\\
\hline\hline
 ~$\lambda_1$~ & ~$0.74$~ & ~$2.14$~ & ~$0.76$~ & ~$3.36$~ & ~$0.94$~ & ~$8.72$~ & ~$0.84$~ & ~$7.21$~
\\
\hline
 ~$\lambda_2$~ & ~$0.98$~ & ~$4.19$~ & ~$0.75$~ & ~$3.94$~ & ~$0.38$~ & ~$3.81$~ & ~$0.98$~ & ~$4.49$~
\\
\hline
 ~$a_1[\rm{TeV}]$~ & ~$21.584$~ & ~$50.927$~ & ~$7.8217$~ & ~$97.109$~ & ~$16.168$~ & ~$2.6443$~ & ~$16.039$~ & ~$99.894$~
\\
\hline
 ~$a_2[\rm{TeV}]$~ & ~$21.369$~ & ~$92.145$~ & ~$13.089$~ & ~$45.996$~ & ~$6.4091$~ & ~$61.504$~ & ~$15.595$~ & ~$40.330$~
\\
\hline
 ~$m_{X}^2[\rm{TeV}^2]$~ & ~$3.6921$~ & ~$1.3270$~ & ~$2.0396$~ & ~$1.7003$~ & ~$8.3105$~ & ~$1.3887$~ & ~$19.376$~ & ~$6.8184$~
\\
\hline
 ~$m_{Y}^2[\rm{TeV}^2]$~ & ~$1738.7$~ & ~$615.01$~ & ~$187.74$~ & ~$536.27$~ & ~$932.00$~ & ~$927.96$~ & ~$22.229$~ & ~$16.492$~
\\
\hline
 ~$m_{Z}^2[\rm{TeV}^2]$~ & ~$45.075$~ & ~$65.873$~ & ~$29.698$~ & ~$14.692$~ & ~$19.147$~ & ~$1.3642$~ & ~$1.2364$~ & ~$2.6549$~
\\
\hline
 ~$m_{S_1}[\rm{TeV}]$~ & ~$10.066$~ & ~$8.6362$~ & ~$6.2595$~ & ~$3.2893$~ & ~$7.7607$~ & ~$10.662$~ & ~$5.1166$~ & ~$2.9237$~
\\
\hline
 ~$m_{\tilde{a}_1}[\rm{TeV}]$~ & ~$2.3640$~ & ~$2.5596$~ & ~$2.0590$~ & ~$2.5547$~ & ~$2.8342$~ & ~$2.1732$~ & ~$4.8160$~ & ~$3.3864$~
\\
\hline
 ~$f_A[10^{9}~\rm{GeV}]$~ & ~$160.99$~ & ~$78.286$~ & ~$25.913$~ & ~$10.330$~ & ~$19.117$~ & ~$6.4615$~ & ~$26.829$~ & ~$8.0113$~
\\
\hline
\end{tabular}
\end{center}
\end{table}
\subsubsection{Numerical analysis for the models $\rm{B_{I}}$, $\rm{B_{II}}$ and $\rm{B_{III}}$}
In base models $\rm{B_{I}}$, $\rm{B_{II}}$ and $\rm{B_{III}}$, the axion decay constant $f_A$ is changed to be $\frac{v_A}{2N}$, where $N=3$. The heaviest PQ-charged and gauge-charged matter supermultiplets are the MSSM Higgs doublets $H_u$ and $H_d$, so the yields of saxion and axino in such types of models, following Ref.~\cite{Kyu Jung Bae 2 one}, are given by
\begin{eqnarray}
Y_{S_i}^{\mathrm{TP}} & \approx & 10^{-7} |C_{S_i}|^2\left(\frac{ \mu}{\mathrm{TeV}}\right)^{2}\left(\frac{10^{12} ~\mathrm{GeV}}{f_{A}}\right)^{2}, \\
Y_{\tilde{a}_i}^{\mathrm{TP}} & \approx & 10^{-7} |C_{\tilde{a}_i}|^2\left(\frac{\mu}{\mathrm{TeV}}\right)^{2}\left(\frac{10^{12} ~\mathrm{GeV}}{f_{A}}\right)^{2}.
\end{eqnarray}
Similar to the $\rm{E_I}$ case, we present the viable parameter space of the parameter $\lambda_1$ versus the parameter $\lambda_2$ in Fig.~\ref{fig:L1L2} ($\rm{B_{I}}$), ($\rm{B_{II}}$) and ($\rm{B_{III}}$) for models $\rm{B_{I}}$, $\rm{B_{II}}$ and $\rm{B_{III}}$ respectively. The parameter space scanned and corresponding restrictions are the same as in the $\rm{E_I}$ case. From these figures, we find that there remain large areas that can offer appropriate $f_A$ in all these base models, and the lower limits of the available $f_A$ can also be roughly estimated. In Table \ref{sample points}, we present some sample points so that more information about these models can be conveyed to the reader.
\begin{figure}[!htb]
\begin{center}
\includegraphics[width=3.5in]{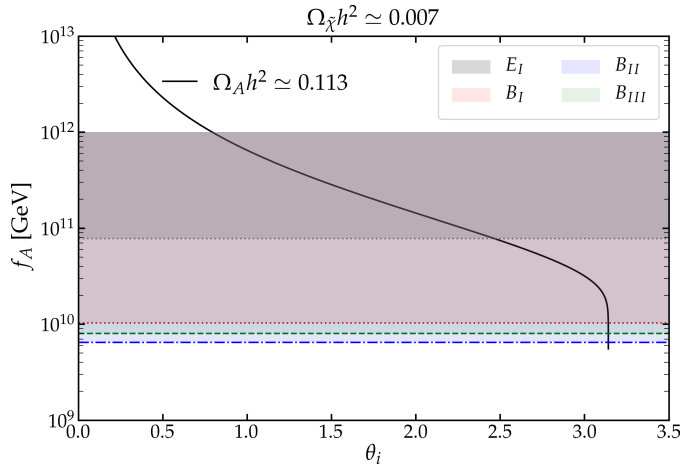}
\caption[]{\label{fig:PREIA} The misalignment angle required to match the relic density of axion $ \Omega_{A} h^{2} \simeq 0.113 $ in pre-inflationary scenario. The grey, red, blue and green lines represent $f_A$ being the values that belong to case 2 of the models $\rm{E_{I}}$, $\rm{B_{I}}$, $\rm{B_{II}}$ and $\rm{B_{III}}$ in the Table \ref{sample points} respectively, and the solid black line gives the relic density of axion $ \Omega_{A} h^{2} \simeq 0.113 $. We have filled the gaps between the upper boundary ($f_A=10^{12}~\rm{GeV}$) and the lower limits (the colored lines) for these four models in different transparent colors.}
\end{center}
\end{figure}

Based on these analysis, for the relic density of axion $\Omega_A h^2\simeq0.113$, the required initial condition $\theta_i$ in pre-inflationary scenario is shown in Fig.~\ref{fig:PREIA}, and the available $f_A$ in each model has been marked with different transparent colors. From this figure, it is clear that all these models can provide an appropriate $f_A$ that gives the correct axion dark matter relic density. For base models, a $f_A$ as small as $10^{10}~\rm{GeV}$ can be derived. The initial condition $\theta_i$, meanwhile, should be close to $\pi$. On the other hand, $\theta_i\lesssim2.5$ is required for model $\rm{E_I}$ due to the lower limit of $f_A$.
\subsubsection{Discussion on the cosmological moduli problem}
When we embed the SUSY setup into the string framework, light moduli are generally present and this may lead to the so-called cosmological moduli problem~\cite{Coughlan:1983ci,Banks:1993en,deCarlos:1993wie,Endo:2006zj}. In the era of inflation, the field value of the lightest modulus $\phi$ may initially be driven to a large value, $\phi_0\sim M_P$. When the expansion rate is close to the mass of the modulus field, Hubble friction ceases, and coherent oscillations (CO) set in. Thereafter, the modulus may quickly come to dominate the energy density of the universe. This may disrupt the successful prediction of light element abundances during Big-Bang nucleosynthesis (BBN) since the modulus could potentially decay after BBN. The gravitino $\psi_{3/2}$, as one of the decay products of the modulus, may also decay after BBN. Additionally, the modulus or gravitino may overproduce LSP dark matter if they decay after the freeze-out of the LSP.

The modulus can decay into gauge sector via the Moroi-Randall (MR) operator~\cite{Moroi:1999zb}
\begin{eqnarray}
\mathcal{L}_{gauge}=\int d^{2}\theta\frac{\lambda_{gauge}}{M_{P}}\phi W^{\alpha}W_{\alpha}+\mathrm{h.c.},
\end{eqnarray}
where the modulus (expanding into the real and imaginary components) $\phi=\frac{1}{\sqrt{2}}(\phi_{R}+i\phi_{I})$ and this operator would lead to helicity-suppressed decay into gauginos. Concretely, the decay widths of modulus into gauginos are estimated to be $\Gamma(\phi_R \rightarrow gauginos)\sim m_{gauginos}^2 m_{\phi_R}/M_{P}^2$, and the unsuppressed decay widths into gauge fields are given by~\cite{Bae:2022okh}
\begin{eqnarray}
&&\Gamma_{\phi_R\to W^+W^-}=\frac{\lambda^2_{\mathrm{SU}(2)}}{4\pi}\frac{m_{\phi_R}^3}{M_P^2}\left(1-4\frac{m_W^2}{m_{\phi_R}^2}+6\frac{m_W^4}{m_{\phi_R}^4}\right)\lambda^{1/2}\left(1,\frac{m_W^2}{m_{\phi_R}^2},\frac{m_W^2}{m_{\phi_R}^2}\right),\nonumber\\
&&\Gamma_{\phi_R\to Z^0Z^0}=\frac{\left(\lambda_{\mathrm{SU}(2)}\cos^2\theta_W+\lambda_{\mathrm{U}(1)}\sin^2\theta_W\right)^2}{8\pi}\frac{m_{\phi_R}^3}{M_P^2}\left(1-4\frac{m_Z^2}{m_{\phi_R}^2}+6\frac{m_Z^4}{m_{\phi_R}^4}\right)\nonumber\\
&&{\ }{\ }{\ }{\ }{\ }{\ }{\ }{\ }{\ }{\ }{\ }{\ }{\ }{\ }{\ }{\ }\times\lambda^{1/2}\left(1,\frac{m_Z^2}{m_{\phi_R}^2},\frac{m_Z^2}{m_{\phi_R}^2}\right),\nonumber\\
&&\Gamma_{\phi_{R}\rightarrow\gamma\gamma}=\frac{\left(\lambda_{\mathrm{SU}(2)}\sin^2\theta_{W}+\lambda_{\mathrm{U}(1)}\cos^2\theta_{W}\right)^2}{8\pi}\frac{m_{\phi_R}^3}{M_{P}^2},\nonumber\\
&&\Gamma_{\phi_R\to Z^0\gamma}=\frac{\sin^22\theta_W\left(\lambda_{\mathrm{SU}(2)}-\lambda_{\mathrm{U}(1)}\right)^2}{16\pi}\frac{m_{\phi_R}^3}{M_P^2}\left(1-\frac{m_Z^2}{m_{\phi_R}^2}\right)^3,\nonumber\\
&&\Gamma_{\phi_R\to gg}=\frac{\lambda_{\mathrm{SU}(3)}}{\pi}\frac{m_{\phi_R}^{3}}{M_{P}^{2}}.\label{phidecay1}
\end{eqnarray}
The modulus can couple to the Higgs sector via operator~\cite{Moroi:1999zb}
\begin{eqnarray}
\mathcal{L}_H=\frac{\lambda_H}{M_P}\int d^4\theta\hat{\phi}\hat{H}_d^*\hat{H}_u^*+\mathrm{h.c.},
\end{eqnarray}
however, this term would be explicitly forbidden by the PQ symmetry if we adopt that the PQ charge of modulus superfield is zero. Hence, we take $\lambda_H=0$. For the modulus-matter interactions which may similarly be forbidden, they may have the form as following~\cite{Moroi:1999zb}:
\begin{eqnarray}
\mathcal{L}_{\mathrm{Q^{\dag}Q}}=\int d^{4}\theta\frac{\lambda_{\mathrm{Q^{\dag}Q}}}{M_{P}}\hat{\phi} \hat{Q}^{\dag}\hat{Q}+\mathrm{h.c.},
\end{eqnarray}
we also assume $\lambda_{\mathrm{Q^{\dag}Q}}=0$ for simplicity in this work. For that the modulus superfield $\hat{\phi}$ couples to the PQ sector, the following operators can be allowed in the K$\rm{\ddot{a}}$hler potential~\cite{Baer:2023bbn}:
\begin{eqnarray}
\mathcal{L}_{PQ}&&\supset\int d^4\theta\left[\frac{\lambda_X}{M_P}\hat{\phi}\hat{X}^\dagger\hat{X}+\frac{\lambda_Y}{M_P}\hat{\phi}\hat{Y}^\dagger\hat{Y}+
\frac{\lambda_Z}{M_P}\hat{\phi}\hat{Z}^\dagger\hat{Z}+\mathrm{h.c.}\right]\nonumber\\
&&\supset\frac{1}{2M_P}\lambda_{PQ}\int d^4\theta\left(\hat{\phi}+\hat{\phi}^\dagger\right)\left(\hat{A}+\hat{A}^\dagger\right)^2,
\end{eqnarray}
where the dimensionless coupling $\lambda_{PQ}=\frac{\lambda_XQ_X^2v_X^2+\lambda_YQ_Y^2v_Y^2+\lambda_ZQ_Z^2v_Z^2}{v_A^2}$. The interactions between the modulus and the PQ sector then can be given by:
\begin{eqnarray}
\mathcal{L}_{\hat{\phi} \hat{A}\hat{A}}\supset-\frac{\lambda_{PQ}}{2\sqrt{2}M_P}\left[SS\partial^2{\phi_R}+2{\phi_R} S\partial^2S-AA\partial^2{\phi_R}+2{\phi_R} A\partial^2A\right]+i\frac{\lambda_{PQ}}{\sqrt{2}}{\phi_R}\overline{\tilde{a}}\gamma_\mu\partial^\mu\tilde{a}.
\end{eqnarray}
Therefore, the corresponding decay widths all can be expressed as
\begin{eqnarray}
&&\Gamma\left({\phi_R}\rightarrow AA\right) =\frac{\lambda_{PQ}^2}{64\pi}\frac{m_{\phi_R}^3}{M_P^2}\left(1-2\frac{m_A^2}{m_{\phi_R}^2}\right)^2\lambda^{1/2}\left(1,\frac{m_A^2}{m_{\phi_R}^2},\frac{m_A^2}{m_{\phi_R}^2}\right)  \nonumber\\&&\Gamma\left({\phi_R}\rightarrow {S_i} {S_i}\right) =\frac{\lambda_{PQ}^2 |C_{S_i}|^2}{64\pi}\frac{m_{\phi_R}^3}{M_P^2}\left(1+2\frac{m_{S_i}^2}{m_{\phi_R}^2}\right)^2\lambda^{1/2}\left(1,\frac{m_{S_i}^2}{m_{\phi_R}^2},\frac{m_{S_i}^2}{m_{\phi_R}^2}\right)
\nonumber\\&&\Gamma\left({\phi_R}\rightarrow\overline{\tilde{a}_i}\tilde{a}_i\right) =\frac{\lambda_{PQ}^{2} |C_{\tilde{a}_i}|^2}{8\pi}\frac{m_{\tilde{a}_i}^{2} m_{\phi_R}}{M_{P}^{2}}\left(1-4\frac{m_{\tilde{a}_i}^{2}}{m_{\phi_R}^{2}}\right)\lambda^{1/2}\left(1,\frac{m_{\tilde{a}_i}^{2}}{m_{\phi_R}^{2}},\frac{m_{\tilde{a}_i}^{2}}{m_{\phi_R}^{2}}\right).
\end{eqnarray}
where $\lambda(a,b,c)=a^2+b^2+c^2-2(ab+bc+ca)$.
The modulus can decay into gravitino pairs as well with a helicity-suppressed width~\cite{Endo:2006zj,Nakamura:2006uc,Dine:2006ii}
\begin{eqnarray}
\Gamma(\phi_R\to\psi_{3/2}\psi_{3/2})=\frac{1}{288\pi}d_{3/2}^2\frac{m_{3/2}^{2}m_{\phi_R}}{M_P^2}\left(1-4m_{3/2}^2/m_{\phi_R}^2\right)^{1/2},
\end{eqnarray}
where $d_{3/2}$ is a dimensionless constant expected to be of order unity. On the other hand, the decay of the unstable gravitino is also important and has been well studied~\cite{Kohri:2005wn,Baer:2011uz,Bae:2014rfa}. Typically, the decay width of $\psi_{3/2}$ is $\Gamma_{3/2}\simeq c/4\pi(m_{3/2}^{3}/M_{P}^{2})$, where $c$ is a constant of order unity. In Ref.~\cite{Baer:2023bbn}, it points out that, in a $\phi \text{PQMSSM}$ model, the dominant decay mode is $\psi_{3/2}\rightarrow \text{gauge boson+gauginos}$ and the branching fractions into PQ sector are below the $1\%$ level, which also applies to our work.

Several crucial moments in the evolution of the universe are associated with the modulus, and the following formulae for estimating temperatures can all be found in Ref.~\cite{Moroi:1999zb}. The first critical temperature is the temperature at which the modulus begins to oscillate, denoted as $T_{osc}^{\phi_R}$:
\begin{equation}
T^{\phi_R}_{\text{osc}}\simeq\left\{
\begin{array}{cl}
\left(10/(\pi^2g_*(T^{\phi_R}_{\text{osc}}))\right)^{1/4}\sqrt{M_Pm_{\phi_R}}&(T^{\phi_R}_{\text{osc}}\leq T_R)\\
\left(10g_*(T_R)/(\pi^2g_*^2(T^{\phi_R}_{\text{osc}}))\right)^{1/8}\left(T_R^2M_Pm_{\phi_R}\right)^{1/4}&(T^{\phi_R}_{\text{osc}}>T_R).\\
\end{array} \right.
\end{equation}
which corresponds to that $3H(T^{\phi_R}_{\mathrm{osc}})\sim m_{\phi_R}(T^{\phi_R}_{\mathrm{osc}})$. Shortly after oscillation, the modulus begins to dominate the energy density of the universe at a temperature $T^{\phi_R}_e$:
\begin{equation}
T^{\phi_R}_e\equiv\left\{
\begin{array}{cl}
\left(15/(\pi^2g_*(T^{\phi_R}_e))\right)^{1/4}\sqrt{m_{\phi_R}\phi_0}&(T^{\phi_R}_{\mathrm{osc}}<T^{\phi_R}_e<T_R)\\
(3/2M_P^2)\phi_0^2\sqrt{M_Pm_{\phi_R}}\left(10/(\pi^2g_*(T^{\phi_R}_e))\right)^{1/4}&(T^{\phi_R}_e<T^{\phi_R}_{\mathrm{osc}}<T_R)\\
(3/2M_P^2)\phi_0^2T_R&(T^{\phi_R}_e<T_R<T^{\phi_R}_{\mathrm{osc}}).
\end{array} \right.
\end{equation}
Finally, modulus $\phi_R$ would decay at a temperature $T^{\phi_R}_D\simeq\sqrt{\Gamma_{\phi_R} M_P}/(\pi^2g_*/90)^{1/4}$, where $\Gamma_{\phi_R}$ is the total decay width of the modulus given above. Similarly, the decay temperature of gravitino $\psi_{3/2}$ is $T^{3/2}_D\simeq\sqrt{\Gamma_{3/2}M_P}/(\pi^2g_*/90)^{1/4}$.

In the following discussion, we assume that gravitino decay occurs before BBN, ensuring that it does not disrupt the successful BBN. The decay temperature of the gravitino $T_D^{3/2}$ must be greater than $T_{BBN}\sim 3-5 ~\rm{MeV}$. This, in turn, sets a lower bound on the gravitino mass, which is roughly estimated to be $50 ~\rm{TeV}$. Therefore, we fix $m_{3/2}=50 ~\rm{TeV}$.

For the the parameter $\phi_0$, we fix $\phi_0=\sqrt{2/3}M_P$. As declared in Ref.~\cite{Moroi:1999zb}, this is the maximum value of $\phi_0$ that is consistent with a radiation-dominated universe at $T = T_R$. The coupling constants $\lambda_{{SU}(3)},\lambda_{{SU}(2)}$ and $\lambda_{{U}(1)}$ in Eq.~(\ref{phidecay1}) are assumed to be unity, and the parameter $\lambda_{PQ}\rightarrow0$, as a limit case,
has been used. Consequently, the modulus mainly decays into the $\text{gauge boson+gauge boson}$, and the branching fractions into PQ sector can be ignored.

\begin{figure}[!htb]
\begin{center}
\begin{minipage}[c]{0.48\textwidth}
\includegraphics[width=3in]{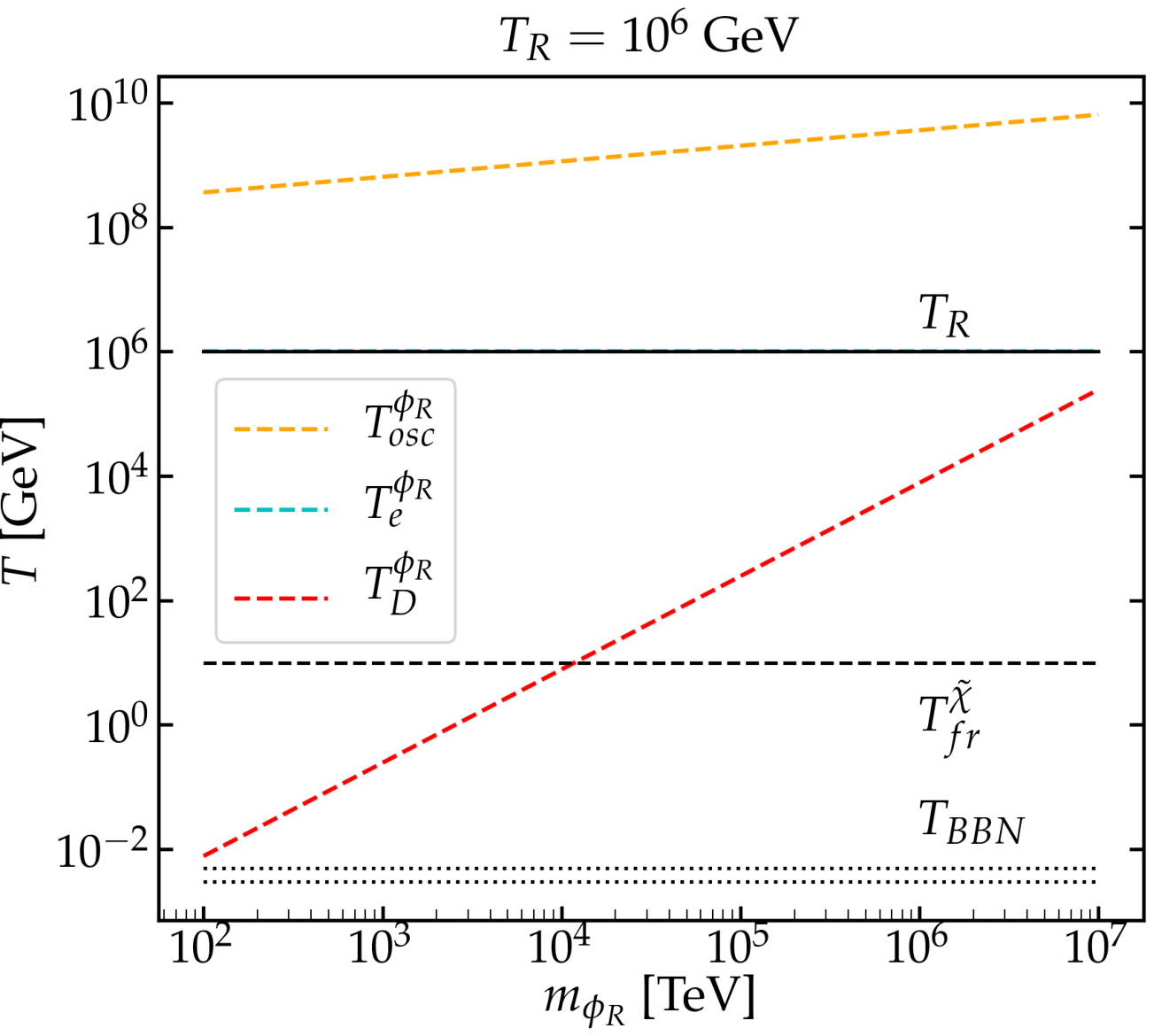}
\end{minipage}%
\begin{minipage}[c]{0.48\textwidth}
\includegraphics[width=3in]{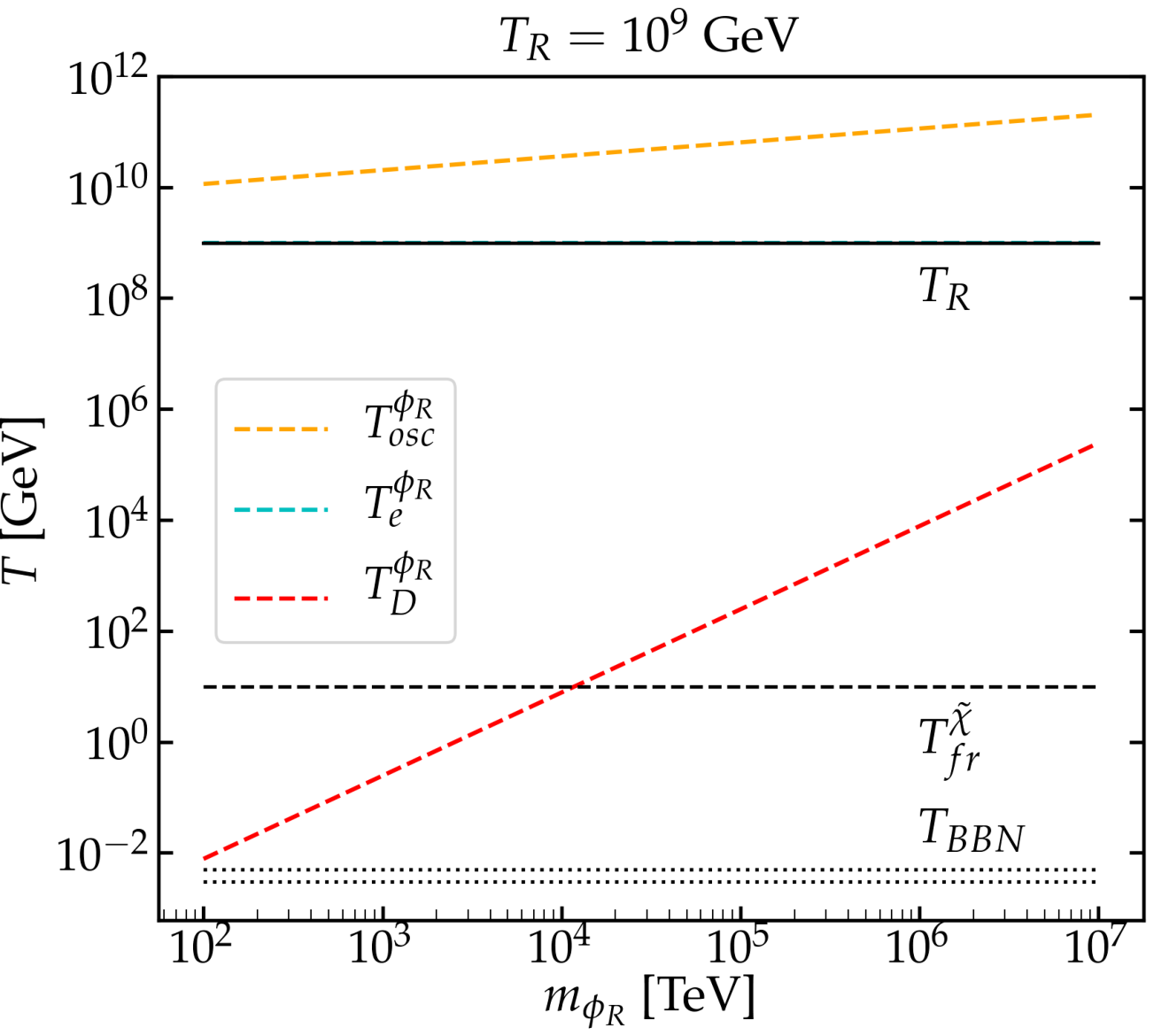}
\end{minipage}
\caption[]{\label{fig:Tmphi} The dependence of critical temperatures on $m_{\phi_R}$, with $T_R=10^6~ \rm{GeV}$ (left panel) and $T_R=10^9~ \rm{GeV}$ (right panel) respectively. Lines colored orange, cyan and red represent the critical temperatures $T_{osc}^{\phi_R}$, $T_{e}^{\phi_R}$ and $T_{D}^{\phi_R}$ respectively. The black solid line is the reheating temperature $T_R$, and the black dashed line represents the LSP freeze-out temperature $T_{fr}^{\tilde{\chi}}$ which is approximately $10~\rm{GeV}$. The two dotted lines represent that $T_{BBN}=3~ \rm{MeV}$ and $T_{BBN}=5 ~\rm{MeV}$ respectively.}
\end{center}
\end{figure}

We expect that the decay of modulus occurs before the freeze-out of the LSP, in which case the abundance of the LSP we obtained above may not be affected. In order to obtain $T_D^{\phi_R}\gtrsim T_{fr}^{\tilde{\chi}}$, we plot the critical temperatures versus the mass of modulus (with different reheating temperatures) in Fig.~\ref{fig:Tmphi}.
From the figure, the mass of the modulus $m_{\phi_R}$ needs to be larger than about $10^4~ \rm{TeV}$ if $T_D^{\phi_R}\gtrsim T_{fr}^{\tilde{\chi}}$ is the true story, and this can be easily achieved in some string framework. In KKLT-type model, for example, a mass hierarchy $m_{\phi_R}\gg m_{3/2}\gg m_{\mathrm{soft}}$ can naturally be realized~\cite{Choi:2005ge}. Here we also give a comment on the effect of the parameter $\lambda_{PQ}$, for a larger $\lambda_{PQ}$, the modulus would have a larger decay width into the PQ sector. The resulted DP saxions are highly relativistic due to that the modulus is far heavier than the saxions, so they may alive much longer than the cold saxions. In this case, the relativistic saxions may potentially decay after freeze-out of the LSP and, as a result, change the abundance of the LSP (this would lead to parts of parameter space, which satisfy the LZ-2022 bound as in Fig. \ref{gm2fig1}-\ref{gm2fig3}, being ruled out since the rescaled $\sigma_{P}^{\rm{SI}}$ is also changed synchronously). The way by which this change may be avoided typically resort to a large enough $T_D^{\phi_R}$ or equivalently a large enough $m_{\phi_R}$. Additionally, the hot axion produced by modulus decay can contribute to the dark radiation, which may be strongly restricted by the experimental measurement of $\Delta N_{\rm{eff}}$ (especially, for a large $\lambda_{PQ}\gtrsim1$~\cite{Baer:2023bbn}). Therefore, a smaller $\lambda_{PQ}$ is more safer and of interest.

In order to investigate the effect of the gravitino decay on the abundance of the LSP, we need the gravitino yield, which is defined by $Y_{3/2}=\frac{n_{3/2}}{s}$, where $n_{3/2}$ is the gravitino number density and $s$ is the entropy density. The gravitino yield can be produced via thermal production (TP) as well as decay production (DP). The TP gravitino is not a problem since the modulus decay releases a large amount of entropy, diluting the TP gravitino to a low level. The gravitino yield produced by the modulus decay, denoted as $Y_{3/2}^{\rm{DP}}$, can be estimated as~\cite{Nakamura:2006uc}
\begin{eqnarray}
Y_{3/2}^{\rm{DP}}\approx\frac{3}{2}B_{3/2}\frac{T_D^{\phi_R}}{m_{\phi_R}},
\end{eqnarray}
where $B_{3/2}$ is the modulus branching fractions into gravitino. Given that the modulus has a helicity-suppressed decay width into gravitino, for $m_{\phi_R}\gtrsim3\times10^4 ~\rm{TeV}$, $Y_{3/2}^{\rm{DP}}$ is smaller than $10^{-14}$. This yield is also much smaller than the yield of the LSP (which, in this work, is approximately equal to $1.1\times10^{-13}$), so gravitino decay would feed into the neutralino abundance with a very small contribution, which can be safely ignored. For smaller $m_{\phi_R}$, however, gravitino decay could potentially alter the relic density of LSP, and the parameter space once considered viable may be ruled out. Given that the entropy dilution resulting from modulus decay is quantified by $T_e^{\phi_R}/T_D^{\phi_R}$, it is also worth noting that the TP gravitino yield may not be diluted away if the mass of modulus $m_{\phi_R}$ is to be very large. Certainly, it is safe for $m_{\phi_R}\lesssim10^5 ~\rm{TeV}$ which is relatively easy to realize.

In the string framework, light modulus can result in a modulus-dominated era, and its decay can release an enormous amount of entropy. The $Numerical~analysis$ performed above may not be accurate enough due to the entropy dilution. However, this dilution would relax the cosmological constraints as long as the decay of the modulus into the PQ sector is very weak.
\subsection{Constraints from $g_{A\gamma\gamma}$ on $f_A$}
\begin{figure}[htb]
\begin{center}
\includegraphics[width=3.5in]{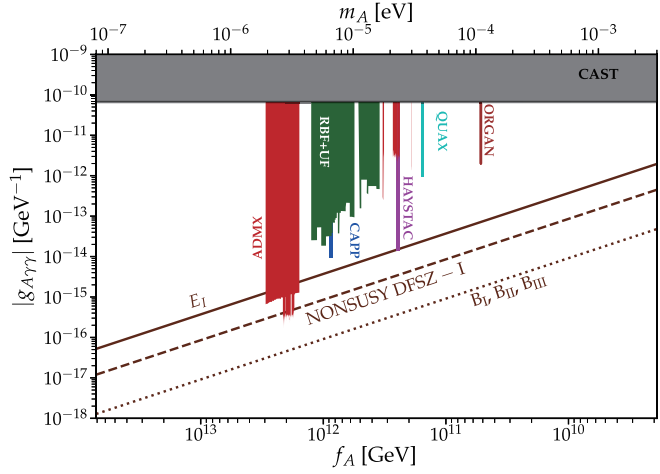}
\caption[]{\label{fig:gA1} Model predictions for $g_{A\gamma\gamma}$ as a function of $m_A$ and $f_A$ in different models, with current bounds from various experiments.}
\end{center}
\end{figure}

Before closing this section, we consider the experimental limitations from $g_{A\gamma\gamma}$ on $f_A$. For the current bounds, we include limits from the helioscope CAST \cite{CAST 1,CAST 2}, ADMX \cite{ADMX 1,ADMX 2,ADMX 3,ADMX 4}, CAPP \cite{CAPP}, RBF \cite{RBF}, UF \cite{UF}, HAYSTAC \cite{HAYSTAC 1,HAYSTAC 2}, QUAX \cite{QUAX}, and ORGAN \cite{ORGAN}. We also consider the projections for future sensitivity from the helioscope IAXO \cite{IAXO PRO} and haloscopes (ADMX \cite{ADMX PRO}, KLASH \cite{KLASH PRO}, MADMAX \cite{MADMAX PRO}, plasma haloscope \cite{Plasma Haloscopes PRO}, and TOORAD \cite{TOORAD PRO}). We plot Fig.~\ref{fig:gA1} using the axion limit data collected in Ref. \cite{Axion Data}.
\begin{figure}[htb]
\begin{center}
\includegraphics[width=3.5in]{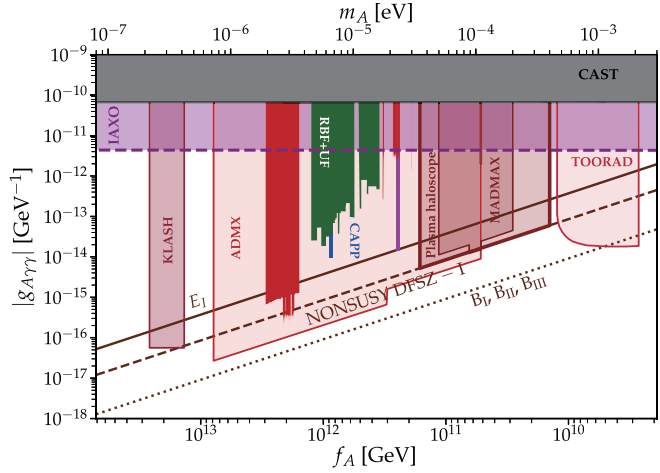}
\caption[]{\label{fig:gA2} Model predictions for $g_{A\gamma\gamma}$ as a function of $m_A$ and $f_A$ in different models, with projections for future experiments.}
\end{center}
\end{figure}

In Fig.~\ref{fig:gA1}, we show the coupling constant $g_{A\gamma\gamma}$ as a function of the axion mass $m_A$ and decay constant $f_A$ in the base models (which are the same as in the SUSY DFSZ-I~\cite{SUSY DFSZ-I} model), and in the extension $\rm{E_I}$. NONSUSY DFSZ-I~\cite{DFSZ2} model for comparison is also plotted in this figure. All these three base models have $E/N=2$ which is close to canceling the contribution $-1.92$ and, therefore, have suppressed coupling compared to the NONSUSY case. This implies that these models fortunately have not been ruled out and unfortunately are unlikely to be verified or tested by future experiments. Model $\rm{E_I}$, however, with the $N_{\rm{DW}}=1$ that avoids the cosmological DW problem provides the smallest $|N|$ and, therefore, can raise this low-energy effective coupling, in this way making the detection more possible. For $f_{A}\leq10^{12}~\rm{GeV}$, only one value of $f_A$ which is near $2.4\times10^{11}~\rm{GeV}$ (corresponding $m_A\approx2.4\times10^{-5}~\rm{eV}$) has been excluded. For pre-inflationary scenario, there remains a large parameter space that is viable. For post-inflationary scenario, we expect a value of $f_A$ is $1.11(2.59)\times10^{11} ~\rm{GeV}$ corresponding to $\langle \theta_i^2 \rangle\simeq 2.96^2(1.81^2)$, and this is also viable under these constraints. We also show the projections for future experiments in the Fig.~\ref{fig:gA2}, which turns out that our model $\rm{E_I}$ could be tested in the future.

\section{Conclusion}\label{Sec IV}
In this paper, we introduce four SUSY models in which the $\mu$ problem is solved by the Kim-Nilles mechanism and the Strong CP problem is addressed by imposing a global $\rm{U(1)_{PQ}}$ symmetry. Under the PQ symmetry, a pseudo-Nambu-Goldstone boson axion $A$ which is also a popular candidate of cold dark matter appears. With the enrichment of model content, however, axion physics imposes many constraints on the SUSY model. For example, the domain wall problem and quality problem should be taken into account. To tackle the domain wall problem, one can assume that the PQ symmetry is broken before or during inflation. Alternatively, the DW number $N_{\rm{DW}}$ of model needs to be 1. In three base model $\rm{B_{I,II,III}}$, $N_{\rm{DW}}\neq1$ and, therefore, we discuss the axion production mechanism in pre-inflationary scenario. In model $\rm{E_{I}}$, which is obtained by adding some Vectorlike pairs of chiral superfields $\Phi+\bar{\Phi}$ to the superpotential of the base model $\rm{B_{I}}$, $N_{\rm{DW}}=1$ can be achieved, so that there remains the possibility that axion is produced in post-inflationary scenario. Meanwhile, a global symmetry is not respected by quantum gravitational effects, which leads to the quality problem. We assume that our model has a $Z_{n}^{R}$ discrete symmetry which can be a subgroup of an anomaly-free continuous $\rm{U(1)}$ symmetry, so that the global $\rm{U(1)_{PQ}}$ symmetry could be seen as an accidental consequence of the discrete symmetry. We consider that $ Z_{n}^{R} $ symmetries may be made consistent with $\rm{ S U(5) }$ or Pati-Salam $\rm{ S U(4)_C\otimes S U(2)_L\otimes U(1)_R}$ embedding and find out that the latter can work well for our models.

We assume that, in this paper, the LSP $\tilde{\chi}$, which is the lightest neutralino, and axion are the dark matter used to fit the Planck measurement. For the sake of naturalness that we discussed in Sec.~\ref{Sec III}, we prefer that the parameter $\mu$ of the SUSY theory is relatively small and, therefore, the Higgsino-like neutralino is the LSP. We plot the viable parameter space that can well explain the latest experimental data of anomalous magnetic moment of the muon $a_\mu$. In these regions, the deviation between experiment and theoretical prediction within $2\sigma$ can be realized. Meanwhile, some regions are not detected by direct detection experiments such as LZ-2022, but can be tested by future experiments. As a conservative choice of $\mu = 250 ~\rm{GeV}$, we obtain that the LSP contributes about $5.8\%$ of the total dark matter relic density.

For the axion as another part of dark matter, we review the conventional misalignment mechanism and give a discussion about the kinetic misalignment mechanism. It turns out that the saxions and axinos, in the KMM case, would receive too light masses, and this would cause LSP decay, which would be contrary to our previous assumptions. Therefore, for that masses of these particles are heavy (e.g., above $2~ \rm{TeV}$), the way to produce axion would be through the CMM.

In order to discuss the constraints from the cosmology of saxions and axinos, we analyze the parameter space of all four models by imposing specific relationships on parameters or directly scanning all the parameters involved. For model $\rm{E_I}$, we plot the diagrams that compare the decay temperature $T_D$, LSP freeze-out temperature $T_{fr}^{\tilde{\chi}}$ and saxion/axino-radiation equality temperature $T_e$ with a reheating temperature $T_R=10^{9}~\rm{GeV}$. It turns out that there remains a large parameter space in which saxions- or axinos- dominanted era would never occur, and saxions or axinos all can decay before LSP freeze-out in the considered region of $f_A$. In addition, we consider the constraints from the dark radiation and find that the effective number of neutrinos $\Delta N_{\rm{eff}}$ is always under the current bound for $T_R\leq10^9 ~\rm{GeV}$. For base models, we present the results by scanning parameter space, and these models all can give a smaller lower limit of $f_A$ compared to that of model $\rm{E_I}$. We can summarize the lower limits of $f_A$ here which are $1.03\times10^{10},6.46\times10^{9},8.01\times10^{9}$ and $7.83\times10^{10}~\rm{GeV}$ for models $\rm{B_{I,II,III}}$ and $\rm{E_{I}}$ respectively. We also give a discussion on the cosmological moduli problem when the SUSY setup is embedded into string framework. The relic density of LSP would not be affected with some assumptions, and the entropy dilution originated from the decay of the modulus can make the cosmological constraints less severe.

Finally, combining the constraints from experiments that focus on $g_{A\gamma \gamma}$, we give the analysis of axion production via CMM for our models. In base models, since the DW number $N_{\rm{DW}}\neq1$, the PQ symmetry should be broken before or during inflation and, therefore, all the available $f_A$ that are smaller than $10^{12}~\rm{GeV}$ can provide the appropriate dark matter relic density $\Omega_{A} h^2\simeq 0.113$. On the other hand, $N_{\rm{DW}}=1$ can be achieved in model $\rm{E_I}$, so that CMM can work in both the pre-inflationary scenario and the post-inflationary scenario. In pre-inflationary scenario, for $f_A\in[7.83\times10^{10}~\rm{GeV},10^{12}~\rm{GeV}]$, only one value of $f_A$ which is near $2.4\times10^{11}~\rm{GeV}$ (corresponding $m_A\approx2.4\times10^{-5}~\rm{eV}$) has been excluded. For the latter scenario, $f_A=1.11(2.59)\times10^{11}~ \rm{GeV}$ corresponding to $\langle \theta_i^2 \rangle\simeq 2.96^2(1.81^2)$ is almost a fixed value that can also be realized in model $\rm{E_I}$.

In summary, within any framework of the base models and the extension $\rm{E_I}$, the LSP and the QCD axion can together give the total dark matter relic density observed by the Plank satellite. The base models may still be invisible to future proposed direct axion searches due to the suppressed axion-photon-photon couplings. However, the model $\rm{E_I}$ possesses an enhanced coupling, which is more likely to be detected by future axion searches. Axion, as a particle that appears very naturally in a strong CP free theory, is very charming in particle physics and cosmology, and we are looking forward to future experimental signals that can uncover its mystery.\\

\noindent{\bf Acknowledgements:} We are very grateful to professor Shu-Min Zhao and associate professor Xin-Yi Zhang from Hebei University, for giving us many useful discussions. This work is supported in part by National Natural Science Foundation of China (NNSFC) under Grant No.12075074, No.12235008, No.A2023201041, No.12347101, No.12175025 and No.11705045, and by the Chongqing Graduate Research and Innovation Foundation under Grant No.ydstd1912.
\appendix
\begin{center}
\Large{{\bf Appendix}}
\end{center}
\vspace{-8mm}

\section{Tadpole equations and some mass matrices}\label{APPENDIX A}
In our model, the scalar potential involving $X$, $Y$ and $Z$ can approximatively be expressed as
\begin{eqnarray}
V^{XYZ} \simeq&& m_{X}^{2}|X|^{2}+m_{Y}^{2}|Y|^{2}+m_{Z}^{2}|Z|^{2}-(\frac{a_{1}}{M_{P}} X^{{\alpha_{\rm{w}}}} Y^{4-{\alpha_{\rm{w}}}}+\frac{a_{2}}{M_{P}} X^{{\beta_{\rm{w}}}} Z^{4-{\beta_{\rm{w}}}}+\text { h.c. }) \nonumber\\
&&+|\frac{\lambda_{1}}{M_{P}} X^{{\alpha_{\rm{w}}}-1} Y^{4-{\alpha_{\rm{w}}}}+\frac{\lambda_{2}}{M_{P}} X^{{\beta_{\rm{w}}}-1} Z^{4-{\beta_{\rm{w}}}}|^{2}+|\frac{\lambda_{1}}{M_{P}} X^{{\alpha_{\rm{w}}}} Y^{3-{\alpha_{\rm{w}}}}|^{2} \nonumber\\
&&+|\frac{\lambda_{2}}{M_{P}} X^{{\beta_{\rm{w}}}} Z^{3-{\beta_{\rm{w}}}}|^{2} .\label{VPQXYZ}
\end{eqnarray}
The vacuum, corresponding to the deepest minima of the potential, is the lowest among many stationary points. In order to calculate the VEVs, we first need to identify all the stationary points of the potential. This involves solving the system of equations composed of the first-order partial derivatives of the potential equated to zero, which are commonly referred to as the tadpole equations. The tadpole equations involving $ X, Y $ and $ Z $ can be organized as
\begin{eqnarray}
\frac{\partial V^{XYZ}}{\partial v_X}&=&m_{X}^{2}v_X +(\frac{1}{4 M_{P}^{2} v_{X}^{3} v_{Y}^{2 {\alpha_{\rm{w}}}} v_{Z}^{2 {\beta_{\rm{w}}}}})\left\{\lambda_{1}^{2} v_{X}^{2 {\alpha_{\rm{w}}}} v_{Y}^{6} v_{Z}^{2 {\beta_{\rm{w}}}} {\alpha_{\rm{w}}}\left[v_{X}^{2}(-4+{\alpha_{\rm{w}}})^{2}+v_{Y}^{2}(-1+{\alpha_{\rm{w}}}) {\alpha_{\rm{w}}}\right]\right.\nonumber\\
&+&\lambda_{1} \lambda_{2} v_{X}^{{\alpha_{\rm{w}}}+{\beta_{\rm{w}}}} v_{Y}^{4+{\alpha_{\rm{w}}}} v_{Z}^{4+{\beta_{\rm{w}}}} {\alpha_{\rm{w}}} {\beta_{\rm{w}}}(-2+{\alpha_{\rm{w}}}+{\beta_{\rm{w}}})-2 M_{P} v_{X}^{2} v_{Y}^{{\alpha_{\rm{w}}}} v_{Z}^{{\beta_{\rm{w}}}}(a_{1} v_{X}^{{\alpha_{\rm{w}}}} v_{Y}^{4} v_{Z}^{{\beta_{\rm{w}}}} {\alpha_{\rm{w}}}\nonumber \\&+&\left.a_{2} v_{X}^{{\beta_{\rm{w}}}} v_{Y}^{{\alpha_{\rm{w}}}} v_{Z}^{4} {\beta_{\rm{w}}})
+\lambda_{2}^{2} v_{X}^{2 {\beta_{\rm{w}}}} v_{Y}^{2 {\alpha_{\rm{w}}}} v_{Z}^{6} {\beta_{\rm{w}}}\left[v_{X}^{2}(-4+{\beta_{\rm{w}}})^{2}+v_{Z}^{2}(-1+{\beta_{\rm{w}}}) {\beta_{\rm{w}}}\right]\right\}=0,\nonumber\\
\frac{\partial V^{XYZ}}{\partial v_Y}&=&m_{Y}^{2}v_Y -\frac{1}{4 M_{P}^{2}}\left\{v _ { X } ^ { - 2 + {\alpha_{\rm{w}}} } v _ { Y } ^ { 3- 2 {\alpha_{\rm{w}}} } v _ { Z } ^ { - {\beta_{\rm{w}}} } ( - 4 + {\alpha_{\rm{w}}} ) \left[-2 a_{1} M_{P} v_{X}^{2} v_{Y}^{{\alpha_{\rm{w}}}} v_{Z}^{{\beta_{\rm{w}}}}\right.\right.\nonumber\\
&+&\left.\left.\lambda_{1}^{2} v_{X}^{{\alpha_{\rm{w}}}} v_{Y}^{2} v_{Z}^{{\beta_{\rm{w}}}}\left(v_{Y}^{2} {\alpha_{\rm{w}}}^{2}+v_{X}^{2}\left(12-7 {\alpha_{\rm{w}}}+{\alpha_{\rm{w}}}^{2}\right)\right)+\lambda_{1} \lambda_{2} v_{X}^{{\beta_{\rm{w}}}} v_{Y}^{{\alpha_{\rm{w}}}} v_{Z}^{4} {\alpha_{\rm{w}}} {\beta_{\rm{w}}}\right]\right\}=0,\nonumber\\
\frac{\partial V^{XYZ}}{\partial v_Z}&=&m_{Z}^{2}v_Z-\frac{1}{4 M_{P}^{2}}\left\{v _ { X } ^ { - 2 + {\beta_{\rm{w}}} } v _ { Y } ^ { - {\alpha_{\rm{w}}} } v _ { Z } ^ { 3 - 2 {\beta_{\rm{w}}} } ( - 4 + {\beta_{\rm{w}}} ) \left[-2 a_{2} M_{P} v_{X}^{2} v_{Y}^{{\alpha_{\rm{w}}}} v_{Z}^{{\beta_{\rm{w}}}}\right.\right.\nonumber\\
&+&\left.\left.\lambda_{1} \lambda_{2} v_{X}^{{\alpha_{\rm{w}}}} v_{Y}^{4} v_{Z}^{{\beta_{\rm{w}}}} {\alpha_{\rm{w}}} {\beta_{\rm{w}}}+\lambda_{2}^{2} v_{X}^{{\beta_{\rm{w}}}} v_{Y}^{{\alpha_{\rm{w}}}} v_{Z}^{2}\left(v_{Z}^{2} {\beta_{\rm{w}}}^{2}+v_{X}^{2}\left(12-7 {\beta_{\rm{w}}}+{\beta_{\rm{w}}}^{2}\right)\right)\right]\right\}=0 .
\end{eqnarray}
After solving these equations, all the stationary points can be obtained, and the deepest minima (VEVs) can then be determined by picking out the stationary solutions minimizing the potential.

Mass matrix elements $M^2_{a_X a_X},~M^2_{a_Y a_Y},~M^2_{a_Z a_Z},~M^2_{a_X a_Y},~M^2_{a_Y a_X},~M^2_{a_X a_Z},~M^2_{a_Z a_X},~M^2_{a_Y a_Z}$ and $M^2_{a_Z a_Y}$ are
\begin{eqnarray}
 M_{a_{X} a_{X}}^{2}=&&v_{Y}^{-{\alpha_{\rm{w}}}} v_{Z}^{-{\beta_{\rm{w}}}}[2 a_{1} M_{P} v_{X}^{2+{\alpha_{\rm{w}}}} v_{Y}^{4} v_{Z}^{{\beta_{\rm{w}}}} {\alpha_{\rm{w}}}^{2}-v_{X}^{{\beta_{\rm{w}}}} v_{Z}^{4} {\beta_{\rm{w}}}(\lambda_{1} \lambda_{2} v_{X}^{{\alpha_{\rm{w}}}} v_{Y}^{4} {\alpha_{\rm{w}}}({\alpha_{\rm{w}}}-{\beta_{\rm{w}}})^{2}\nonumber\\&&-2 a_{2} M_{P} v_{X}^{2} v_{Y}^{{\alpha_{\rm{w}}}} {\beta_{\rm{w}}})]/(4 M_{P}^{2} v_{X}^{4}),\nonumber\\
 M_{a_{Y} a_{Y}}^{2} =&&\frac{v_{X}^{-2+{\alpha_{\rm{w}}}} v_{Y}^{2-{\alpha_{\rm{w}}}} v_{Z}^{-{\beta_{\rm{w}}}}(-4+{\alpha_{\rm{w}}})^{2}\left(2 a_{1} M_{P} v_{X}^{2} v_{Z}^{{\beta_{\rm{w}}}}-\lambda_{1} \lambda_{2} v_{X}^{{\beta_{\rm{w}}}} v_{Z}^{4} {\alpha_{\rm{w}}} {\beta_{\rm{w}}}\right)}{4 M_{P}^{2}},\nonumber \\
M_{a_{Z} a_{Z}}^{2} =&&\frac{v_{X}^{-2+{\beta_{\rm{w}}}} v_{Y}^{-{\alpha_{\rm{w}}}} v_{Z}^{2-{\beta_{\rm{w}}}}(-4+{\beta_{\rm{w}}})^{2}\left(2 a_{2} M_{P} v_{X}^{2} v_{Y}^{{\alpha_{\rm{w}}}}-\lambda_{1} \lambda_{2} v_{X}^{{\alpha_{\rm{w}}}} v_{Y}^{4} {\alpha_{\rm{w}}} {\beta_{\rm{w}}}\right)}{4 M_{P}^{2}}, \nonumber\\
M_{a_{X} a_{Y}}^{2}=&&M_{a_{Y} a_{X}}^{2}=v_{X}^{-3+{\alpha_{\rm{w}}}} v_{Y}^{3-{\alpha_{\rm{w}}}} v_{Z}^{-{\beta_{\rm{w}}}}(-4+{\alpha_{\rm{w}}}) {\alpha_{\rm{w}}}[-2 a_{1} M_{P} v_{X}^{2} v_{Z}^{{\beta_{\rm{w}}}}\nonumber \\&&-\lambda_{1} \lambda_{2} v_{X}^{{\beta_{\rm{w}}}} v_{Z}^{4} {\beta_{\rm{w}}}(-{\alpha_{\rm{w}}}+{\beta_{\rm{w}}})]/(4 M_{P}^{2}),\nonumber \\
M_{a_{X} a_{Z}}^{2}=&&M_{a_{Z} a_{X}}^{2}=v_{X}^{-3+{\beta_{\rm{w}}}} v_{Y}^{-{\alpha_{\rm{w}}}} v_{Z}^{3-{\beta_{\rm{w}}}}(-4+{\beta_{\rm{w}}}) {\beta_{\rm{w}}}[-2 a_{2} M_{P} v_{X}^{2} v_{Y}^{{\alpha_{\rm{w}}}}\nonumber\\&&-\lambda_{1} \lambda_{2} v_{X}^{{\alpha_{\rm{w}}}} v_{Y}^{4} {\alpha_{\rm{w}}}({\alpha_{\rm{w}}}-{\beta_{\rm{w}}})]/(4 M_{P}^{2}),\nonumber\\
M_{a_{Y} a_{Z}}^{2}=&&M_{a_{Z} a_{Y}}^{2}=\frac{\lambda_{1} \lambda_{2} v_{X}^{-2+{\alpha_{\rm{w}}}+{\beta_{\rm{w}}}} v_{Y}^{3-{\alpha_{\rm{w}}}} v_{Z}^{3-{\beta_{\rm{w}}}}(-4+{\alpha_{\rm{w}}}) {\alpha_{\rm{w}}}(-4+{\beta_{\rm{w}}}) {\beta_{\rm{w}}}}{4 M_{P}^{2}}.
\end{eqnarray}

Mass matrix elements $ M_{\rho_{X} \rho_{X}}^{2}, M_{\rho_{X} \rho_{Y}}^{2}, M_{\rho_{X} \rho_{Z}}^{2}, M_{\rho_{Y} \rho_{X}}^{2}, M_{\rho_{Y} \rho_{Y}}^{2}, M_{\rho_{Y} \rho_{Z}}^{2}, M_{\rho_{Z} \rho_{X}}^{2}, M_{\rho_{Z} \rho_{Y}}^{2} $ and $ M_{\rho_{Z} \rho_{Z}}^{2} $ are
\begin{eqnarray}
M_{\rho_{X} \rho_{X}}^{2}=&& v_{Y}^{-2 {\alpha_{\rm{w}}}} v_{Z}^{-2 {\beta_{\rm{w}}}}\big\{2 \lambda_{1}^{2} v_{X}^{2 {\alpha_{\rm{w}}}} v_{Y}^{6} v_{Z}^{2 {\beta_{\rm{w}}}}(-1+{\alpha_{\rm{w}}}) {\alpha_{\rm{w}}}(v_{X}^{2}(-4+{\alpha_{\rm{w}}})^{2}+v_{Y}^{2}(-2+{\alpha_{\rm{w}}}) {\alpha_{\rm{w}}})\nonumber\\
&&+\lambda_{1} \lambda_{2} v_{X}^{{\alpha_{\rm{w}}}+{\beta_{\rm{w}}}} v_{Y}^{4+{\alpha_{\rm{w}}}} v_{Z}^{4+{\beta_{\rm{w}}}} {\alpha_{\rm{w}}} {\beta_{\rm{w}}}(8+{\alpha_{\rm{w}}}^{2}+2 {\alpha_{\rm{w}}}(-3+{\beta_{\rm{w}}})-6 {\beta_{\rm{w}}}+{\beta_{\rm{w}}}^{2})\nonumber \\
&&+2 v_{Y}^{{\alpha_{\rm{w}}}}[-a_{1} M_{P} v_{X}^{2+{\alpha_{\rm{w}}}} v_{Y}^{4} v_{Z}^{2 {\beta_{\rm{w}}}}(-2+{\alpha_{\rm{w}}}) {\alpha_{\rm{w}}}\nonumber\\&&+v_{X}^{{\beta_{\rm{w}}}} v_{Y}^{{\alpha_{\rm{w}}}} v_{Z}^{4} {\beta_{\rm{w}}}(-a_{2} M_{P} v_{X}^{2} v_{Z}^{{\beta_{\rm{w}}}}(-2+{\beta_{\rm{w}}})
+\lambda_{2}^{2} v_{X}^{{\beta_{\rm{w}}}} v_{Z}^{2}(-1+{\beta_{\rm{w}}})(v_{X}^{2}(-4+{\beta_{\rm{w}}})^{2}\nonumber\\&&+v_{Z}^{2}(-2+{\beta_{\rm{w}}}) {\beta_{\rm{w}}}))]\big\} /(4 M_{P}^{2} v_{X}^{4})\nonumber \\
M_{\rho_{Y} \rho_{Y}}^{2}=&&\left\{v _ { X } ^ { - 2 + {\alpha_{\rm{w}}} } v _ { Y } ^ { 2 - 2 {\alpha_{\rm{w}}} } v _ { Z } ^ { - {\beta_{\rm{w}}} } ( - 4 + {\alpha_{\rm{w}}} ) \left[-2 a_{1} M_{P} v_{X}^{2} v_{Y}^{{\alpha_{\rm{w}}}} v_{Z}^{{\beta_{\rm{w}}}}(-2+{\alpha_{\rm{w}}})\right.\right.\nonumber\\
&&+2 \lambda_{1}^{2} v_{X}^{{\alpha_{\rm{w}}}} v_{Y}^{2} v_{Z}^{{\beta_{\rm{w}}}}(-3+{\alpha_{\rm{w}}})\left(v_{Y}^{2} {\alpha_{\rm{w}}}^{2}+v_{X}^{2}\left(8-6 {\alpha_{\rm{w}}}+{\alpha_{\rm{w}}}^{2}\right)\right) \nonumber\\
&&\left.\left.+\lambda_{1} \lambda_{2} v_{X}^{{\beta_{\rm{w}}}} v_{Y}^{{\alpha_{\rm{w}}}} v_{Z}^{4}(-2+{\alpha_{\rm{w}}}) {\alpha_{\rm{w}}} {\beta_{\rm{w}}}\right]\right\} /\left(4 M_{P}^{2}\right)\nonumber \\
M_{\rho_{Z} \rho_{Z}}^{2}=&&\left\{v _ { X } ^ { - 2 + {\beta_{\rm{w}}} } v _ { Y } ^ { - {\alpha_{\rm{w}}} } v _ { Z } ^ { 2 - 2 {\beta_{\rm{w}}} } ( - 4 + {\beta_{\rm{w}}} ) \left[-2 a_{2} M_{P} v_{X}^{2} v_{Y}^{{\alpha_{\rm{w}}}} v_{Z}^{{\beta_{\rm{w}}}}(-2+{\beta_{\rm{w}}})\right.\right.\nonumber\\
&&+\lambda_{1} \lambda_{2} v_{X}^{{\alpha_{\rm{w}}}} v_{Y}^{4} v_{Z}^{{\beta_{\rm{w}}}} {\alpha_{\rm{w}}}(-2+{\beta_{\rm{w}}}) {\beta_{\rm{w}}}\nonumber \\
&&\left.\left.+2 \lambda_{2}^{2} v_{X}^{{\beta_{\rm{w}}}} v_{Y}^{{\alpha_{\rm{w}}}} v_{Z}^{2}(-3+{\beta_{\rm{w}}})\left(v_{Z}^{2} {\beta_{\rm{w}}}^{2}+v_{X}^{2}\left(8-6 {\beta_{\rm{w}}}+{\beta_{\rm{w}}}^{2}\right)\right)\right]\right\} /\left(4 M_{P}^{2}\right),\nonumber \\
M_{\rho_{X} \rho_{Y}}^{2}=&& M_{\rho_{Y} \rho_{X}}^{2}=-\left\{v _ { X } ^ { - 3 + {\alpha_{\rm{w}}} } v _ { Y } ^ { 3 - 2 {\alpha_{\rm{w}}} } v _ { Z } ^ { - {\beta_{\rm{w}}} } ( - 4 + {\alpha_{\rm{w}}} ) {\alpha_{\rm{w}}} \left[-2 a_{1} M_{P} v_{X}^{2} v_{Y}^{{\alpha_{\rm{w}}}} v_{Z}^{{\beta_{\rm{w}}}}\right.\right.\nonumber\\
&&+2 \lambda_{1}^{2} v_{X}^{{\alpha_{\rm{w}}}} v_{Y}^{2} v_{Z}^{{\beta_{\rm{w}}}}\left(v_{Y}^{2}(-1+{\alpha_{\rm{w}}}) {\alpha_{\rm{w}}}+v_{X}^{2}\left(12-7 {\alpha_{\rm{w}}}+{\alpha_{\rm{w}}}^{2}\right)\right)\nonumber\\
&&\left.\left.+\lambda_{1} \lambda_{2} v_{X}^{{\beta_{\rm{w}}}} v_{Y}^{{\alpha_{\rm{w}}}} v_{Z}^{4} {\beta_{\rm{w}}}(-2+{\alpha_{\rm{w}}}+{\beta_{\rm{w}}})\right]\right\} /\left(4 M_{P}^{2}\right), \nonumber\\
M_{\rho_{X} \rho_{Z}}^{2}=&& M_{\rho_{Z} \rho_{X}}^{2}=-\left\{v _ { X } ^ { - 3 + {\beta_{\rm{w}}} } v _ { Y } ^ { - } {\alpha_{\rm{w}}} v _ { Z } ^ { 3 - 2 {\beta_{\rm{w}}} } ( - 4 + {\beta_{\rm{w}}} ) {\beta_{\rm{w}}} \left[-2 a_{2} M_{P} v_{X}^{2} v_{Y}^{{\alpha_{\rm{w}}}} v_{Z}^{{\beta_{\rm{w}}}}\right.\right.\nonumber\\
&&+\lambda_{1} \lambda_{2} v_{X}^{{\alpha_{\rm{w}}}} v_{Y}^{4} v_{Z}^{{\beta_{\rm{w}}}} {\alpha_{\rm{w}}}(-2+{\alpha_{\rm{w}}}+{\beta_{\rm{w}}})+2 \lambda_{2}^{2} v_{X}^{{\beta_{\rm{w}}}} v_{Y}^{{\alpha_{\rm{w}}}} v_{Z}^{2}\left(v_{Z}^{2}(-1+{\beta_{\rm{w}}}) {\beta_{\rm{w}}}\right.\nonumber\\
&&\left.\left.\left.+v_{X}^{2}\left(12-7 {\beta_{\rm{w}}}+{\beta_{\rm{w}}}^{2}\right)\right)\right]\right\} /\left(4 M_{P}^{2}\right),\nonumber \\
M_{\rho_{Y} \rho_{Z}}^{2}=&&M_{\rho_{Z} \rho_{Y}}^{2}= \frac{\lambda_{1} \lambda_{2} v_{X}^{-2+{\alpha_{\rm{w}}}+{\beta_{\rm{w}}}} v_{Y}^{3-{\alpha_{\rm{w}}}} v_{Z}^{3-{\beta_{\rm{w}}}}(-4+{\alpha_{\rm{w}}}) {\alpha_{\rm{w}}}(-4+{\beta_{\rm{w}}}) {\beta_{\rm{w}}}}{\left(4 M_{P}^{2}\right) }.
\end{eqnarray}

Mass matrix elements $ M_{\tilde{a}_{X} \tilde{a}_{X}}, M_{\tilde{a}_{X} \tilde{a}_{Y}}, M_{\tilde{a}_{X} \tilde{a}_{Z}}, M_{\tilde{a}_{Y} \tilde{a}_{X}}, M_{\tilde{a}_{Y} \tilde{a}_{Y}}, M_{\tilde{a}_{Y} \tilde{a}_{Z}}, M_{\tilde{a}_{Z} \tilde{a}_{X}}, M_{\tilde{a}_{Z} \tilde{a}_{Y}} $ and $ M_{\tilde{a}_{Z} \tilde{a}_{Z}} $ are
\begin{eqnarray}
M_{\tilde{a}_{X} \tilde{a}_{X}}&=&\frac{\lambda_{1} v_{X}^{-2+{\alpha_{\rm{w}}}} v_{Y}^{4-{\alpha_{\rm{w}}}}(-1+{\alpha_{\rm{w}}}) {\alpha_{\rm{w}}}+\lambda_{2} v_{X}^{-2+{\beta_{\rm{w}}}} v_{Z}^{4-{\beta_{\rm{w}}}}(-1+{\beta_{\rm{w}}}) {\beta_{\rm{w}}}}{2 M_{P}}, \nonumber\\
M_{\tilde{a}_{Y} \tilde{a}_{Y}}&=&\frac{\lambda_{1} v_{X}^{{\alpha_{\rm{w}}}} v_{Y}^{2-{\alpha_{\rm{w}}}}(3-{\alpha_{\rm{w}}})(4-{\alpha_{\rm{w}}})}{2 M_{P}}, \nonumber\\
M_{\tilde{a}_{Z} \tilde{a}_{Z}}&=&\frac{\lambda_{2} v_{X}^{{\beta_{\rm{w}}}} v_{Z}^{2-{\beta_{\rm{w}}}}(3-{\beta_{\rm{w}}})(4-{\beta_{\rm{w}}})}{2 M_{P}}, \nonumber\\
M_{\tilde{a}_{X} \tilde{a}_{Y}}&=&M_{\tilde{a}_{Y} \tilde{a}_{X}}=\frac{\lambda_{1} v_{X}^{-1+{\alpha_{\rm{w}}}} v_{Y}^{3-{\alpha_{\rm{w}}}}(4-{\alpha_{\rm{w}}}) {\alpha_{\rm{w}}}}{2 M_{P}}, \nonumber\\
M_{\tilde{a}_{X} \tilde{a}_{Z}}&=&M_{\tilde{a}_{Z} \tilde{a}_{X}}=\frac{\lambda_{2} v_{X}^{-1+{\beta_{\rm{w}}}} v_{Z}^{3-{\beta_{\rm{w}}}}(4-{\beta_{\rm{w}}}) {\beta_{\rm{w}}}}{2 M_{P}}, \nonumber\\
M_{\tilde{a}_{Y} \tilde{a}_{Z}}&=&M_{\tilde{a}_{Z} \tilde{a}_{Y}}=0 .
\end{eqnarray}
From this matrix, we can easily check that there would be a massless axino if $ \alpha_{\rm{w}}=\beta_{\rm{w}}=3 $.
\section{Misalignment mechanism}\label{APPENDIX B}
Following the dilute instanton gas approximation \cite{diga}, the axion potential can be expressed as
\begin{eqnarray}
&&V_A(T)=m_A^2(T) f_A^2 \left( 1-\cos\frac{A}{f_A}\right),
m_A(T)\mid_{T>\Lambda_{\rm{QCD}}}=\zeta m_A (\frac{\Lambda_{\rm{QCD}}}{T})^4.\label{V_A(T)}
\end{eqnarray}
Here $m_A(T)$ is the temperature-dependent mass of axion, and $m_A$ is the zero-temperature axion mass given by Eq.~(\ref{eq:mAintermsoffA}). For temperature $T<\Lambda_{\rm{QCD}}$, the temperature-dependent axion mass $m_A(T)$ is equal to the zero-temperature axion mass $m_A$. In this study, we assume that the QCD scale $\Lambda_{\rm{QCD}}$ is approximately $160 \rm{MeV}$ and the parameter $\zeta$ is equal to $0.026$ \cite{LMDQCD}. Such a potential can result in a non-thermal relic density for axion as the cold dark matter.

The equation of motion of the axion in an expanding Universe is
\begin{eqnarray}
\ddot{A}+3H(T)\dot{A} +V_A^{\prime}(T)=0,
\end{eqnarray}
where the Hubble rate $H(T)$ can be expressed as $H(T)=(\frac{\pi^2}{90}g_{*}(T) \frac{T^4}{M_{P}^2})^{\frac{1}{2}}$ in a radiation-dominated Universe, and $V_A^{\prime}(T)$ is the first derivative of the potential with respect to the axion field $A$. Simplifying this equation of motion by using $\theta(x)\equiv\frac{A(x)}{f_A}$ and $V_A(T) \approx \frac{1}{2}m_A^2(T) A^2$ (which can be obtained by expanding the potential $V_A(T)$ around $\frac{A}{f_A} = 0$), we get
\begin{eqnarray}
\ddot{\theta}+3H(T)\dot{\theta} +m_A^2(T) \theta=0.
\end{eqnarray}
This is the damped vibration differential equation, and the Hubble friction damps the evolution of the axion field as long as the Hubble rate is significantly larger than the mass of axion. We can set $\dot{\theta}_i=0$, since the Hubble friction freezes the motion of axion. However, the initial value $\theta_i$ is not necessary to stay at the minima in early Universe. As the Universe temperature decreases and the axion mass switches on, the axion field starts to oscillate at a temperature $T_{osc}$ at which $m_A(T_{osc})=3H(T_{osc})$. With a value of  $g_*(\Lambda_{\rm{QCD}})$ is about $26$, we can determine that $T_{osc}>\Lambda_{\rm{QCD}}$ due to $m_A(\Lambda_{\rm{QCD}})>3H(\Lambda_{\rm{QCD}})$.

When $T<\Lambda_{\rm{QCD}}$, the energy density of axion can be computed by $\rho_{A}(T)=m_A n_A(T)$, where $n_A(T)$ is the axion number density at temperature $T$. This axion number density is related to a conserved yield
\begin{eqnarray}
\frac{n_A(T)}{s(T)}=\frac{n_A(T_{osc})}{s(T_{osc})},
\end{eqnarray}
where $s(T)=\frac{2\pi^2}{45}g_s(T)T^3$ is the entropy density at temperature $T$, and $n_A(T_{osc})= \frac{V_A(T_{osc})}{m_A(T_{osc})}=\frac{1}{2}m_A(T_{osc}) f_A^2 \langle \theta_i^2 \rangle$ can be realized in the post-inflationary scenario~\cite{landscape}. The parameter $\langle \theta_i^2 \rangle$ is given by
\begin{eqnarray}
\langle \theta_i^2 \rangle=\frac{1}{2\pi}\int_{-\pi}^{\pi}d\theta_if(\theta_i)\theta_i^2,
\end{eqnarray}
where the angle brackets represent the value of the initial condition averaged over $[-\pi,\pi)$. For $f(\theta_i)=1$, $\langle \theta_i^2 \rangle=\frac{\pi^2}{3}\simeq1.81^2$ can be easily obtained. In the case $f(\theta_i)=(\log[\frac{e}{(1 - \frac{\theta_i^2}{\pi^2})}])^{\frac{7}{6}}$~\cite{Ftheta} that accounting for the anharmonic corrections, this parameter would be $\langle \theta_i^2 \rangle\simeq2.96^2$. The mass of axion at the temperature $T_{osc}$ is $m_A(T_{osc})\simeq\frac{\sqrt{g_s(T_{osc})}\pi T_{osc}^2}{\sqrt{10}M_P}$, and the oscillation temperature $T_{osc}$ could be computed here by using condition $m_A(T_{osc})=3H(T_{osc})$:
\begin{eqnarray}
T_{osc}\simeq (\frac{\sqrt{10} \zeta m_A M_P \Lambda_{QCD}^4}{\pi g_s(T_{osc})^{\frac{1}{2}}})^{\frac{1}{6}},
\end{eqnarray}
where $g_*(T_{osc})=g_s(T_{osc})\simeq61.75$ have been used as a reference. Then, the axion relic density in terms of $f_A$ could be expressed as
\begin{eqnarray}
\Omega_A h^2= \frac{\rho_{A}(T_0)}{\rho_{tot}(T_0)}h^2\simeq1.6679\times 10^{-15} (\frac{f_A}{\rm{GeV}})^{\frac{7}{6}}\langle \theta_i^2 \rangle.\label{OmegaPIA}
\end{eqnarray}
For that axion is responsible for the rest of total relic density of dark matter or equivalently $\Omega_A h^2 = 0.113$, we need $f_A \simeq 1.11(2.59)\times10^{11} \rm{GeV}$ with a parameter $\langle \theta_i^2 \rangle\simeq 2.96^2(1.81^2)$.

In the pre-inflationary scenario~\cite{landscape}, on the other hand, the initial misalignment angle $\theta_i$ has a homogeneous value, so that the expression of relic density of axion would be changed to
\begin{eqnarray}
\Omega_A h^2= \frac{\rho_{A}(T_0)}{\rho_{tot}(T_0)}h^2\simeq1.6679\times 10^{-15} (\frac{f_A}{\rm{GeV}})^{\frac{7}{6}} f(\theta_i)\theta_i^2.
\end{eqnarray}
In addition, there would be no Domain Wall problem even if the Domain Wall number $N_{\rm{DW}}\neq1$, since topological defects are inflated away and do not contribute to the axion energy density.

Nonzero $\dot{\theta}_i$ is also possible and well motivated. In this case, the mechanism for producing axion relic density is called the kinetic misalignment mechanism (KMM), and it is also very appealing in axion studies. In our model, the scalar potential containing only $X$, $Y$ and $Z$ is given by
\begin{eqnarray}
V_{P Q}^{X \& Y \& Z} \simeq&& m_{X}^{2}|X|^{2}+m_{Y}^{2}|Y|^{2}+m_{Z}^{2}|Z|^{2}-(\frac{a_{1}}{M_{P}} X^{{\alpha_{\rm{w}}}} Y^{4-{\alpha_{\rm{w}}}}+\frac{a_{2}}{M_{P}} X^{{\beta_{\rm{w}}}} Z^{4-{\beta_{\rm{w}}}}+\text { h.c. }) \nonumber\\
&&+|\frac{\lambda_{1}}{M_{P}} X^{{\alpha_{\rm{w}}}-1} Y^{4-{\alpha_{\rm{w}}}}+\frac{\lambda_{2}}{M_{P}} X^{{\beta_{\rm{w}}}-1} Z^{4-{\beta_{\rm{w}}}}|^{2}+|\frac{\lambda_{1}}{M_{P}} X^{{\alpha_{\rm{w}}}} Y^{3-{\alpha_{\rm{w}}}}|^{2} \nonumber\\
&&+|\frac{\lambda_{2}}{M_{P}} X^{{\beta_{\rm{w}}}} Z^{3-{\beta_{\rm{w}}}}|^{2} .\label{VPQXYZ}
\end{eqnarray}
If this potential is sufficiently flat, the radial components can be initially displaced from their minima and driven to large field values during inflation. As the temperature of Universe drop, they becomes free to oscillate and the higher-dimensional PQ-violating terms of potential can give a nonzero $\dot{\theta}_i$. Under the assumption that parameters in the potential fulfill formulae $ m_{X}=m_{Y}=m_{Z}=m_{\rm{x y z}} $, $ \lambda_{1}=\lambda_{2}=\lambda $ and $ a_{1}=a_{2}=a_{\lambda} $, we roughly found that axion contributing a relic density $ \Omega_{A} h^{2} \simeq 0.113 $ with $ f_{A} \sim 10^{11} ~ \mathrm{GeV} $ requires the parameter $ \lambda $ to be of the order of $ 10^{-13} \sim 10^{-10} $. In this case, both of the parameters $ m_{\rm{x y z}}, a_{\lambda} $ are required to be unacceptably small. To clarify this, we temporarily assume that the VeVs satisfy $ \langle X\rangle=\langle Y\rangle=\langle Z\rangle=S $, so the scalar potential can be rewritten as
\begin{eqnarray}
V & = & 3 m_{\rm{x y z}}^{2} S^{2}-\frac{4 a_{\lambda}}{M_{P}} S^{4}+30\left(\frac{\lambda}{M_{P}}\right)^{2} S^{6} .
\end{eqnarray}
There would be a local minimum when $ \left|\frac{a_{\lambda}}{\lambda}\right|^{2}>\frac{135}{8} m_{\rm{x y z}}^{2} $, and $ S $ can be solved as
\begin{eqnarray}
S^{2} & = & \frac{8 a_{\lambda}+8 \sqrt{a_{\lambda}^{2}-\frac{135}{8} \lambda^{2} m_{\rm{x y z}}^{2}}}{180 \lambda^{2}} M_{P} .\label{S2}
\end{eqnarray}
For that $ \lambda $ is of the order of $ 10^{-11} $ and $ f_{A} \sim S \sim 10^{11} ~ \mathrm{GeV} $, parameter $ a_{\lambda} $ is required to be of the order of $ 10^{-17} ~ \mathrm{GeV} $, and $ m_{\rm{x y z}} $ is also extremely small (about $ 10^{-7} ~ \mathrm{GeV} $). It then turns out that the masses of saxions and axinos are also too small to hold the higgsino-like neutralino is the candidate of dark matter. In a word, parameter space for the KMM at play is unfortunately not self-consistent with the previous assumption that $ \Omega_{\tilde{\chi}} h^{2} \simeq 0.007 $.


\begin{thebibliography}{99}


\bibitem{125GeV1}
G.~Aad \textit{et al.} [ATLAS],
Observation of a new particle in the search for the Standard Model Higgs boson with the ATLAS detector at the LHC,
Phys. Lett. B \textbf{716}, 1-29 (2012)
doi:10.1016/j.physletb.2012.08.020
[arXiv:1207.7214 [hep-ex]].

\bibitem{125GeV2}
S.~Chatrchyan \textit{et al.} [CMS],
Observation of a New Boson at a Mass of 125 GeV with the CMS Experiment at the LHC,
Phys. Lett. B \textbf{716}, 30-61 (2012)
doi:10.1016/j.physletb.2012.08.021
[arXiv:1207.7235 [hep-ex]].

\bibitem{PQM}
R.~D.~Peccei and H.~R.~Quinn,
CP Conservation in the Presence of Instantons,
Phys. Rev. Lett. \textbf{38}, 1440-1443 (1977)
doi:10.1103/PhysRevLett.38.1440

\bibitem{review1}
M.~S.~Turner,
Windows on the Axion,
Phys. Rept. \textbf{197}, 67-97 (1990)
doi:10.1016/0370-1573(90)90172-X

\bibitem{review2}
P.~Sikivie,
The Pool table analogy to axion physics,
doi:10.1063/1.881573
[arXiv:hep-ph/9506229 [hep-ph]].

\bibitem{review3}
M.~Dine,
TASI lectures on the strong CP problem,
[arXiv:hep-ph/0011376 [hep-ph]].

\bibitem{review4}
R.~D.~Peccei,
The Strong CP problem and axions,
Lect. Notes Phys. \textbf{741}, 3-17 (2008)
doi:10.1007/978-3-540-73518-2\_1
[arXiv:hep-ph/0607268 [hep-ph]].

\bibitem{review5}
J.~E.~Kim and G.~Carosi,
Axions and the Strong CP Problem,
Rev. Mod. Phys. \textbf{82}, 557-602 (2010)
[erratum: Rev. Mod. Phys. \textbf{91}, no.4, 049902 (2019)]
doi:10.1103/RevModPhys.82.557
[arXiv:0807.3125 [hep-ph]].

\bibitem{review6}
Y.~K.~Semertzidis and S.~Youn,
Axion dark matter: How to see it?,
Sci. Adv. \textbf{8}, no.8, abm9928 (2022)
doi:10.1126/sciadv.abm9928
[arXiv:2104.14831 [hep-ph]].

\bibitem{review7}
A.~Hook,
TASI Lectures on the Strong CP Problem and Axions,
PoS \textbf{TASI2018}, 004 (2019)
[arXiv:1812.02669 [hep-ph]].

\bibitem{review8}
P.~Sikivie,
Invisible Axion Search Methods,
Rev. Mod. Phys. \textbf{93}, no.1, 015004 (2021)
doi:10.1103/RevModPhys.93.015004
[arXiv:2003.02206 [hep-ph]].

\bibitem{AC}
D.~J.~E.~Marsh,
Axion Cosmology,
Phys. Rept. \textbf{643}, 1-79 (2016)
doi:10.1016/j.physrep.2016.06.005
[arXiv:1510.07633 [astro-ph.CO]].

\bibitem{landscape}
L.~Di Luzio, M.~Giannotti, E.~Nardi and L.~Visinelli,
The landscape of QCD axion models,
Phys. Rept. \textbf{870}, 1-117 (2020)
doi:10.1016/j.physrep.2020.06.002
[arXiv:2003.01100 [hep-ph]].

\bibitem{PQWW1}
S.~Weinberg,
A New Light Boson?,
Phys. Rev. Lett. \textbf{40}, 223-226 (1978)
doi:10.1103/PhysRevLett.40.223

\bibitem{PQWW2}
F.~Wilczek,
Problem of Strong  $P$  and  $T$  Invariance in the Presence of Instantons,
Phys. Rev. Lett. \textbf{40}, 279-282 (1978)
doi:10.1103/PhysRevLett.40.279

\bibitem{DFSZ1}
A.~R.~Zhitnitsky,
On Possible Suppression of the Axion Hadron Interactions. (In Russian),
Sov. J. Nucl. Phys. \textbf{31}, 260 (1980)

\bibitem{DFSZ2}
M.~Dine, W.~Fischler and M.~Srednicki,
A Simple Solution to the Strong CP Problem with a Harmless Axion,
Phys. Lett. B \textbf{104}, 199-202 (1981)
doi:10.1016/0370-2693(81)90590-6

\bibitem{KSVZ1}
J.~E.~Kim,
Weak Interaction Singlet and Strong CP Invariance,
Phys. Rev. Lett. \textbf{43}, 103 (1979)
doi:10.1103/PhysRevLett.43.103

\bibitem{KSVZ2}
M.~A.~Shifman, A.~I.~Vainshtein and V.~I.~Zakharov,
Can Confinement Ensure Natural CP Invariance of Strong Interactions?,
Nucl. Phys. B \textbf{166}, 493-506 (1980)
doi:10.1016/0550-3213(80)90209-6

\bibitem{KNM}
J.~E.~Kim and H.~P.~Nilles,
The mu Problem and the Strong CP Problem,
Phys. Lett. B \textbf{138}, 150-154 (1984)
doi:10.1016/0370-2693(84)91890-2

\bibitem{QP1}
H.~M.~Georgi, L.~J.~Hall and M.~B.~Wise,
Grand Unified Models With an Automatic {Peccei-Quinn} Symmetry,
Nucl. Phys. B \textbf{192}, 409-416 (1981)
doi:10.1016/0550-3213(81)90433-8

\bibitem{QP2}
M.~Dine and N.~Seiberg,
String Theory and the Strong {CP} Problem,
Nucl. Phys. B \textbf{273}, 109-124 (1986)
doi:10.1016/0550-3213(86)90043-X

\bibitem{QP3}
S.~M.~Barr and D.~Seckel,
Planck scale corrections to axion models,
Phys. Rev. D \textbf{46}, 539-549 (1992)
doi:10.1103/PhysRevD.46.539

\bibitem{QP4}
M.~Kamionkowski and J.~March-Russell,
Planck scale physics and the Peccei-Quinn mechanism,
Phys. Lett. B \textbf{282}, 137-141 (1992)
doi:10.1016/0370-2693(92)90492-M
[arXiv:hep-th/9202003 [hep-th]].

\bibitem{QP5}
R.~Holman, S.~D.~H.~Hsu, T.~W.~Kephart, E.~W.~Kolb, R.~Watkins and L.~M.~Widrow,
Solutions to the strong CP problem in a world with gravity,
Phys. Lett. B \textbf{282}, 132-136 (1992)
doi:10.1016/0370-2693(92)90491-L
[arXiv:hep-ph/9203206 [hep-ph]].

\bibitem{QP6}
S.~Ghigna, M.~Lusignoli and M.~Roncadelli,
Instability of the invisible axion,
Phys. Lett. B \textbf{283}, 278-281 (1992)
doi:10.1016/0370-2693(92)90019-Z

\bibitem{QGE1}
S.~W.~Hawking, D.~N.~Page and C.~N.~Pope,
THE PROPAGATION OF PARTICLES IN SPACE-TIME FOAM,
Phys. Lett. B \textbf{86}, 175-178 (1979)
doi:10.1016/0370-2693(79)90812-8

\bibitem{QGE2}
M.~J.~Perry,
Tp Inversion in Quantum Gravity,
Phys. Rev. D \textbf{19}, 1720 (1979)
doi:10.1103/PhysRevD.19.1720

\bibitem{Planck}
N.~Aghanim \textit{et al.} [Planck],
Planck 2018 results. VI. Cosmological parameters,
Astron. Astrophys. \textbf{641}, A6 (2020)
[erratum: Astron. Astrophys. \textbf{652}, C4 (2021)]
doi:10.1051/0004-6361/201833910
[arXiv:1807.06209 [astro-ph.CO]].

\bibitem{Muong-2:2006rrc}
G.~W.~Bennett \textit{et al.} [Muon g-2],
Final Report of the Muon E821 Anomalous Magnetic Moment Measurement at BNL,
Phys. Rev. D \textbf{73}, 072003 (2006)
doi:10.1103/PhysRevD.73.072003
[arXiv:hep-ex/0602035 [hep-ex]].

\bibitem{g-2 SM EXP}
B.~Abi \textit{et al.} [Muon g-2],
Measurement of the Positive Muon Anomalous Magnetic Moment to 0.46 ppm,
Phys. Rev. Lett. \textbf{126}, no.14, 141801 (2021)
doi:10.1103/PhysRevLett.126.141801
[arXiv:2104.03281 [hep-ex]].

\bibitem{Kotov:2023wug}
A.~Kotov [Budapest-Marseille-Wuppertal],
Intermediate window observable for the muon g-2 from overlap valence quarks on staggered ensembles,
PoS \textbf{LATTICE2022}, 320 (2023)
doi:10.22323/1.430.0320

\bibitem{Toth:2022lsa}
B.~C.~Toth, S.~Borsanyi, Z.~Fodor, J.~Guenther, C.~Hoelbling, S.~D.~Katz, L.~Lellouch, T.~Lippert, K.~Miura and L.~Parato, \textit{et al.}
Muon g-2: BMW calculation of the hadronic vacuum polarization contribution,
PoS \textbf{LATTICE2021}, 005 (2022)
doi:10.22323/1.396.0005

\bibitem{BMW:2022rdk}
L.~Parato \textit{et al.} [BMW],
QED and strong isospin corrections in the hadronic vacuum polarization contribution to the anomalous magnetic moment of the muon,
PoS \textbf{LATTICE2021}, 358 (2022)
doi:10.22323/1.396.0358
[arXiv:2202.05807 [hep-lat]].

\bibitem{Budapest-Marseille-Wuppertal:2017sdk}
T.~Kawanai \textit{et al.} [Budapest-Marseille-Wuppertal],
Disconnected hadronic contribution to the muon magnetic moment at the physical point,
PoS \textbf{LATTICE2016}, 171 (2017)
doi:10.22323/1.256.0171

\bibitem{Budapest-Marseille-Wuppertal:2018ivi}
S.~Borsanyi \textit{et al.} [Budapest-Marseille-Wuppertal],
Lattice QCD results for the HVP contribution to the anomalous magnetic moments of leptons,
EPJ Web Conf. \textbf{175}, 06016 (2018)
doi:10.1051/epjconf/201817506016

\bibitem{Coughlan:1983ci}
G.~D.~Coughlan, W.~Fischler, E.~W.~Kolb, S.~Raby and G.~G.~Ross,
Cosmological Problems for the Polonyi Potential,
Phys. Lett. B \textbf{131}, 59-64 (1983)
doi:10.1016/0370-2693(83)91091-2

\bibitem{Banks:1993en}
T.~Banks, D.~B.~Kaplan and A.~E.~Nelson,
Cosmological implications of dynamical supersymmetry breaking,
Phys. Rev. D \textbf{49}, 779-787 (1994)
doi:10.1103/PhysRevD.49.779
[arXiv:hep-ph/9308292 [hep-ph]].

\bibitem{deCarlos:1993wie}
B.~de Carlos, J.~A.~Casas, F.~Quevedo and E.~Roulet,
Model independent properties and cosmological implications of the dilaton and moduli sectors of 4-d strings,
Phys. Lett. B \textbf{318}, 447-456 (1993)
doi:10.1016/0370-2693(93)91538-X
[arXiv:hep-ph/9308325 [hep-ph]].

\bibitem{Endo:2006zj}
M.~Endo, K.~Hamaguchi and F.~Takahashi,
Moduli-induced gravitino problem,
Phys. Rev. Lett. \textbf{96}, 211301 (2006)
doi:10.1103/PhysRevLett.96.211301
[arXiv:hep-ph/0602061 [hep-ph]].

\bibitem{CMM1}
L.~F.~Abbott and P.~Sikivie,
A Cosmological Bound on the Invisible Axion,
Phys. Lett. B \textbf{120}, 133-136 (1983)
doi:10.1016/0370-2693(83)90638-X

\bibitem{CMM2}
M.~Dine and W.~Fischler,
The Not So Harmless Axion,
Phys. Lett. B \textbf{120}, 137-141 (1983)
doi:10.1016/0370-2693(83)90639-1

\bibitem{CMM3}
J.~Preskill, M.~B.~Wise and F.~Wilczek,
Cosmology of the Invisible Axion,
Phys. Lett. B \textbf{120}, 127-132 (1983)
doi:10.1016/0370-2693(83)90637-8

\bibitem{KMM1}
R.~T.~Co, L.~J.~Hall and K.~Harigaya,
Axion Kinetic Misalignment Mechanism,
Phys. Rev. Lett. \textbf{124}, no.25, 251802 (2020)
doi:10.1103/PhysRevLett.124.251802
[arXiv:1910.14152 [hep-ph]].

\bibitem{KMM2}
R.~T.~Co, L.~J.~Hall, K.~Harigaya, K.~A.~Olive and S.~Verner,
Axion Kinetic Misalignment and Parametric Resonance from Inflation,
JCAP \textbf{08}, 036 (2020)
doi:10.1088/1475-7516/2020/08/036
[arXiv:2004.00629 [hep-ph]].


\bibitem{Bae:2019dgg}
K.~J.~Bae, H.~Baer, V.~Barger and D.~Sengupta,
Revisiting the SUSY $\mu$ problem and its solutions in the LHC era,
Phys. Rev. D \textbf{99}, no.11, 115027 (2019)
doi:10.1103/PhysRevD.99.115027
[arXiv:1902.10748 [hep-ph]].

\bibitem{Babu:2002ic}
K.~S.~Babu, I.~Gogoladze and K.~Wang,
Stabilizing the axion by discrete gauge symmetries,
Phys. Lett. B \textbf{560}, 214-222 (2003)
doi:10.1016/S0370-2693(03)00411-8
[arXiv:hep-ph/0212339 [hep-ph]].

\bibitem{Baer:2018avn}
H.~Baer, V.~Barger and D.~Sengupta,
Gravity safe, electroweak natural axionic solution to strong $CP$ and SUSY $\mu$ problems,
Phys. Lett. B \textbf{790}, 58-63 (2019)
doi:10.1016/j.physletb.2019.01.007
[arXiv:1810.03713 [hep-ph]].

\bibitem{S P M}
P.~N.~Bhattiprolu and S.~P.~Martin,
High-quality axions in solutions to the \ensuremath{\mu} problem,
Phys. Rev. D \textbf{104}, no.5, 055014 (2021)
doi:10.1103/PhysRevD.104.055014
[arXiv:2106.14964 [hep-ph]].

\bibitem{Georgi:1974sy}
H.~Georgi and S.~L.~Glashow,
Unity of All Elementary Particle Forces,
Phys. Rev. Lett. \textbf{32}, 438-441 (1974)
doi:10.1103/PhysRevLett.32.438

\bibitem{DW}
A.~Ernst, A.~Ringwald and C.~Tamarit,
Axion Predictions in $SO(10)\times U(1)_{\rm PQ}$ Models,
JHEP \textbf{02}, 103 (2018)
doi:10.1007/JHEP02(2018)103
[arXiv:1801.04906 [hep-ph]].

\bibitem{Choi:2022fha}
G.~Choi and T.~T.~Yanagida,
High quality axion in supersymmetric models,
JHEP \textbf{12}, 067 (2022)
doi:10.1007/JHEP12(2022)067
[arXiv:2209.09290 [hep-ph]].

\bibitem{Babu:2002tx}
K.~S.~Babu, I.~Gogoladze and K.~Wang,
Natural R parity, $\mu$-term, and fermion mass hierarchy from discrete gauge symmetries,
Nucl. Phys. B \textbf{660}, 322-342 (2003)
doi:10.1016/S0550-3213(03)00258-X
[arXiv:hep-ph/0212245 [hep-ph]].

\bibitem{charge n}
L.~M.~Krauss and F.~Wilczek,
Discrete Gauge Symmetry in Continuum Theories,
Phys. Rev. Lett. \textbf{62}, 1221 (1989)
doi:10.1103/PhysRevLett.62.1221

\bibitem{GSM}
M.~B.~Green and J.~H.~Schwarz,
Anomaly Cancellation in Supersymmetric D=10 Gauge Theory and Superstring Theory,
Phys. Lett. B \textbf{149}, 117-122 (1984)
doi:10.1016/0370-2693(84)91565-X

\bibitem{Pati:1974yy}
J.~C.~Pati and A.~Salam,
Lepton Number as the Fourth Color,
Phys. Rev. D \textbf{10}, 275-289 (1974)
[erratum: Phys. Rev. D \textbf{11}, 703-703 (1975)]
doi:10.1103/PhysRevD.10.275

\bibitem{SAA}
E.~J.~Chun and A.~Lukas,
Axino mass in supergravity models,
Phys. Lett. B \textbf{357}, 43-50 (1995)
doi:10.1016/0370-2693(95)00881-K
[arXiv:hep-ph/9503233 [hep-ph]].

\bibitem{LZ:2022lsv}
J.~Aalbers \textit{et al.} [LZ],
First Dark Matter Search Results from the LUX-ZEPLIN (LZ) Experiment,
Phys. Rev. Lett. \textbf{131}, no.4, 041002 (2023)
doi:10.1103/PhysRevLett.131.041002
[arXiv:2207.03764 [hep-ex]].

\bibitem{Iwamoto:2021aaf}
S.~Iwamoto, T.~T.~Yanagida and N.~Yokozaki,
Wino-Higgsino dark matter in MSSM from the g-2 anomaly,
Phys. Lett. B \textbf{823}, 136768 (2021)
doi:10.1016/j.physletb.2021.136768
[arXiv:2104.03223 [hep-ph]].

\bibitem{SOMMERFELD}
A.~Sommerfeld,
\"Uber die Beugung und Bremsung der Elektronen,
Annalen Phys. \textbf{403}, no.3, 257-330 (1931)
doi:10.1002/andp.19314030302

\bibitem{g-2 SM 1}
T.~Aoyama, N.~Asmussen, M.~Benayoun, J.~Bijnens, T.~Blum, M.~Bruno, I.~Caprini, C.~M.~Carloni Calame, M.~C\`e and G.~Colangelo, \textit{et al.}
The anomalous magnetic moment of the muon in the Standard Model,
Phys. Rept. \textbf{887}, 1-166 (2020)
doi:10.1016/j.physrep.2020.07.006
[arXiv:2006.04822 [hep-ph]].

\bibitem{g-2 SM 2}
T.~Blum \textit{et al.} [RBC and UKQCD],
Calculation of the hadronic vacuum polarization contribution to the muon anomalous magnetic moment,
Phys. Rev. Lett. \textbf{121}, no.2, 022003 (2018)
doi:10.1103/PhysRevLett.121.022003
[arXiv:1801.07224 [hep-lat]].

\bibitem{g-2 SM 3}
H.~Davoudiasl and W.~J.~Marciano,
Tale of two anomalies,
Phys. Rev. D \textbf{98}, no.7, 075011 (2018)
doi:10.1103/PhysRevD.98.075011
[arXiv:1806.10252 [hep-ph]].

\bibitem{Baer}
H.~Baer, V.~Barger and P.~Huang,
Hidden SUSY at the LHC: the light higgsino-world scenario and the role of a lepton collider,
JHEP \textbf{11}, 031 (2011)
doi:10.1007/JHEP11(2011)031
[arXiv:1107.5581 [hep-ph]].

\bibitem{Chan}
K.~L.~Chan, U.~Chattopadhyay and P.~Nath,
Naturalness, weak scale supersymmetry and the prospect for the observation of supersymmetry at the Tevatron and at the CERN LHC,
Phys. Rev. D \textbf{58}, 096004 (1998)
doi:10.1103/PhysRevD.58.096004
[arXiv:hep-ph/9710473 [hep-ph]].

\bibitem{BaerTata}
H.~Baer and X.~Tata,
Weak scale supersymmetry: From superfields to scattering events,
Cambridge University Press, 2006,
ISBN 978-0-521-29031-9, 978-0-511-19011-7, 978-0-521-29031-9, 978-0-521-85786-4

\bibitem{Choi}
K.~Y.~Choi and H.~M.~Lee,
Axino abundances in high-scale supersymmetry,
Phys. Dark Univ. \textbf{22}, 202-207 (2018)
doi:10.1016/j.dark.2018.11.003
[arXiv:1810.00293 [hep-ph]].

\bibitem{Porod:2011nf}
W.~Porod and F.~Staub,
SPheno 3.1: Extensions including flavour, CP-phases and models beyond the MSSM,
Comput. Phys. Commun. \textbf{183}, 2458-2469 (2012)
doi:10.1016/j.cpc.2012.05.021
[arXiv:1104.1573 [hep-ph]].

\bibitem{Porod:2003um}
W.~Porod,
SPheno, a program for calculating supersymmetric spectra, SUSY particle decays and SUSY particle production at e+ e- colliders,
Comput. Phys. Commun. \textbf{153}, 275-315 (2003)
doi:10.1016/S0010-4655(03)00222-4
[arXiv:hep-ph/0301101 [hep-ph]].

\bibitem{GM2}
P.~Athron, M.~Bach, H.~G.~Fargnoli, C.~Gnendiger, R.~Greifenhagen, J.~h.~Park, S.~Pa\ss{}ehr, D.~St\"ockinger, H.~St\"ockinger-Kim and A.~Voigt,
GM2Calc: Precise MSSM prediction for $(g - 2)$ of the muon,
Eur. Phys. J. C \textbf{76}, no.2, 62 (2016)
doi:10.1140/epjc/s10052-015-3870-2
[arXiv:1510.08071 [hep-ph]].

\bibitem{MOMG1}
G.~Belanger, F.~Boudjema, A.~Pukhov and A.~Semenov,
MicrOMEGAs: A Program for calculating the relic density in the MSSM,
Comput. Phys. Commun. \textbf{149}, 103-120 (2002)
doi:10.1016/S0010-4655(02)00596-9
[arXiv:hep-ph/0112278 [hep-ph]].

\bibitem{MOMG2}
G.~Belanger, F.~Boudjema, A.~Pukhov and A.~Semenov,
MicrOMEGAs 2.0: A Program to calculate the relic density of dark matter in a generic model,
Comput. Phys. Commun. \textbf{176}, 367-382 (2007)
doi:10.1016/j.cpc.2006.11.008
[arXiv:hep-ph/0607059 [hep-ph]].

\bibitem{MOMG3}
G.~Belanger, F.~Boudjema, A.~Pukhov and A.~Semenov,
micrOMEGAs 2.0.7: A program to calculate the relic density of dark matter in a generic model,
Comput. Phys. Commun. \textbf{177}, 894-895 (2007)
doi:10.1016/j.cpc.2007.08.002

\bibitem{MOMG4}
G.~Belanger, F.~Boudjema, A.~Pukhov and A.~Semenov,
micrOMEGAs$\_$3: A program for calculating dark matter observables,
Comput. Phys. Commun. \textbf{185}, 960-985 (2014)
doi:10.1016/j.cpc.2013.10.016
[arXiv:1305.0237 [hep-ph]].

\bibitem{Athron:2016fuq}
P.~Athron, J.~h.~Park, T.~Steudtner, D.~St\"ockinger and A.~Voigt,
Precise Higgs mass calculations in (non-)minimal supersymmetry at both high and low scales,
JHEP \textbf{01}, 079 (2017)
doi:10.1007/JHEP01(2017)079
[arXiv:1609.00371 [hep-ph]].

\bibitem{Allanach:2018fif}
B.~C.~Allanach and A.~Voigt,
Uncertainties in the Lightest $CP$ Even Higgs Boson Mass Prediction in the Minimal Supersymmetric Standard Model: Fixed Order Versus Effective Field Theory Prediction,
Eur. Phys. J. C \textbf{78}, no.7, 573 (2018)
doi:10.1140/epjc/s10052-018-6046-z
[arXiv:1804.09410 [hep-ph]].

\bibitem{Bahl:2019hmm}
H.~Bahl, S.~Heinemeyer, W.~Hollik and G.~Weiglein,
Theoretical uncertainties in the MSSM Higgs boson mass calculation,
Eur. Phys. J. C \textbf{80}, no.6, 497 (2020)
doi:10.1140/epjc/s10052-020-8079-3
[arXiv:1912.04199 [hep-ph]].

\bibitem{g-2 one}
M.~Chakraborti, S.~Heinemeyer and I.~Saha,
Improved $(g-2)_\mu$ Measurements and Supersymmetry,
Eur. Phys. J. C \textbf{80}, no.10, 984 (2020)
doi:10.1140/epjc/s10052-020-08504-8
[arXiv:2006.15157 [hep-ph]].

\bibitem{g-2 two}
M.~Chakraborti, S.~Heinemeyer and I.~Saha,
Improved ${(g-2)_\mu }$ measurements and wino/higgsino dark matter,
Eur. Phys. J. C \textbf{81}, no.12, 1069 (2021)
doi:10.1140/epjc/s10052-021-09814-1
[arXiv:2103.13403 [hep-ph]].

\bibitem{Chxx}
J.~Hisano, S.~Matsumoto, M.~M.~Nojiri and O.~Saito,
Direct detection of the Wino and Higgsino-like neutralino dark matters at one-loop level,
Phys. Rev. D \textbf{71}, 015007 (2005)
doi:10.1103/PhysRevD.71.015007
[arXiv:hep-ph/0407168 [hep-ph]].

\bibitem{ATLAS:2019lff}
G.~Aad \textit{et al.} [ATLAS],
Search for electroweak production of charginos and sleptons decaying into final states with two leptons and missing transverse momentum in $\sqrt{s}=13$ TeV $pp$ collisions using the ATLAS detector,
Eur. Phys. J. C \textbf{80}, no.2, 123 (2020)
doi:10.1140/epjc/s10052-019-7594-6
[arXiv:1908.08215 [hep-ex]].

\bibitem{ATLAS:2019lng}
G.~Aad \textit{et al.} [ATLAS],
Searches for electroweak production of supersymmetric particles with compressed mass spectra in $\sqrt{s}=$ 13 TeV $pp$ collisions with the ATLAS detector,
Phys. Rev. D \textbf{101}, no.5, 052005 (2020)
doi:10.1103/PhysRevD.101.052005
[arXiv:1911.12606 [hep-ex]].


\bibitem{gae1}
F.~Capozzi and G.~Raffelt,
Axion and neutrino bounds improved with new calibrations of the tip of the red-giant branch using geometric distance determinations,
Phys. Rev. D \textbf{102}, no.8, 083007 (2020)
doi:10.1103/PhysRevD.102.083007
[arXiv:2007.03694 [astro-ph.SR]].

\bibitem{gae2}
O.~Straniero, C.~Pallanca, E.~Dalessandro, I.~Dominguez, F.~R.~Ferraro, M.~Giannotti, A.~Mirizzi and L.~Piersanti,
The RGB tip of galactic globular clusters and the revision of the axion-electron coupling bound,
Astron. Astrophys. \textbf{644}, A166 (2020)
doi:10.1051/0004-6361/202038775
[arXiv:2010.03833 [astro-ph.SR]].

\bibitem{Kyu Jung Bae 2 one}
K.~J.~Bae, H.~Baer and E.~J.~Chun,
Mixed axion/neutralino dark matter in the SUSY DFSZ axion model,
JCAP \textbf{12}, 028 (2013)
doi:10.1088/1475-7516/2013/12/028
[arXiv:1309.5365 [hep-ph]].

\bibitem{Kyu Jung Bae 2 two}
K.~J.~Bae, H.~Baer and E.~J.~Chun,
Mainly axion cold dark matter from natural supersymmetry,
Phys. Rev. D \textbf{89}, no.3, 031701 (2014)
doi:10.1103/PhysRevD.89.031701
[arXiv:1309.0519 [hep-ph]].

\bibitem{Kyu Jung Bae 1}
K.~J.~Bae, H.~Baer and A.~Lessa,
Dark Radiation Constraints on Mixed Axion/Neutralino Dark Matter,
JCAP \textbf{04}, 041 (2013)
doi:10.1088/1475-7516/2013/04/041
[arXiv:1301.7428 [hep-ph]].

\bibitem{TDCP1}
K.~Rajagopal, M.~S.~Turner and F.~Wilczek,
Cosmological implications of axinos,
Nucl. Phys. B \textbf{358}, 447-470 (1991)
doi:10.1016/0550-3213(91)90355-2

\bibitem{TDCP2}
K.~Y.~Choi, J.~E.~Kim and L.~Roszkowski,
Review of axino dark matter,
J. Korean Phys. Soc. \textbf{63}, 1685-1695 (2013)
doi:10.3938/jkps.63.1685
[arXiv:1307.3330 [astro-ph.CO]].

\bibitem{Baer:2011hx}
H.~Baer, A.~Lessa, S.~Rajagopalan and W.~Sreethawong,
Mixed axion/neutralino cold dark matter in supersymmetric models,
JCAP \textbf{06}, 031 (2011)
doi:10.1088/1475-7516/2011/06/031
[arXiv:1103.5413 [hep-ph]].

\bibitem{BBN}
B.~D.~Fields, K.~A.~Olive, T.~H.~Yeh and C.~Young,
Big-Bang Nucleosynthesis after Planck,
JCAP \textbf{03}, 010 (2020)
[erratum: JCAP \textbf{11}, E02 (2020)]
doi:10.1088/1475-7516/2020/03/010
[arXiv:1912.01132 [astro-ph.CO]].

\bibitem{Kyu Jung Bae 3 one}
K.~J.~Bae, K.~Choi and S.~H.~Im,
Effective Interactions of Axion Supermultiplet and Thermal Production of Axino Dark Matter,
JHEP \textbf{08}, 065 (2011)
doi:10.1007/JHEP08(2011)065
[arXiv:1106.2452 [hep-ph]].

\bibitem{Kyu Jung Bae 3 two}
K.~J.~Bae, E.~J.~Chun and S.~H.~Im,
Cosmology of the DFSZ axino,
JCAP \textbf{03}, 013 (2012)
doi:10.1088/1475-7516/2012/03/013
[arXiv:1111.5962 [hep-ph]].

\bibitem{YTP}
A.~Brandenburg and F.~D.~Steffen,
Axino dark matter from thermal production,
JCAP \textbf{08}, 008 (2004)
doi:10.1088/1475-7516/2004/08/008
[arXiv:hep-ph/0405158 [hep-ph]].



\bibitem{Moroi:1999zb}
T.~Moroi and L.~Randall,
Wino cold dark matter from anomaly mediated SUSY breaking,
Nucl. Phys. B \textbf{570}, 455-472 (2000)
doi:10.1016/S0550-3213(99)00748-8
[arXiv:hep-ph/9906527 [hep-ph]].

\bibitem{Bae:2022okh}
K.~J.~Bae, H.~Baer, V.~Barger and R.~W.~Deal,
The cosmological moduli problem and naturalness,
JHEP \textbf{02}, 138 (2022)
doi:10.1007/JHEP02(2022)138
[arXiv:2201.06633 [hep-ph]].

\bibitem{Baer:2023bbn}
H.~Baer, V.~Barger and R.~Wiley Deal,
Dark matter and dark radiation from the early universe with a modulus coupled to the PQMSSM,
JHEP \textbf{06}, 083 (2023)
doi:10.1007/JHEP06(2023)083
[arXiv:2301.12546 [hep-ph]].

\bibitem{Nakamura:2006uc}
S.~Nakamura and M.~Yamaguchi,
Gravitino production from heavy moduli decay and cosmological moduli problem revived,
Phys. Lett. B \textbf{638}, 389-395 (2006)
doi:10.1016/j.physletb.2006.05.078
[arXiv:hep-ph/0602081 [hep-ph]].

\bibitem{Dine:2006ii}
M.~Dine, R.~Kitano, A.~Morisse and Y.~Shirman,
Moduli decays and gravitinos,
Phys. Rev. D \textbf{73}, 123518 (2006)
doi:10.1103/PhysRevD.73.123518
[arXiv:hep-ph/0604140 [hep-ph]].

\bibitem{Kohri:2005wn}
K.~Kohri, T.~Moroi and A.~Yotsuyanagi,
Big-bang nucleosynthesis with unstable gravitino and upper bound on the reheating temperature,
Phys. Rev. D \textbf{73}, 123511 (2006)
doi:10.1103/PhysRevD.73.123511
[arXiv:hep-ph/0507245 [hep-ph]].

\bibitem{Baer:2011uz}
H.~Baer, A.~Lessa and W.~Sreethawong,
Coupled Boltzmann calculation of mixed axion/neutralino cold dark matter production in the early universe,
JCAP \textbf{01}, 036 (2012)
doi:10.1088/1475-7516/2012/01/036
[arXiv:1110.2491 [hep-ph]].

\bibitem{Bae:2014rfa}
K.~J.~Bae, H.~Baer, A.~Lessa and H.~Serce,
Coupled Boltzmann computation of mixed axion neutralino dark matter in the SUSY DFSZ axion model,
JCAP \textbf{10}, 082 (2014)
doi:10.1088/1475-7516/2014/10/082
[arXiv:1406.4138 [hep-ph]].

\bibitem{Choi:2005ge}
K.~Choi, A.~Falkowski, H.~P.~Nilles and M.~Olechowski,
Soft supersymmetry breaking in KKLT flux compactification,
Nucl. Phys. B \textbf{718}, 113-133 (2005)
doi:10.1016/j.nuclphysb.2005.04.032
[arXiv:hep-th/0503216 [hep-th]].

\bibitem{CAST 1}
S.~Andriamonje \textit{et al.} [CAST],
An Improved limit on the axion-photon coupling from the CAST experiment,
JCAP \textbf{04}, 010 (2007)
doi:10.1088/1475-7516/2007/04/010
[arXiv:hep-ex/0702006 [hep-ex]].

\bibitem{CAST 2}
V.~Anastassopoulos \textit{et al.} [CAST],
New CAST Limit on the Axion-Photon Interaction,
Nature Phys. \textbf{13}, 584-590 (2017)
doi:10.1038/nphys4109
[arXiv:1705.02290 [hep-ex]].

\bibitem{ADMX 1}
C.~Bartram \textit{et al.} [ADMX],
Axion dark matter experiment: Run 1B analysis details,
Phys. Rev. D \textbf{103}, no.3, 032002 (2021)
doi:10.1103/PhysRevD.103.032002
[arXiv:2010.06183 [astro-ph.CO]].

\bibitem{ADMX 2}
C.~Boutan \textit{et al.} [ADMX],
Piezoelectrically Tuned Multimode Cavity Search for Axion Dark Matter,
Phys. Rev. Lett. \textbf{121}, no.26, 261302 (2018)
doi:10.1103/PhysRevLett.121.261302
[arXiv:1901.00920 [hep-ex]].

\bibitem{ADMX 3}
N.~Du \textit{et al.} [ADMX],
A Search for Invisible Axion Dark Matter with the Axion Dark Matter Experiment,
Phys. Rev. Lett. \textbf{120}, no.15, 151301 (2018)
doi:10.1103/PhysRevLett.120.151301
[arXiv:1804.05750 [hep-ex]].

\bibitem{ADMX 4}
S.~J.~Asztalos \textit{et al.} [ADMX],
A SQUID-based microwave cavity search for dark-matter axions,
Phys. Rev. Lett. \textbf{104}, 041301 (2010)
doi:10.1103/PhysRevLett.104.041301
[arXiv:0910.5914 [astro-ph.CO]].

\bibitem{CAPP}
S.~Lee, S.~Ahn, J.~Choi, B.~R.~Ko and Y.~K.~Semertzidis,
Axion Dark Matter Search around 6.7 $\mu$eV,
Phys. Rev. Lett. \textbf{124}, no.10, 101802 (2020)
doi:10.1103/PhysRevLett.124.101802
[arXiv:2001.05102 [hep-ex]].

\bibitem{RBF}
S.~De Panfilis, A.~C.~Melissinos, B.~E.~Moskowitz, J.~T.~Rogers, Y.~K.~Semertzidis, W.~Wuensch, H.~J.~Halama, A.~G.~Prodell, W.~B.~Fowler and F.~A.~Nezrick,
Limits on the Abundance and Coupling of Cosmic Axions at 4.5-Microev \ensuremath{<} m(a) \ensuremath{<} 5.0-Microev,
Phys. Rev. Lett. \textbf{59}, 839 (1987)
doi:10.1103/PhysRevLett.59.839

\bibitem{UF}
C.~Hagmann, P.~Sikivie, N.~S.~Sullivan and D.~B.~Tanner,
Results from a search for cosmic axions,
Phys. Rev. D \textbf{42}, 1297-1300 (1990)
doi:10.1103/PhysRevD.42.1297

\bibitem{HAYSTAC 1}
K.~M.~Backes \textit{et al.} [HAYSTAC],
A quantum-enhanced search for dark matter axions,
Nature \textbf{590}, no.7845, 238-242 (2021)
doi:10.1038/s41586-021-03226-7
[arXiv:2008.01853 [quant-ph]].

\bibitem{HAYSTAC 2}
L.~Zhong \textit{et al.} [HAYSTAC],
Results from phase 1 of the HAYSTAC microwave cavity axion experiment,
Phys. Rev. D \textbf{97}, no.9, 092001 (2018)
doi:10.1103/PhysRevD.97.092001
[arXiv:1803.03690 [hep-ex]].

\bibitem{QUAX}
D.~Alesini, C.~Braggio, G.~Carugno, N.~Crescini, D.~D'Agostino, D.~Di Gioacchino, R.~Di Vora, P.~Falferi, S.~Gallo and U.~Gambardella, \textit{et al.}
Galactic axions search with a superconducting resonant cavity,
Phys. Rev. D \textbf{99}, no.10, 101101 (2019)
doi:10.1103/PhysRevD.99.101101
[arXiv:1903.06547 [physics.ins-det]].

\bibitem{ORGAN}
B.~T.~McAllister, G.~Flower, J.~Kruger, E.~N.~Ivanov, M.~Goryachev, J.~Bourhill and M.~E.~Tobar,
The ORGAN Experiment: An axion haloscope above 15 GHz,
Phys. Dark Univ. \textbf{18}, 67-72 (2017)
doi:10.1016/j.dark.2017.09.010
[arXiv:1706.00209 [physics.ins-det]].

\bibitem{IAXO PRO}
I.~Shilon, A.~Dudarev, H.~Silva and H.~H.~J.~ten Kate,
Conceptual Design of a New Large Superconducting Toroid for IAXO, the New International AXion Observatory,
IEEE Trans. Appl. Supercond. \textbf{23}, no.3, 4500604 (2013)
doi:10.1109/TASC.2013.2251052
[arXiv:1212.4633 [physics.ins-det]].

\bibitem{ADMX PRO}
I.~Stern,
ADMX Status,
PoS \textbf{ICHEP2016}, 198 (2016)
doi:10.22323/1.282.0198
[arXiv:1612.08296 [physics.ins-det]].

\bibitem{KLASH PRO}
D.~Alesini, D.~Babusci, D.~Di Gioacchino, C.~Gatti, G.~Lamanna and C.~Ligi,
The KLASH Proposal,
[arXiv:1707.06010 [physics.ins-det]].

\bibitem{MADMAX PRO}
S.~Beurthey, N.~B\"ohmer, P.~Brun, A.~Caldwell, L.~Chevalier, C.~Diaconu, G.~Dvali, P.~Freire, E.~Garutti and C.~Gooch, \textit{et al.}
MADMAX Status Report,
[arXiv:2003.10894 [physics.ins-det]].

\bibitem{Plasma Haloscopes PRO}
M.~Lawson, A.~J.~Millar, M.~Pancaldi, E.~Vitagliano and F.~Wilczek,
Tunable axion plasma haloscopes,
Phys. Rev. Lett. \textbf{123}, no.14, 141802 (2019)
doi:10.1103/PhysRevLett.123.141802
[arXiv:1904.11872 [hep-ph]].

\bibitem{TOORAD PRO}
J.~Sch\"utte-Engel, D.~J.~E.~Marsh, A.~J.~Millar, A.~Sekine, F.~Chadha-Day, S.~Hoof, M.~N.~Ali, K.~C.~Fong, E.~Hardy and L.~\v{S}mejkal,
Axion quasiparticles for axion dark matter detection,
JCAP \textbf{08}, 066 (2021)
doi:10.1088/1475-7516/2021/08/066
[arXiv:2102.05366 [hep-ph]].

\bibitem{Axion Data}C.O'Hare, cajohare/AxionLimits: AxionLimits (Version v1.0). Zenodo (2020). https://doi.org/10.5281/zenodo.3932430.

\bibitem{SUSY DFSZ-I}
K.~J.~Bae, H.~Baer and H.~Serce,
Prospects for axion detection in natural SUSY with mixed axion-higgsino dark matter: back to invisible?,
JCAP \textbf{06}, 024 (2017)
doi:10.1088/1475-7516/2017/06/024
[arXiv:1705.01134 [hep-ph]].

\bibitem{diga}
D.~J.~Gross, R.~D.~Pisarski and L.~G.~Yaffe,
QCD and Instantons at Finite Temperature,
Rev. Mod. Phys. \textbf{53}, 43 (1981)
doi:10.1103/RevModPhys.53.43

\bibitem{LMDQCD}
A.~Bazavov, T.~Bhattacharya, M.~Cheng, C.~DeTar, H.~T.~Ding, S.~Gottlieb, R.~Gupta, P.~Hegde, U.~M.~Heller and F.~Karsch, \textit{et al.}
The chiral and deconfinement aspects of the QCD transition,
Phys. Rev. D \textbf{85}, 054503 (2012)
doi:10.1103/PhysRevD.85.054503
[arXiv:1111.1710 [hep-lat]].

\bibitem{Ftheta}
L.~Visinelli and P.~Gondolo,
Dark Matter Axions Revisited,
Phys. Rev. D \textbf{80}, 035024 (2009)
doi:10.1103/PhysRevD.80.035024
[arXiv:0903.4377 [astro-ph.CO]].
\end{thebibliography}
\end{document}